\documentclass[12pt]{article} 
\pdfoutput=1
\usepackage[utf8]{inputenc}
\usepackage[T1]{fontenc}
\usepackage{lmodern, amsmath, amssymb, slashed, graphicx, comment,adjustbox}
\usepackage[dvipsnames]{xcolor}
\usepackage[numbers, elide, compress]{natbib}
\usepackage[colorlinks=true, citecolor=blue, urlcolor=blue, linkcolor=blue, breaklinks=true, pdfpagelabels=false]{hyperref}

\makeatletter\g@addto@macro\bfseries{\boldmath}\makeatother

\def\figureautorefname~#1\null{Fig.\,#1\null}

\def\equationautorefname~#1\null{Eq.\,(#1)\null}

\numberwithin{equation}{section}

\allowdisplaybreaks

\DeclareMathOperator{\br}{BR}
\DeclareMathOperator{\sign}{sgn}
\let\d\undefined
\DeclareMathOperator{\d}{d}
\DeclareMathOperator{\GDP}{GDP}

\newcommand{\inab}{\,{\rm ab}^{-1}}
\newcommand{\infb}{\,{\rm fb}^{-1}}

\newcommand{\ee}{e^+e^-}
\newcommand{\eehz}{e^+e^- \to hZ}
\newcommand{\eevvh}{e^+e^- \to \nu \bar{\nu} h}
\newcommand{\eeww}{e^+e^- \to WW}
\newcommand{\eehzvvh}{e^+e^- \to hZ / \nu \bar{\nu} h}
\newcommand{\vvh}{\nu \bar{\nu} h}

\newcommand{\La}{\mathcal{L} }

\newcommand{\eetth}{e^+e^- \to t \bar{t} h}
\newcommand{\tth}{t \bar{t} h}

\newcommand{\bpm}{\begin{pmatrix}}
\newcommand{\epm}{\end{pmatrix}}

\newcommand{\hl}[1]{{\color{BlueGreen} #1}}

\begin{document}

\begin{flushright}
DESY 17-018
\end{flushright}

\vspace*{1.5cm}

\begin{center}

{\Large\bf
The leptonic future of the Higgs
\par}
\vspace{9mm}

{\bf Gauthier~Durieux,$^{a}$ Christophe~Grojean,$^{a,b}$ \footnote{On leave from Instituci\'o Catalana de Recerca i Estudis Avan\c cats, 08010 Barcelona, Spain} Jiayin~Gu,$^{a,c}$ Kechen~Wang$\, ^{a,c}$}\\ [4mm]
{\small\it
$^a$ DESY, Notkestra{\ss}e 85, D-22607 Hamburg, Germany \\[2mm]
$^b$ Institut f\"ur Physik, Humboldt-Universit\"at zu Berlin, D-12489 Berlin, Germany  \\[2mm]
$^c$ Center for Future High Energy Physics, Institute of High Energy Physics, \\ Chinese Academy of Sciences, Beijing 100049, China
\par}
\vspace{.5cm}
\centerline{\tt \small gauthier.durieux@desy.de, christophe.grojean@desy.de, jiayin.gu@desy.de, kechen.wang@desy.de}

\end{center}

\begin{abstract}
Precision study of electroweak symmetry breaking strongly motivates the
construction of a lepton collider with center-of-mass energy of at least
$240$\,GeV. Besides Higgsstrahlung ($\eehz$), such a collider would measure weak
boson pair production ($\eeww$) with an astonishing precision. The
weak-boson-fusion production process ($\eevvh$) provides an increasingly
powerful handle at higher center-of-mass energies. High energies also benefit
the associated top-Higgs production ($e^+e^-\to t\bar th$) that is crucial to
constrain directly the top Yukawa coupling. The impact and complementarity of
differential measurements, at different center-of-mass energies and for several
beam polarization configurations, are studied in a global effective-field-theory
framework. We define a \emph{global determinant parameter} (GDP) which
characterizes the overall strengthening of constraints independently of the
choice of operator basis. The reach of the CEPC, CLIC, FCC-ee, and ILC designs
is assessed.
\end{abstract}

\newpage
{\small 
\tableofcontents}

\setcounter{footnote}{0}

\pagebreak
\section{Introduction}

With the discovery of a scalar whose properties are compatible with that of the
standard model (SM) Higgs boson, the first run of the LHC has found an essential
ingredient for the deep understanding of matter and has revealed a fascinating
and complex structure of the vacuum. As its second run is proceeding at an
increased center-of-mass energy, no unambiguous sign of new physics (NP) has
been found yet. Direct exploration of this energy frontier will continue for a
couple of decades but a detailed understanding of electroweak symmetry breaking
and the indirect search for NP via precision measurement would benefit from the
cleaner environment of a lepton collider. An $e^+e^-$ machine running at a
center-of-mass energy of $240$--$250$\,GeV, close to the maximum of the $\eehz$
Higgsstrahlung cross section would indeed determine the Higgs couplings with
exquisite precision. Several proposals of such \emph{Higgs factories} have been
made, including the Circular Electron Positron Collider (CEPC) in
China~\cite{CEPC-SPPCStudyGroup:2015csa}, the Future Circular Collider with
$\ee$ (FCC-ee) at CERN, previously known as TLEP~\cite{Gomez-Ceballos:2013zzn},
and the International Linear Collider (ILC) in Japan~\cite{Baer:2013cma}. The
Compact Linear Collider (CLIC) at CERN~\cite{CLIC:2016zwp} could also run at
higher center-of-mass energies. The Higgs coupling measurements have been widely
studied in the corresponding design studies through global fits in the so-called
\emph{kappa} framework~\cite{LHCHiggsCrossSectionWorkingGroup:2012nn}.

As new physics is being constrained to lie further and further above the
electroweak scale, the description of its effects at future lepton colliders
seems to fall in a low-energy regime. Effective field theories (EFTs) therefore
look like prime exploration tools~\cite{Buchmuller:1985jz, Giudice:2007fh,
Grzadkowski:2010es, Contino:2013kra, Henning:2014wua, deFlorian:2016spz}. Given
that the parity of an operator dimension is that of $(\Delta B-\Delta
L)/2$~\cite{Kobach:2016ami}, all operators conserving baryon and lepton numbers
are of even dimension:
\begin{equation}
\mathcal{L}_{\rm EFT} =
	\mathcal{L}_{\rm SM} + 
	\sum_i \frac{c^{(6)}_i}{\Lambda^2} \mathcal{O}^{(6)}_i + 
	\sum_j \frac{c^{(8)}_j}{\Lambda^4} \mathcal{O}^{(8)}_j + 		\cdots
\end{equation}
where $\Lambda$ is a mass scale and $c_i^{(d)}$ are the dimensionless
coefficients of the $\mathcal{O}^{(d)}_i$ operators of canonical dimension $d$.
The standard-model effective field theory (SMEFT) allows for a systematic
exploration of the theory space in direct vicinity of the standard model,
encoding established symmetry principles. As a genuine quantum field theory,
its predictions are also perturbatively improvable. It therefore relies on much 
firmer theoretical bases than the \emph{kappa}
framework. While very helpful in
illustrating the precision reach of Higgs measurements, the latter can in
particular miss interactions of Lorentz structure different from that of the
standard model, or correlations deriving from gauge invariance, notably between
Higgs couplings to different gauge bosons.

Many effective-field-theory studies have been performed, for Higgs measurements
at LHC~\cite{Elias-Miro:2013mua, Pomarol:2013zra, Ellis:2014dva,
Falkowski:2015fla, Butter:2016cvz}, electroweak (EW) precision observables at
LEP~\cite{Han:2004az, Ciuchini:2013pca, Ciuchini:2014dea, Falkowski:2014tna,
Efrati:2015eaa}, diboson measurements at both LEP~\cite{Falkowski:2015jaa} and
LHC~\cite{Falkowski:2016cxu, Zhang:2016zsp}, or the combination of 
measurements in several sectors~\cite{Ellis:2014jta, Berthier:2015gja}.
Among the studies performed in the context of future Higgs
factories~\cite{Craig:2014una, Beneke:2014sba, Henning:2014gca, Craig:2015wwr,
Ellis:2015sca, Ge:2016zro, deBlas:2016ojx, Ellis:2017kfi, Khanpour:2017cfq}, many
estimated constraints on individual dimension-six operators. A challenge related
to the consistent use of the EFT framework is indeed the simultaneous inclusion
of all operators up to a given dimension. It is required for this approach to
retain its power and generality. As a result, various observables have to be
combined to constrain efficiently all directions of the multidimensional space
of effective-operator coefficients. The first few measurements included
bring the more significant improvements by lifting large approximate degeneracies.
Besides Higgsstrahlung production and decay rates in different channels, angular
distributions contain additional valuable
information~\cite{Beneke:2014sba,Craig:2015wwr}. Our knowledge about
differential distributions could also be exploited more extensively through
statistically optimal observables~\cite{Atwood:1991ka, Diehl:1993br}.
Higgs production through weak-boson fusion provides complementary information of
increasing relevance at higher center-of-mass energies. Direct constraints on
the top Yukawa coupling can moreover only be obtained through Higgs production
in association with a pair of tops. Measurements at
$\sqrt{s}=350$\,GeV and above can thus be very helpful. As the sensitivities to
operator coefficients can vary with $\sqrt{s}$, these higher-energy runs would
also constrain different directions of the parameter space and therefore resolve
degeneracies. Beam polarization, more easily implemented at linear colliders,
could be similarly helpful. Finally, the Higgs and anomalous triple gauge
couplings (aTGCs) are related in a gauge-invariant EFT, and a subset of
operators relevant for Higgs physics can be efficiently bounded through diboson
production $\eeww$~\cite{Butter:2016cvz, Falkowski:2015jaa}.

We parametrize deviations from the standard-model in the processes enumerated
above through dimension-six operators, in the so-called Higgs
basis~\cite{Falkowski:2001958}. Translation to other bases is however
straightforward. Our assumption of perfectly standard-model-like electroweak
precision measurements is more easily implemented in that framework. No
deviation in the gauge-boson couplings to fermions, or $W$ mass is permitted.
Given the poor sensitivity expected for the Yukawa couplings of lighter fermion,
we only allow for modifications of the (flavor-conserving) muon, tau, charm,
bottom, and top ones. Neither CP-violating, nor fermion dipole operators are
considered. The potential impact of these assumptions on our results is
carefully discussed in the text. A global effective-field-theory analysis is
then performed, in a twelve-dimensional parameter space, assuming that
measurements coincide with their SM predictions. Prospects for the different
machines are discussed in view of their respective design and run plan.

The rest of this paper is organized as follows. In \autoref{sec:eft}, we lay
down the EFT framework used. In \autoref{sec:measurement}, we detail the
observables included in our study. The results of the global fits are shown in
\autoref{sec:results}. The reach of the different colliders is summarized in
\autoref{fig:fit0}. Our conclusions are drawn in \autoref{sec:con}.
Further details are provided in the appendix. We define our twelve
effective-field-theory parameters and provide their expressions in the SILH'
basis in \autoref{app:basis}. Additional information about the measurement
inputs is provided in \autoref{app:input}. Supplementary figures and results are
available in \autoref{app:more}. In \autoref{app:express}, we provide numerical
expressions for the observables used in terms of our twelve effective-field-theory
parameters. Finally, the numerical results of the global fits are tabulated in
\autoref{app:rho}. They could be used to set limits on specific models while
accounting for the correlations in the full twelve-dimensional parameter space.


\section{Effective-field-theory framework}
\label{sec:eft}

A global effective-field-theory treatment of any process requires to consider
simultaneously all contributing operators appearing in a complete basis, up to a
given dimension. Assuming baryon and lepton number conservations, we restrict
ourselves to dimension-six operators. As mentioned in the introduction, we would
like to model the following processes:
\begin{itemize} 
\item Higgsstrahlung production: $\eehz$ (rates and distributions),
\par followed by Higgs decays in various channels,
\item Higgs production through weak-boson fusion: $\eevvh$,
\item Higgs production in association with top quarks: $\eetth$,
\item weak-boson pair production: $\eeww$ (rate and distributions).
\end{itemize}
Several combinations of operators affecting these processes are however well
constrained by other measurements. As discussed in \autoref{sec:ww},
electroweak precision observables could be constrained to a sufficient level,
although this remains to be established explicitly.
At leading order, CP-violating operators give no linear contribution to the
Higgs rates but could manifest themselves in angular
asymmetries~\cite{Beneke:2014sba, Craig:2015wwr}. They could moreover be well
constrained by dedicated searches. Under restrictive assumptions, indirect
constraints arising from EDM experiments~\cite{Barr:1990vd, Fan:2013qn,
Baron:2013eja} for instance render Higgs CP-violating asymmetries inaccessible
at future colliders~\cite{Craig:2015wwr}, even thought some room may be left in the CP violating Yukawa of the charm and bottom quarks for which the direct and indirect bounds are not that restrictive~\cite{Chien:2015xha}. 
It is also possible for CP violating Yukawa couplings of heavy flavor leptons to evade the constraints from EDM experiments which could be probed in Higgs decays~\cite{Harnik:2013aja}.
As a first working hypothesis, we thus
assume electroweak and CP-violating observables are perfectly constrained to be
standard-model like.

Throughout this paper, we only retain the interferences of
effective-field-theory amplitudes with standard-model ones. The squares of
amplitudes featuring a dimension-six operator insertion are discarded. They are
formally of the same $c^2/\Lambda^4$ order as the interferences of
dimension-eight operators with standard-model amplitudes. The relative
importance of these two kinds of $c^2/\Lambda^4$ contributions can however not
be determined without assuming a definite power counting or referring to a
specific model. Nevertheless, thanks to the high precision to which most
observables are measured at lepton colliders that collect large amount of
integrated luminosity in clean environments, we generically expect the
discarded terms to have small impact on our results. The percent-level
measurement of an observable of schematic
\begin{equation*}
	\frac{O}{O_\text{SM}} = 1
	+ \mathcal{O}(1) \frac{cE^2}{\Lambda^2}
	+ \mathcal{O}(1) \left(\frac{cE^2}{\Lambda^2}\right)^2
\end{equation*}
effective-field-theory dependence (where $E$ is a typical energy scale) will for
instance constrain $c\,E^2/\Lambda^2$ at the percent level. The quadratic
term then only induces a relative percent-level correction to this limit. In
specific cases, the interference of dimension-six operators with standard-model
amplitudes can however suffer accidental suppressions. This could invalidate the
na\"ive hierarchy above between linear and quadratic terms. Helicity selection
rules~\cite{Azatov:2016sqh} can for instance cause significant suppressions of
the linear contribution compared to the quadratic one, at energies higher than
electroweak mass scales. If the standard model and dimension-six operators give
rise to amplitudes with electroweak bosons of different helicities, their
interference is expected to scale as $c\,m_V^2/\Lambda^2$. A measurement
of $O/O_\text{SM}$ with precision $x$ would still imply a limit of order $x$ on
$cm_V^2/\Lambda^2$ at low energies but this bound would receive
corrections scaling as $xE^4/m_V^4$ for increasing $E$. Given $m_V$ of
order $100$\,GeV, only measurements of $10^{-2}$, $10^{-3}$, $10^{-4}$,
$10^{-5}$ and $10^{-6}$ precisions at least are roughly expected to be dominated
by linear effective-field-theory contributions at $250$, $500$, $1000$, $1400$
and $3000$\,GeV energies, respectively. We will comment further on accidental
suppressions and on their possible impact on our results in
\autoref{sec:results}. Light fermion dipole operators also have interferences
with standard-model amplitudes that suffer drastic mass suppressions. As a
consequence, their dominant effects arise at the $c^2/\Lambda^4$ level. We
however leave the study of this family of operators for future work.

Under the above assumptions, together with flavor universality, it was shown
that there are $10$ independent combinations of operators that contribute to
Higgs (excluding its self coupling) and TGC
measurements~\cite{Elias-Miro:2013mua, Pomarol:2013zra, Falkowski:2015fla,
Falkowski:2015jaa}.\footnote{Refs.~\cite{Ellis:2014jta, Ellis:2015sca,
Ellis:2017kfi} additionally set lepton and down-type Yukawa couplings equal
while Ref.~\cite{Butter:2016cvz} focuses on third-generation fermions instead of
assuming flavor universality.} We however lift the flavor universality
requirement and treat separately the top, charm, bottom, tau, and muon Yukawa
couplings. No flavor violation is allowed and we refer to
Refs.~\cite{Goertz:2014qia, Altmannshofer:2015qra, Kagan:2014ila} for studies of
the possible means to probe the light-fermion Yukawas at present and future
experiments. In total, $12$ degrees of freedom are thus considered. While all
non-redundant basis are equivalent, we find the Higgs
basis~\cite{Falkowski:2001958} particularly convenient. It is defined in the
broken electroweak phase and therefore closely related to experimental
observables. Distinguishing the operators contributing to electroweak precision
measurements from the ones of Higgs and TGC measurements is also straightforward
in this basis. The parameters we use are:
\begin{equation}
	\delta c_Z		\,,~~
	c_{ZZ}			\,,~~
	c_{Z\square}		\,,~~
	c_{\gamma\gamma}	\,,~~
	c_{Z\gamma}		\,,~~
	c_{gg}			\,,~~
	\delta y_t		\,,~~
	\delta y_c		\,,~~
	\delta y_b		\,,~~
	\delta y_\tau		\,,~~
	\delta y_\mu		\,,~~
	\lambda_Z		\,.
\label{eq:para10}
\end{equation}
Their exact definitions as well as a correspondence map to the SILH' basis of
gauge-invariant dimension-six operators can be found in \autoref{app:basis}. The
numerical expressions of the various observables we use as functions of these
parameters are given in \autoref{app:express}.

Compared to the widely-used \emph{kappa} framework, an important feature of this
effective field theory is the appearance of Higgs couplings with Lorentz
structures differing from SM ones. In addition to $\delta c_Z\: hZ_\mu
Z^\mu$ which modifies an existing SM coupling, the $c_{ZZ}\;h Z_{\mu\nu}
Z^{\mu\nu}$ and $c_{Z\square}\; h Z_\mu \partial_\nu Z^{\mu\nu}$ interactions
are for instance also generated by gauge-invariant dimension-six operators. The $\eehz$ rate,
at a given center-of-mass energy and for a fixed beam polarization, depends on
one combination of these parameters. Runs at various energies, with different
beam polarizations, as well as additional measurements are therefore crucial to
constrain all other orthogonal directions. Measurements at higher center-of-mass
energies have an enhanced sensitivity to $c_{ZZ}$ and $c_{Z\square}$. Angular
asymmetries in $\eehz$, weak-boson-fusion production rate, weak-boson pair
production, or the $h\to ZZ^*$ and $h\to WW^*$ decays, each play a role. The
measurement of the $h\to Z\gamma$ decay is crucial too. The $c_{Z\gamma}$
coupling which contributes to the Higgsstrahlung process otherwise remains
loosely constrained and weakens the whole fit.

The treatment of the $h\to gg$, $\gamma\gamma$, and $Z\gamma$ decays requires
some special attention. Given that they are loop-level generated in the standard
model, one may wish to include their loop-level dependence in effective
parameters like $\delta y_t$, $\delta y_b$, $\delta c_W$ which rescale
standard-model interactions, or $c_{ZZ}$, $c_{Z\square}$, etc. which do not.
Complete effective-field-theory results at that order are however not currently
available for the above processes (see Ref.~\cite{Hartmann:2015aia} for the
treatment of $h\to \gamma\gamma$). The computation of next-to-leading-order
effective-field-theory contributions to processes that are not loop-level
generated in the standard model would also be needed to ensure a consistent
global treatment. Misleading results can otherwise be obtained. Let us
illustrate this point with the dependence of the $h\to \gamma\gamma$ partial
width on $c_{\gamma\gamma}$ and $\delta y_t$, at tree- and loop-level,
respectively. The Higgsstrahlung, weak-boson fusion, and weak-boson pair
production processes also depend at tree level on $c_{\gamma\gamma}$ and receive
loop corrections proportional to $\delta y_t$. A combination of these two
parameters similar to the one entering in the $h\to\gamma\gamma$ partial width
may moreover be expected. Including the dependence of this partial width on
$\delta y_t$, but not that of the $\eehz$, $\eevvh$, and $\eeww$ cross sections,
one would artificially render their constraints orthogonal. Tight bounds on
$\delta y_t$ would then be obtained. Consistently including all one-loop
dependences on these parameters might however still leave a combination of
$c_{\gamma\gamma}$ and $\delta y_t$ at least nearly unconstrained. To avoid such
a pitfall, we choose not to include any loop-level dependence on
effective-field-theory parameters in the $h\to \gamma\gamma$ and $Z\gamma$
partial widths. Once direct constraints on the top Yukawa coupling (from the LHC
or from $e^+e^-\to t\bar th$) are included, we however checked that including
the whole loop dependence of the $h\to \gamma\gamma$ branching fraction has only
marginal effects on our results.\footnote{We used the numerical expressions
derived from the results of Ref.~\cite{Hartmann:2015aia} in the appendix of
Ref.~\cite{DiVita:2017eyz}.} For our purpose, it is on the contrary safe to
account for the loop-level $\delta y_t$ and $\delta y_b$ dependences of the
$h\to gg$ partial width. It remains to be examined whether the loop-level
dependence on $\delta y_t$ in processes measured at lepton collider, below the
$t\bar th$ threshold, could serve to improve on the high-luminosity LHC
constraints. A similar question, asked for the trilinear Higgs
coupling~\cite{McCullough:2013rea} could be further investigated.

Absorbing also, for convenience, a standard-model normalization factor into
barred effective parameters, we thus obtain:
\begin{equation}
\frac{\Gamma_{\gamma\gamma}}{\Gamma^{\rm SM}_{\gamma\gamma}} 
	\simeq  1-2 \bar{c}_{\gamma\gamma}
	\,,  \hspace{1cm}
\frac{\Gamma_{Z\gamma}}{\Gamma^{\rm SM}_{Z\gamma}}
	\simeq 1-2 \bar{c}_{Z\gamma}
	\,,
\label{eq:barcvv}
\end{equation} 
and
\begin{equation}
\frac{\Gamma_{gg}}{\Gamma^{\rm SM}_{gg}} 
	~\simeq~
	1 + 2\bar{c}^{\rm \,eff}_{gg}
	~\simeq~
	1+ 2 \, \bar{c}_{gg} + 2.10 \, \delta y_t -0.10 \, \delta y_b
	\,,
\label{eq:barcgg}
\end{equation}
at the linear order.
Compared to the standard Higgs-basis effective parameters, our normalization is the following:
\begin{equation}
\bar{c}_{\gamma\gamma}	\simeq \frac{c_{\gamma\gamma}}{8.3\times 10^{-2}} \,, \hspace{1cm} 
\bar{c}_{Z\gamma}	\simeq \frac{c_{Z\gamma}}{5.9\times 10^{-2}} \,, \hspace{1cm}  
\bar{c}_{gg}		\simeq \frac{c_{gg}}{8.3\times 10^{-3}} \,.  \label{eq:cnorm}
\end{equation}
We will sometimes display results in terms of the $\bar{c}^{\rm \,eff}_{gg}$
parameter that is directly probed by the $h\to gg$ branching fraction. It is
particularly informative to do so when $c_{gg}$ and $\delta y_t$
are only poorly constrained individually.

Measurement of the $h\to ZZ^*$ rate relies on its fermionic decay products and
has some sensitivity on $c_{\gamma\gamma}$ and $c_{Z\gamma}$, in addition to
$\delta c_Z$, $c_{ZZ}$ and $c_{Z\square}$. Higgs decays to off-shell photons can
indeed produce the same final state. Each fermionic decay channel actually has a
somewhat different sensitivity which depends strongly on the invariant mass of
fermion pairs. Loosened cuts would provide increased sensitivities to
$c_{\gamma\gamma}$ and $c_{Z\gamma}$~\cite{Chen:2015iha}.\footnote{See also
Ref.~\cite{Boselli:2017pef} for a recent EFT study of the Higgs decay into four
charged leptons exploiting both the rates and kinematic distributions.}
For simplicity, we however neglect the contributions of those two
effective-field-theory parameters to $h\to ZZ^*$. Standard invariant mass cuts
together with the constraints on $c_{\gamma\gamma}$ and $c_{Z\gamma}$ arising
from the direct measurements of $h\to Z\gamma$ and $h \to \gamma\gamma$ decays
should be sufficient to limit the impact of this approximation on our results.

The standard-model effective field theory we use specifically assumes the
absence of new states below the electroweak scale. It does therefore not account
for possible invisible decays of the Higgs. The corresponding branching fraction
would nevertheless be significantly constrained at future lepton colliders. An
integrated luminosity of $5\inab$ collected at $240$\,GeV would for instance
bound $\sigma(hZ)\times{\rm BR}(h\to \,{\rm inv})$ to be smaller than $0.28\%$
of $\sigma(hZ)$ at the $95\%$\,CL~\cite{CEPC-SPPCStudyGroup:2015csa}. Other
exotic Higgs decays not modeled in a SMEFT framework would also be constrained
very well at future lepton colliders~\cite{Liu:2016zki}. We do therefore not
expect an effective field theory modified to include such decays to lead to
results widely different from the ones we obtain.


\section{Measurements and fit}
\label{sec:measurement}

To the best of our knowledge, the most updated run plans of each machine are the
following:
\begin{itemize}

\item According to its preCDR, the CEPC would collect $5\inab$ of integrated
luminosity at $240$\,GeV. Recently, the reference circumference of its tunnel
has been fixed to 100\,km~\cite{CEPCcern}. A run at 350\,GeV could therefore be
envisioned. The luminosity to expect at that center-of-mass energy however
depends on the machine design and is currently unknown. To study the impact of
the measurements at 350\,GeV, we take a conservative benchmark value of
$200\infb$ and explore a larger range in \autoref{sec:results}.

\item The CDR of the FCC-ee project is expected by the year 2018~\cite{fccplan}
and will supersede the TLEP white paper~\cite{Gomez-Ceballos:2013zzn} that still
contains the most recent results on Higgs physics. The latter document, we rely
on, assumes that $10\inab$ of data would be collected at $240$\,GeV and
$2.6\inab$ at $350$\,GeV.

\item Recent ILC documents suggest that, with a luminosity upgrade, it could
collect $2\inab$ at 250\,GeV, $200\infb$ at 350\,GeV, and $4\inab$ at 500\,GeV
\cite{Fujii:2015jha, Barklow:2015tja}. This significantly extends the plans
presented in its TDR~\cite{Baer:2013cma}. The updated estimations are adopted in
our study. The ILC could also run with longitudinally polarized beams. We follow
Refs.~\cite{Baer:2013cma, Barklow:2015tja} and assume that a maximum
polarization of $\pm80\%$ ($\pm30\%$) can be achieved for the incoming electron
(positron). While collecting $1\inab$ of integrated luminosity at a
center-of-mass energy of $1$\,TeV, with $P(e^-, e^+) = (-0.8,+0.2)$
polarization, is also considered in the TDR~\cite{Baer:2013cma}, we follow
Ref.~\cite{Fujii:2015jha, Barklow:2015tja} and do not take such a run into
account. Nevertheless, results including the 1\,TeV measurements of precision
quoted in Ref.~\cite{Asner:2013psa} are shown in \autoref{app:more}.

\item Recent Ref.~\cite{CLIC:2016zwp} proposed that CLIC would collect
$100\infb$ at the top threshold, $500\infb$ at $380$\,GeV, $1.5\inab$ at
$1.5$\,TeV, and $3\inab$ at $3$\,TeV. The more specific study of Higgs
measurements of Ref.~\cite{Abramowicz:2016zbo} however assumed $500\infb$ at
$350\,$GeV, $1.5\inab$ at $1.4$\,TeV and $2\inab$ at $3$\,TeV. We follow the
latter plan in order to make use of its estimations. While the implementation of
beam polarization is also likely at CLIC, we follow again
Ref.~\cite{Abramowicz:2016zbo} and assume unpolarized beams.

\end{itemize}

In the rest of this section, we summarize the important aspects of each of the
measurements we take into account. We detail the assumptions made in the many
cases where necessary information is not provided in the literature. The
numerical inputs we use are given in \autoref{app:input}.

\subsection{Higgsstrahlung production}
\subsubsection*{Rate measurements}
\label{sec:hzrate}

\begin{figure}[htb]
\centering
\includegraphics[height=3cm]{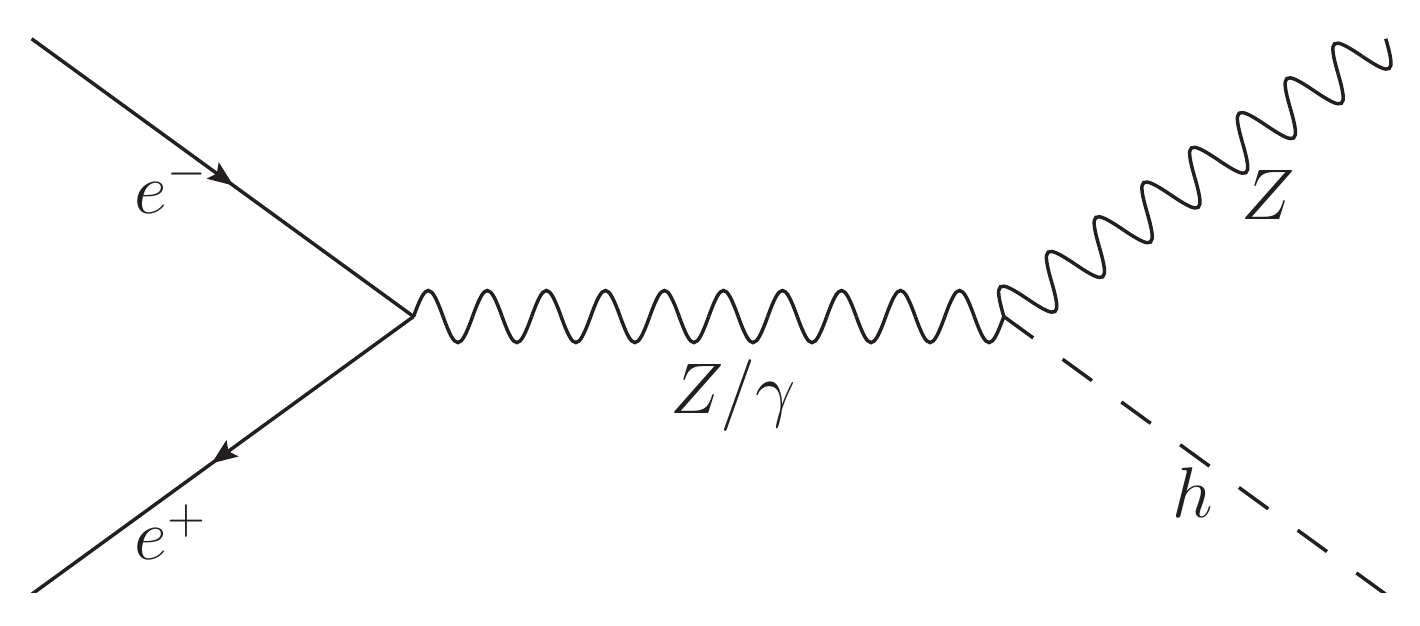}
\caption{Leading-order contribution to the Higgsstrahlung process, $\eehz$.}
\label{fig:feynhz}
\end{figure}

The Higgsstrahlung process (see \autoref{fig:feynhz}) dominates the Higgs
production modes at lepton colliders below center-of-mass energies of about
$450$\,GeV where weak-boson fusion takes over. Its cross section is maximized
around $250$\,GeV but bremsstrahlung makes it more advantageous for circular
colliders to run at $240$\,GeV. At this energy, an integrated luminosity of
$5\inab$ would yield about $1.06\times 10^6$ Higgses. At $250$\,GeV, $2\inab$ of
data collected with $P(e^-,e^+)=(-0.8,+0.3)$ beam polarization would contain
approximatively $6.4\times10^5$ Higgses. The latter polarization configuration
maximizes the $\eehz$ cross section. The recoil mass of the $Z$ gives access to
the inclusive $\eehz$ rate independently of the exclusive Higgs decay channels
measurements. The Higgsstrahlung process can also be measured at higher center-of-mass energies. Despite the smaller cross sections, this allows to probe
different combinations of EFT parameters and is thus helpful for resolving
(approximate) degeneracies among them.
The estimated measurement precisions at each collider and at different energies
are shown in \autoref{tab:higgsinputc}, \ref{tab:higgsinputi} and
\ref{tab:higgsinputclic} of \autoref{app:input}, where further details
are also provided.

A few important comments are in order. As mentioned in \autoref{sec:eft}, the
measurement of the rare $h\to Z\gamma$ decay, while not very constraining for
the SM $hZ\gamma$ coupling, could be very important to resolve the degeneracies
of EFT parameters in the production processes. Therefore, while the estimation
of this measurement is not available for the FCC-ee and ILC, we scale the
precision estimated for the CEPC, assuming the dominance of statistical
uncertainties. Some care must also be taken to avoid potential double counting
between the $\eehz, ~Z\to \nu\bar{\nu}, ~h\to b\bar{b}$ process and the
weak-boson fusion $\eevvh, ~h\to b\bar{b}$, which yield the same final state.
This is further discussed in \autoref{sec:vvh} and \autoref{app:input}. Note
also that the interferences between $s$-channel $Z$ and photon amplitudes are
accidentally suppressed by a factor of $1-4\sin^2\theta_W\simeq 0.06$ in the
total unpolarized cross section. This factor arises from the sum of the
left- and right-handed couplings of the electron to the $Z$,
$\frac{e}{2s_Wc_W}(-1+2s_W^2)$ and $\frac{e}{2s_Wc_W}(2s_W^2)$, respectively.
Beam polarization thus significantly affects the sensitivity of the
Higgsstrahlung rate to operators contributing to the $hZ\gamma$
vertex.\footnote{We thank Michael Peskin for helping us understand this
interesting phenomena.} Numerical expressions in the Higgs basis are provided in
\autoref{eq:hz_numerical}. Introducing $c_{\gamma\square}$ defined in
\autoref{eq:cwtocz} and contributing for an off-shell photon however renders
this effect more transparent. For $P(e^-,e^+)=(0,0)$, $(-0.8,+0.3)$,
$(+0.8,-0.3)$ polarization configurations at $\sqrt{s}=250$\,GeV, we for
instance obtain:
\begin{equation}
\left.  \frac{\sigma_{hZ}}{\sigma_{hZ}^\text{SM}} \right\vert_{250\,{\rm GeV}} ^{\scriptstyle P= \tiny \bpm  (0,0) \\ (-0.8,+0.3) \\  (+0.8,-0.3) \epm }
	\simeq 1
	+ 2   \, \delta c_Z
	+ 1.6 \, c_{ZZ}
	+ 3.5 \, c_{Z\square}
	+ {\scriptstyle \bpm 0.060 \\ 0.82 \\ -0.89 \epm} \, c_{Z\gamma}
	+ {\scriptstyle \bpm 0.16  \\ 2.2  \\ -2.3  \epm} \, c_{\gamma\square}
	\,.
\end{equation}
An increase in the sensitivity magnitude of more than an order of magnitude is
brought by beam polarization. Reversing the polarization also flips the sign of
the $c_{Z\gamma}$ and $c_{\gamma\square}$ prefactors, given the opposite signs
of the left- and right-handed couplings of the $Z$ to electrons.


\subsubsection*{Angular asymmetries}
\label{sec:hzasy}

Three angles and two invariant masses fully characterize the differential
distribution of the $\eehz \to h f\bar f$ process (see \autoref{fig:angle}). It
naturally provides information complementary to that of the total rate alone.
The effective-field-theory contributions to the angular distributions have been
thoroughly studied in Ref.~\cite{Beneke:2014sba}.
At tree level and linear order in the effective-field-theory parameters, they
can all be captured through the following asymmetries:
\begin{figure}[t]
\centering
\includegraphics[width=10cm]{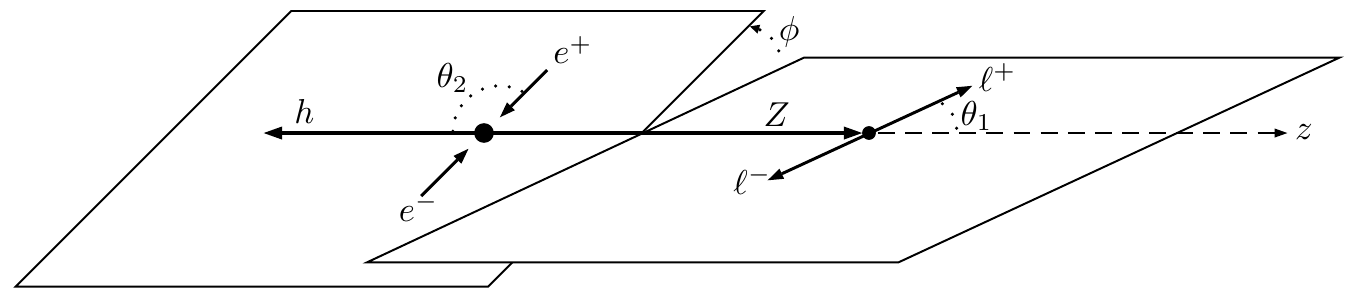}
\caption{Definition of the $\Omega=\{\theta_1,\theta_2,\phi\}$ angles in a
         $\eehz$ event (taken from Ref.~\cite{Craig:2015wwr}). Note the two
         polar angles are respectively defined in the center-of-mass and $Z$
         restframes.}
\label{fig:angle}
\end{figure}
\newcommand{\asym}[1]{
	\frac{1}{\sigma}  \, \int \d \Omega \,
	#1 \,
	\frac{\d\sigma}{\d\Omega}}
\begin{align}
\mathcal{A}_{\rm  \theta_1} =
	&~ \asym{	\sign\{\cos (2\theta_1)\}	} \,, \nonumber \\
\mathcal{A}_{\rm  \phi}^{(1)} =
	&~ \asym{	\sign\{\sin \phi\}	} \,, \nonumber \\
\mathcal{A}_{\rm  \phi}^{(2)} =
	&~ \asym{	\sign\{\sin (2 \phi) \}	} \,, \nonumber \\
\mathcal{A}_{\rm \phi}^{(3)} =
	&~ \asym{	\sign\{\cos \phi \}	} \,, \nonumber \\
\mathcal{A}_{\rm  \phi}^{(4)} =
	&~ \asym{	\sign\{\cos (2 \phi) \}	} \,, \nonumber\\
\mathcal{A}_{c \theta_1, c \theta_2} =
	&~ \asym{	\sign\{\cos \theta_1\cos \theta_2\}}	\,, \label{eq:A}
\end{align}
where $\Omega=\{\theta_1,\theta_2,\phi\}$ and the $\sign$ function gives the
sign of its argument. Among these asymmetries, $\mathcal{A}_{\rm \phi}^{(1)}$
and $\mathcal{A}_{\rm \phi}^{(2)}$ are sensitive to CP-violating parameters (or
absorptive parts of amplitude), while $\mathcal{A}_{\rm \theta_1}$ and
$\mathcal{A}_{\rm \phi}^{(4)}$ depend on the same combination of operator
coefficients. In the absence of CP violation, the angular observables therefore
provide three independent constraints on effective-field-theory parameters.  
The corresponding Higgs-basis expressions are provided in \autoref{app:express}.

A phenomenological study of these angular asymmetries at circular $\ee$
colliders has been performed in Ref.~\cite{Craig:2015wwr}. In particular, it was
shown that the uncertainties on their determination is statistics dominated for
leptonic $Z$ decays. The absolute statistical uncertainty (one standard
deviation) on each asymmetry $\mathcal{A}$ measured with $N$ events is given by~\cite{Craig:2015wwr}
\begin{equation}
	\sigma_{\mathcal{A}}  = 
	\sqrt{\frac{1-{\mathcal{A}}^2}{N}} \approx \frac{1}{\sqrt{N}}
	\,.
\label{eq:sigmaA}
\end{equation}
Following Ref.~\cite{Craig:2015wwr}, we use only the events with Higgs
decays to bottom quarks ($\eehz\,,~Z\to \ell^+ \ell^-\,,~h\to b\bar{b}$) which
has negligible backgrounds. Reference~\cite{Craig:2015wwr} refers to a
preliminary version of the CEPC preCDR which suggests the signal selection
efficiency of this channel at 240\,GeV is around $54\%$. For simplicity, we
assume a universal efficiency of $60\%$ for the event selection of this channel
at all energies for the angular asymmetry analysis.
For the CEPC, with $5\inab$ collected at 240\,GeV, this constitutes a subsample
of approximately $2.7 \times 10^4$ Higgsstrahlung events. For the ILC, the
effects of beam polarizations on the asymmetries is taken into account.
No systematic uncertainty is included. We however expect statistical
uncertainties to be dominant given the fairly rare but clean $Z$ decay to
leptons.


\subsection{Higgs production through weak boson fusion}
\label{sec:vvh}

\begin{figure}[htb]
\centering \hspace{1cm}
\includegraphics[height=3cm]{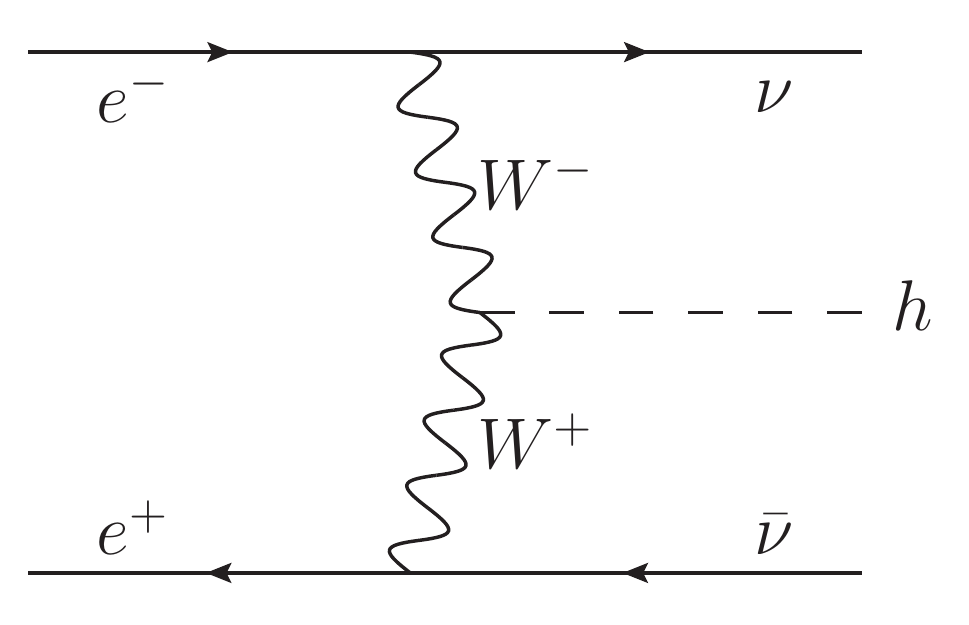}  \hspace{1cm}
\includegraphics[height=3.4cm]{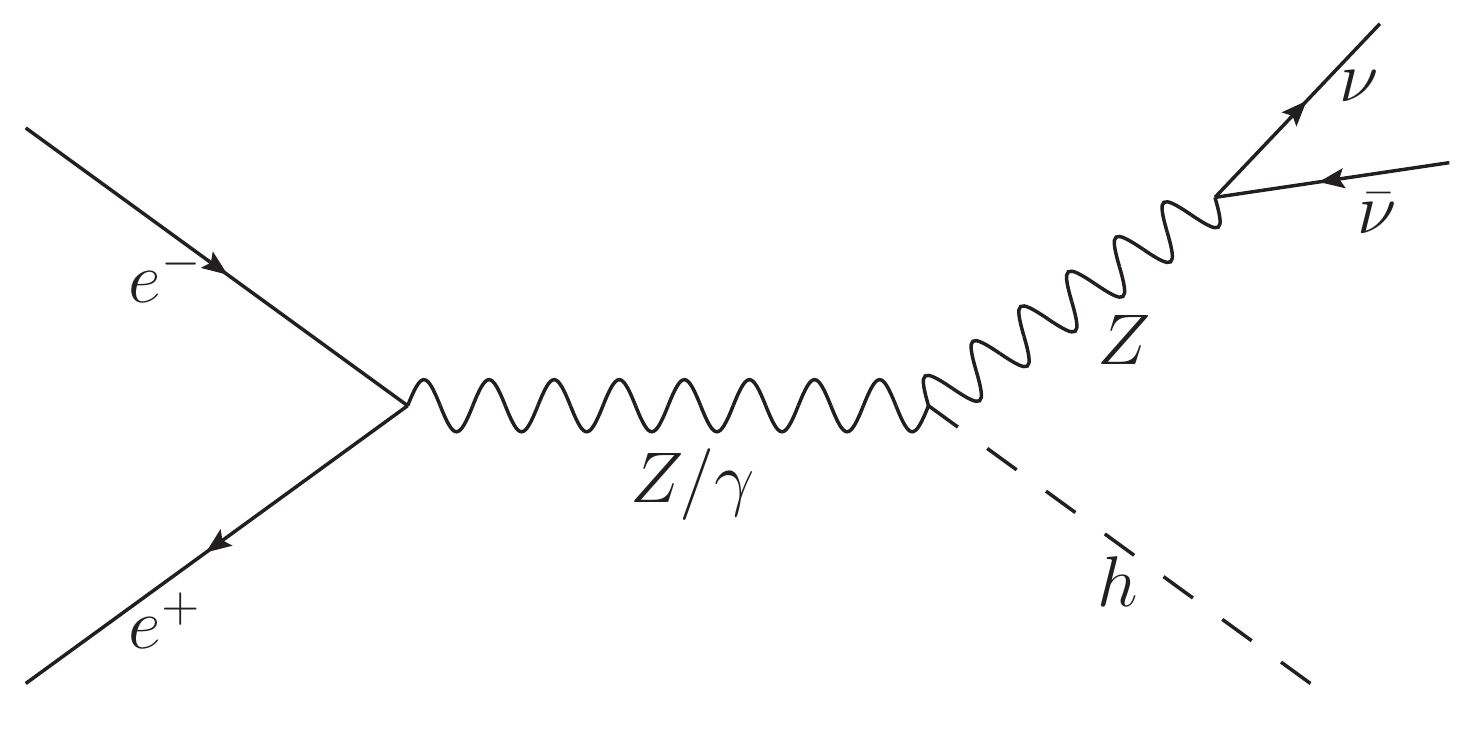}
\caption{Two contributions to the $\eevvh$ process: weak-boson fusion (left),
         and $\eehz, Z\to \nu \bar{\nu}$ (right).
}
\label{fig:feynvvh}
\end{figure}

The Higgs couplings to $W$, $Z$ bosons, and photons are related by $SU(2)_L$
gauge invariance. As such, the measurement of the weak-boson fusion process, first considered in $e^+e^-$ colliders in Ref.~\cite{Jones:1979bq}, is
complementary to that of the Higgsstrahlung process. So, a combination of the
two measurements can efficiently resolve the degeneracy among the EFT parameters
that contribute to the production processes. The weak-boson fusion cross section
grows with energy, so that it is better measured at a center-of-mass energy of
350\,GeV or above. Nevertheless, the measurement at 240\,GeV can still provide
important information, especially if runs at higher energies are not performed.

Importantly, Higgsstrahlung with $Z$ decay to neutrinos ($\eehz, ~Z\to
\nu\bar{\nu}$) yields the same final state as weak-boson fusion (see
\autoref{fig:feynvvh}) and has a rate about six times larger at a center-of-mass
energy of $240$\,GeV (without beam polarization). At this center-of-mass
energy the missing mass distributions for both processes moreover peak at similar
energies (see Fig.~3.16 on page 75 of Ref.~\cite{CEPC-SPPCStudyGroup:2015csa}).
Isolating the weak-boson fusion contribution is therefore difficult. For
the CEPC and FCC-ee at 240\,GeV, we therefore consider an inclusive $\eevvh$
sample to which the two processes contribute, and only use the $h\to b\bar{b}$
channel for which the precision on the $\eevvh$ rate measurement is reported in
the literature. We neglect the contributions of the weak-boson fusion in the
other Higgs decay channels of $\eehz, ~Z\to \nu\bar{\nu}$. For the ILC,
Ref.~\cite{Asner:2013psa} states that a $\chi^2$ fit of the recoil mass
distribution is used to separate the weak-boson-fusion and the Higgsstrahlung
processes. We thus consider that the precision on $\sigma(\eevvh)\times{\rm
BR}(h\to b\bar{b})$ quoted in Ref.~\cite{Barklow:2015tja} applies directly to
the weak-boson fusion contribution. Both processes reach equal rates at a
center-of-mass energy close to $350$\,GeV (without beam polarization). At this
and higher energies, we thus assume that their distinct recoil-mass
distributions are sufficient to efficiently separate them. More details on the
treatment of this measurement can be found in \autoref{app:input}.


\subsection{Higgs production in association with tops}
\label{sec:tth}

\begin{figure}[htb]
\centering
\includegraphics[width=5.5cm]{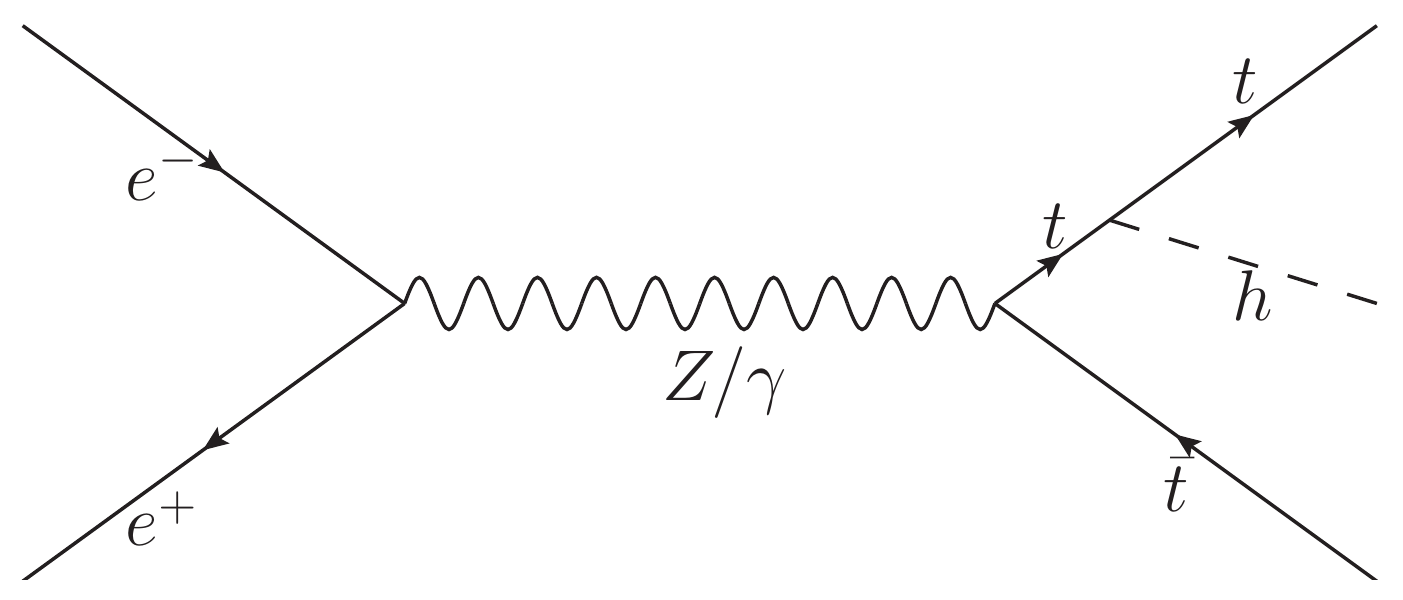}  \hspace{1cm}
\includegraphics[width=5.5cm]{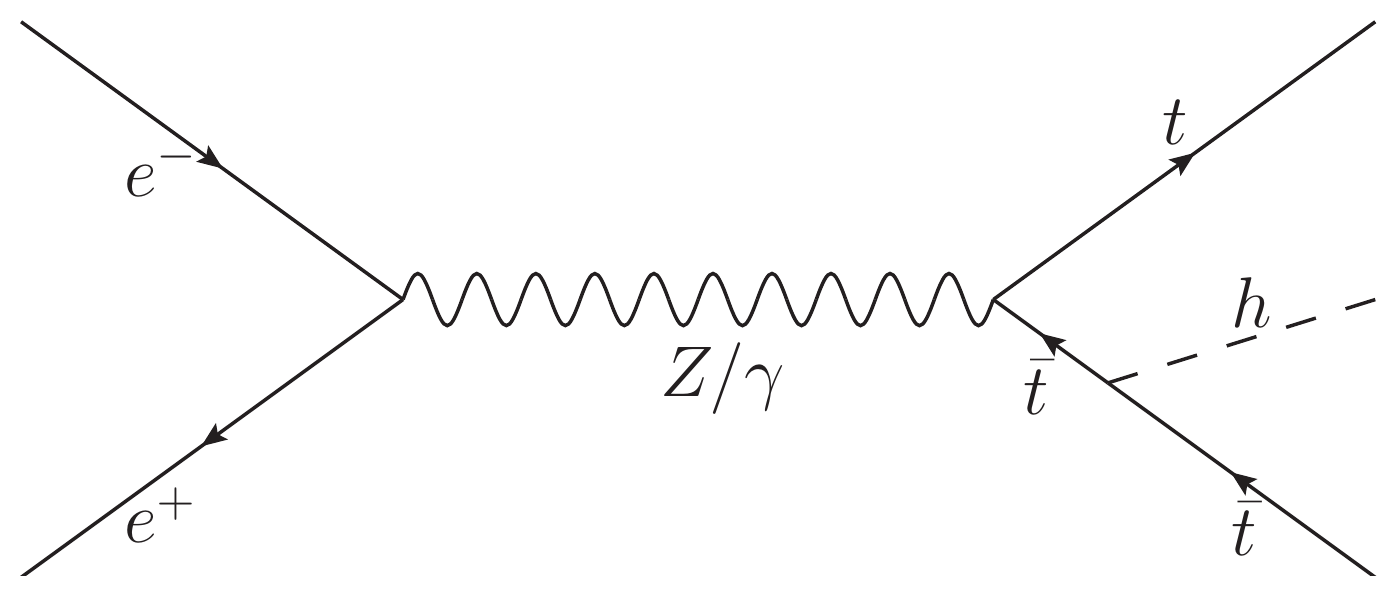}  \\ \vspace{0.7cm}
\includegraphics[width=5.5cm]{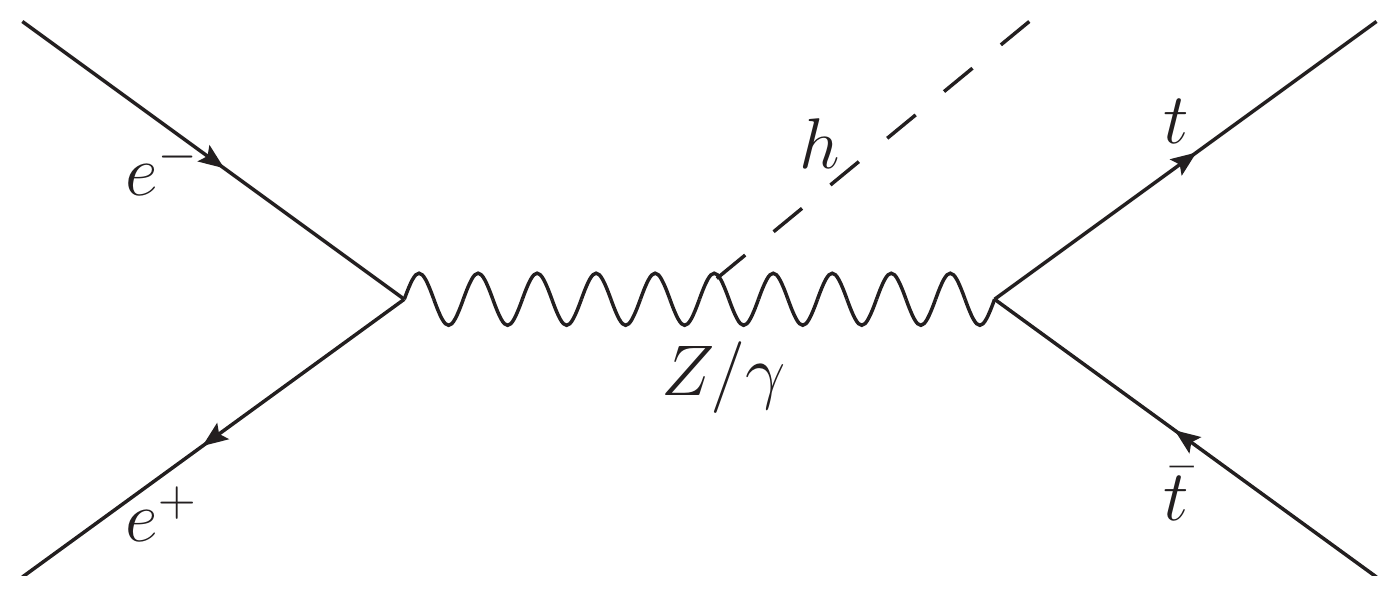}  \hspace{1cm}
\includegraphics[width=5.5cm]{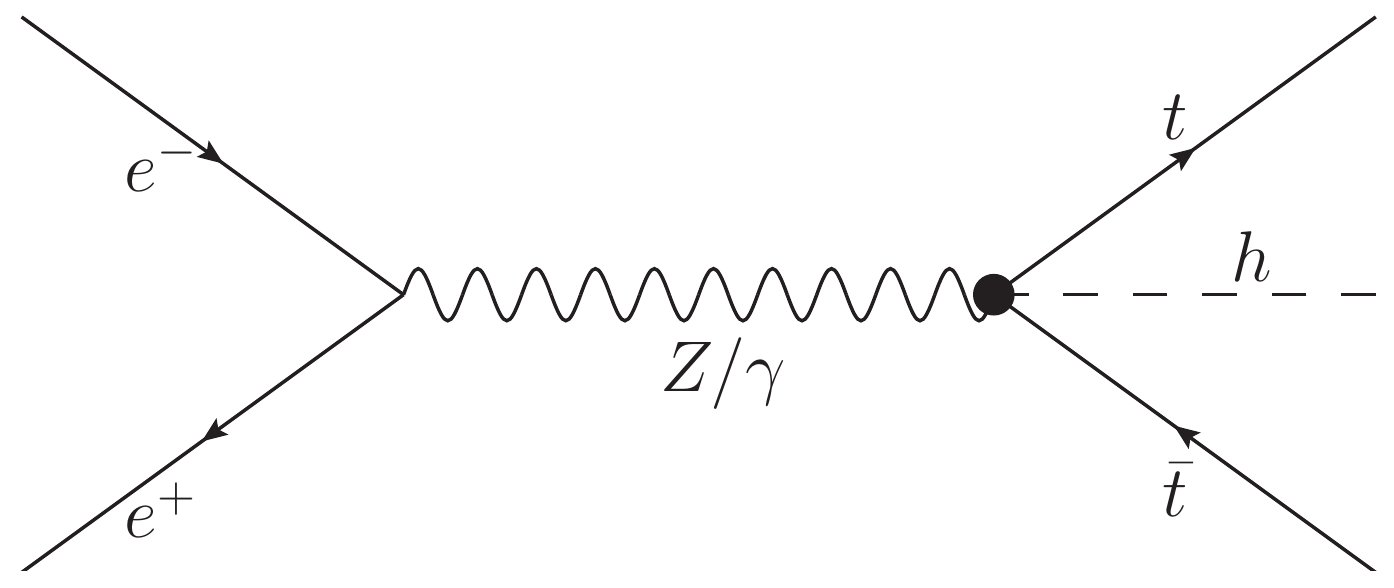}
\caption{Leading-order diagrams for the $\eetth$ process. In the SM, the
         dominant contribution are the ones involving the top Yukawa coupling.
         Other EFT contributions (including that of four-fermion operators, not
         depicted) should be well constrained by other measurements.
}
\label{fig:feyntth}
\end{figure}

The $\eetth$ production of a Higgs boson in association with top quarks (see
\autoref{fig:feyntth}) requires a large center-of-mass energy which is only
achieved at a linear collider. A $10\%$ precision on $\sigma(\tth)\times {\rm
BR}(h\to b\bar{b})$ could be achieved with $4\inab$ of ILC data collected at
$\sqrt{s}=500$\,GeV (scaled from $28\%$ of the $500\infb$ result in
Ref.~\cite{Barklow:2015tja}). At CLIC, $1.5\inab$ of $1.4$\,TeV data should
yield an $8.4\%$ precision \cite{Abramowicz:2016zbo}. In the SM, the dominant contributions to this process
involve a top Yukawa coupling. The radiation of a Higgs from the $s$-channel $Z$
boson is comparatively negligible~\cite{Baer:2013cma}. In the effective field
theory, we only include modifications of the top Yukawa coupling. Other
contributions should be sufficiently constrained by the measurement of top pair
production and other processes. Neither the four-point $Zhtt$ interaction
depicted on \autoref{fig:feyntth} (bottom-right), nor four-fermion operator
contributions are thus accounted for here. This channel could also be used to establish 
the CP properties of the Higgs boson~\cite{BhupalDev:2007ftb}, which we simply assumed to be a 0$^+$  state throughout our analysis.

\subsection{Weak-boson pair production}
\label{sec:ww}

\begin{figure}[htb]
\centering
\includegraphics[height=3cm]{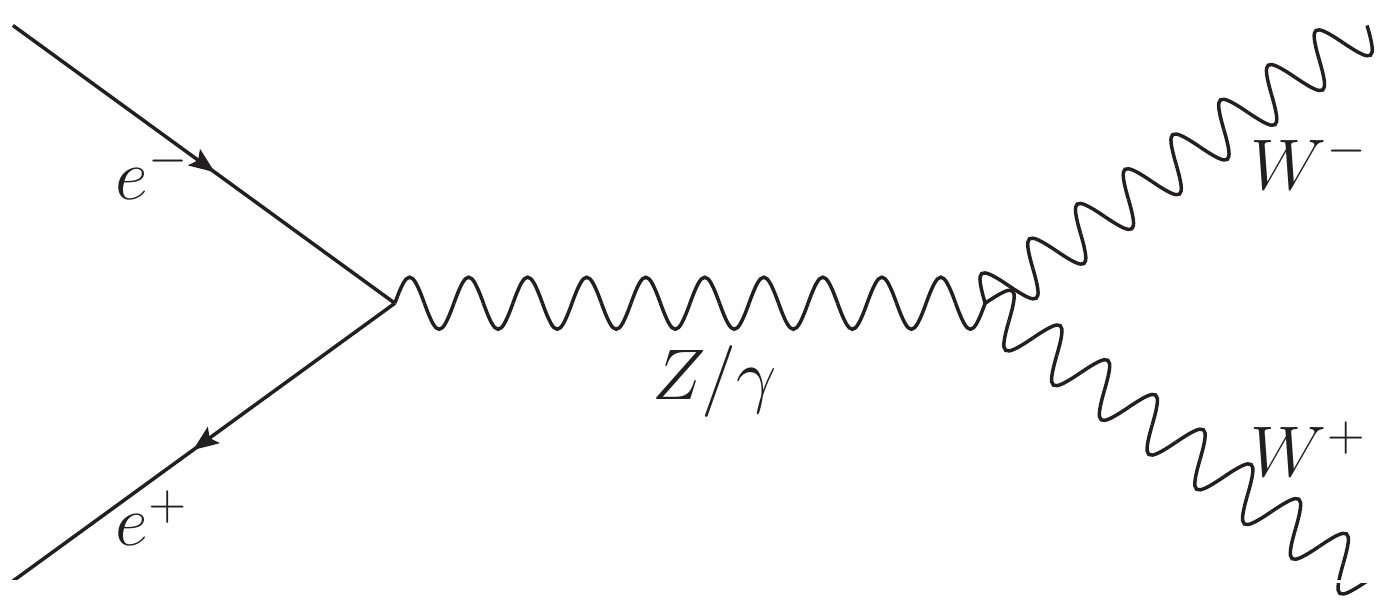}  \hspace{1cm}
\includegraphics[height=3cm]{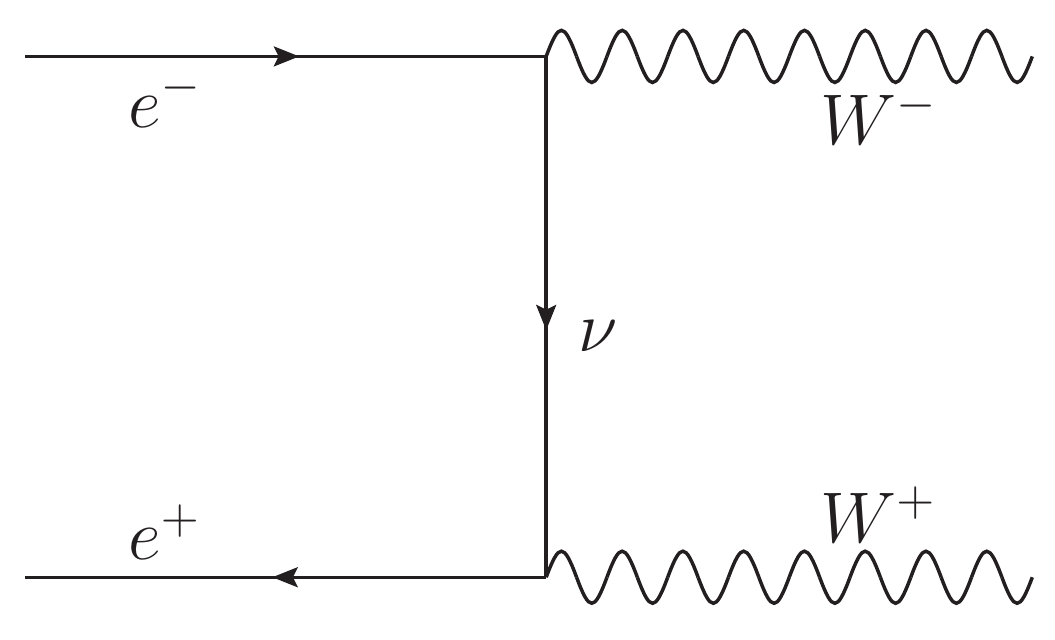}
\caption{Leading-order diagrams contributing to $\eeww$. The $s$-channel diagram
         on the left with an intermediate $Z$ or photon involves a triple gauge
         coupling.}
\label{fig:feyntgc}
\end{figure}
The diagrams contributing to the $\eeww$ process, at leading order, are depicted
in \autoref{fig:feyntgc}. The $s$-channel diagrams with an intermediate $Z$ or
photon involve triple gauge couplings. Considering CP-even dimension-six
operators only, the aTGCs are traditionally parameterized using $\delta
g_{1,Z}$, $\delta \kappa_\gamma$ and $\lambda_Z$~\cite{Hagiwara:1993ck,
Gounaris:1996rz}, defined in \autoref{eq:tgc}. Among them, $\delta g_{1,Z}$ and
$\delta \kappa_\gamma$ are generated by effective operators that also contribute
to Higgs observables. As pointed out in Ref.~\cite{Falkowski:2015jaa}, this
leads to an interesting interplay between Higgs and TGC measurements.

Triple gauge couplings have been measured thoroughly at
LEP2~\cite{Schael:2013ita}. Various studies of future lepton colliders' reach
have also been carried out~\cite{Abe:2001swa, Battaglia:2004mw,
Marchesini:2011aka, Rosca:2016hcq, Wells:2015eba, Bian:2015zha}. At future
circular colliders, most of the $W$ pairs are likely to be produced at
$240$\,GeV, as a byproduct of the Higgs measurement run which requires large
luminosities. At this energy, the $\eeww$ cross section is approximately two
orders of magnitude larger than that of $\eehz$. With $5\inab$, the CEPC would
thus produce about $9\times 10^7$ $\eeww$ events, thereby improving
significantly our knowledge of TGCs. A run at $350$\,GeV, probing a different
combination, could bring further improvement on the constraints. Longitudinal
beam polarization is also very helpful in probing the aTGCs. With $500\infb$
collected at $500$\,GeV and equally shared between four $P(e^-,e^+) =
(\pm80\%,\pm30\%)$ beam polarization configurations, the ILC could constrain the
three TGCs to the $10^{-4}$ level~\cite{Marchesini:2011aka}. Note the runs with
$++$ and $--$ polarizations are mostly meant to provide a simultaneous and
sufficiently accurate polarization magnitude measurement. Comparable results can
be expected for more realistic repartitions of the
luminosities~\cite{Rosca:2016hcq}.

For the CEPC and FCC-ee prospects, we follow Ref.~\cite{Bian:2015zha} which
exploited kinematic distributions in the $\eeww \to 4f$ process. Five angles can
be reconstructed in each such event: the polar angle between the incoming $e^-$
and the outgoing $W^-$, and two angles specifying the kinematics of each $W$
decay products. When both $W$s decay leptonically, the $W$ mass constraints
allow to fully reconstruct the kinematics up to a fourfold ambiguity at most.
Here, we make the optimistic assumption that the correct solution is always
found. In the hadronic $W$ decays, one can not discriminate between the quark
and antiquark. The angular distributions of the $W$ decay products are thus
\emph{folded}. We divide the differential distributions of each angle into $20$
bins ($10$ in folded distributions). Uncorrelated Poisson distributions are
assumed in each bin and their $\chi^2$ are summed over. The total $\chi^2$ is
constructed by summing over the $\chi^2$ of all the angular distributions of all
decay channels. The statistical correlation between angular distributions is
neglected.

Given the huge statistics that would be collected, and although they were
neglected in Ref.~\cite{Bian:2015zha}, the systematic uncertainties could play
an important role. Theoretical uncertainties could also become limiting. At the
moment, there is however no dedicated experimental study of TGC measurements at
future circular colliders. We therefore introduce a benchmark systematic
uncertainty of $1\%$ in each bin of the differential distributions. This
guess is probably too conservative compared to few $10^{-4}$ systematic
uncertainties on the $\delta g_{1,Z}$, $\delta\kappa_\gamma$, and $\lambda_Z$
TGC parameters recently estimated by the ILC collaboration~\cite{JListPrivate}.
We therefore examine the impact of variation of this value in
\autoref{sec:results} and also provide constraints obtained by assuming no
deviation on the TGC from their standard-model values.

For the prospects of the full ILC program, we use the one-sigma statistical
uncertainties obtained in Ref.~\cite{Marchesini:2011aka} ($\Delta \delta g_{1,Z}
= 6.1\times 10^{-4}$, $\Delta \delta \kappa_\gamma = 6.4\times 10^{-4}$ and
$\Delta\lambda_Z = 7.2\times 10^{-4}$), together with their correlations shown
in \autoref{tab:tgcilc} of \autoref{app:input}. We do however not scale
these numbers to higher luminosities, as systematic uncertainties are likely to
become important. The current estimates by the ILC collaboration for systematics
uncertainties are of a few $10^{-4}$~\cite{JListPrivate}. When focusing on the
$250$ and $350$\,GeV runs of the ILC, we use the strategy described above for
the CEPC instead. As a dedicated experimental study of TGC measurements at CLIC
is also missing,\footnote{ For CLIC at $3$\,TeV and an integrated luminosity
of $1\,\inab$, Ref.~\cite{Battaglia:2004mw} bases itself on
Ref.~\cite{Abe:2001swa} which derived individual constraints and quotes
$\Delta\delta\kappa_\gamma = 0.9\times 10^{-4}$, $\Delta\lambda_Z = 1.3\times
10^{-4}$ constraints (we thank Philipp Roloff for pointing out this reference).
These results are however insufficient to serve as input for our global
analysis. A phenomenological study for CLIC based on total $\eeww$ rates only
was also performed in Ref.~\cite{Ellis:2017kfi}. The results in Section\,3.2 and
Eq.\,(4.2) there imply individual constraints rescaled for $1\,\inab$ that are
less than a factor of two better than that of Ref.~\cite{Battaglia:2004mw}. }
we assume a precision similar to the ILC one can be reached there. It should be
noted, however, that the 1.4 and 3\,TeV runs at CLIC could potentially provide
even stronger constraints on the aTGCs due to the increase of sensitivities with
energy~\cite{Ellis:2017kfi}.

Another important issue raised by the significant improvement in the $\eeww\to
4f$ measurement precision concerns the uncertainty on electroweak precision
observables. In the extraction of the constraints on aTGCs, one usually makes
the \emph{TGC dominance assumption} and neglects the impact of new physics on
all other parameters. At LEP, this was justified given the better precision of
$Z$-pole and $W$-mass measurements compared to that of $W$ pair production. In
this work, we also assume that runs at lower energies will give us sufficient
control on such effects. Exploiting diboson data could also be an
alternative if runs at lower energies are not performed. Further investigations
are required in this direction.
The $W$ mass can be measured very well at a Higgs factory by reconstructing the
$W$ decay products in the $\eeww$ process. To leading order, the aTGCs affect
the differential distributions of $\eeww$, but not the $W$ invariant mass. The
two measurements are thus approximatively independent. A precision of $3$\,MeV
could be achieved at the CEPC with this
method~\cite{CEPC-SPPCStudyGroup:2015csa}. A dedicated threshold scan at
center-of-mass energies of $160$--$170\,$GeV could also be performed. As such,
it is reasonable to assume the $W$ mass will be sufficiently well constrained at
future $e^+e^-$ colliders.
The corrections to gauge-boson propagators and fermion gauge couplings could
however have a non-negligible impact on the determination of triple gauge
couplings, especially without a future $Z$ factory to improve their
constraints.\footnote{See also Ref.~\cite{Zhang:2016zsp} for a recent discussion
on this topic in the context of LHC measurements.}
While the CEPC and FCC-ee could perform a run at the $Z$ pole, the interest of
such a $Z$-pole run at the ILC and CLIC is still under investigation. Notably,
the ILC precision on aTGCs quoted above already surpasses the precision obtained
at LEP on the electroweak observables. A global fit including Higgs, TGC and the
$Z$-pole measurements would be instructive but is beyond the scope of this
paper.


\subsection{Global fit and determinant parameter}

Our total $\chi^2$ can be rewritten as the sum of that of the measurements described previously in this section:
\begin{equation}
\chi^2_{\rm tot} = \chi^2_{hZ / \nu\bar{\nu} h ,\, {\rm rates}} + \chi^2_{hZ ,\, {\rm asymmetries}} + \chi^2_{WW} \,,
\end{equation}
where\footnote{Note that we have used the symbol $\sigma$ to denote both cross sections and  standard deviations. What we mean in each case should be clear from the context.}
\begin{align}
	\chi^2_{hZ / \nu\bar{\nu} h ,\, {\rm rates}} =
		&~ \sum_i	\frac{(\mu_i^{\rm NP} - \mu_i^{\rm SM})^2}
				{\sigma^2_{\mu_i}}
		\,, \label{eq:chia}\\
	\chi^2_{hZ ,\, {\rm asymmetries}} =
		&~ \sum_i	\frac{(\mathcal{A}^{\rm NP}_i - \mathcal{A}^{\rm SM}_i)^2}
				{\sigma^2_{\mathcal{A}_i}}
		\,, \label{eq:chib}\\
	\chi^2_{WW \mbox{\tiny (CEPC \& FCC-ee)}} =
		&~ \sum_i	\frac{(n_i^{\rm NP} - n_i^{\rm SM})^2}
				{(\sqrt{n_i}+\sigma_i^{\rm sys})^2} 
		\,. \label{eq:chic}
\end{align}
The $\mu_i$ are the signal strengths (rates normalized to SM predictions) of the
rate measurements, summed over $\sigma(hZ)$, $\sigma (hZ)\times \br$ and
$\sigma(\vvh)\times \br$. The corresponding one-sigma
uncertainties are listed in \autoref{tab:higgsinputc}, \ref{tab:higgsinputi} and
\ref{tab:higgsinputclic} of \autoref{app:input}, for the different colliders.
$\mathcal{A}_i$ are the asymmetries of \autoref{eq:A}, and
$\sigma_{\mathcal{A}_i}$ their uncertainties, given in \autoref{eq:sigmaA}. For
the $\eeww$ measurements at CEPC and FCC-ee, the $\chi^2$ is summed over all
$W$-boson decay channels, over the five angular distributions, and over all
their bins. A systematic uncertainty $\sigma_i^\text{sys}$ is included in each
bin. Unless otherwise specified, we take $\sigma_i^\text{sys}/n_i=1\%$ where
$n_i$ is the number of events in that bin.  
For ILC and CLIC, the
$\chi^2_{WW}$ is directly reconstructed from one-sigma bounds and the
correlation matrix of aTGCs from Ref.~\cite{Marchesini:2011aka} (shown in
\autoref{tab:tgcilc} of \autoref{app:input}).  
Finally, the $\chi^2$ is summed over runs with different energies and beam
polarizations (if applicable).

As we only retained the linear dependence of all observables in terms of
effective-operator coefficients, our $\chi^2$ are quadratic functions:
\begin{equation}
	\chi^2 = \sum_{ij} (c-c_0)_i  \, \sigma^{-2}_{ij} \, (c-c_0)_j
	\,, \hspace{1cm} \mbox{where}  \hspace{0.5cm}   
	\sigma^{-2}_{ij} \equiv \left( \delta c_i \: \rho_{ij} \: \delta c_j  \right)^{-1}
	\,,  \label{eq:chipara}
\end{equation}
where $c_{i=1,\,...\,12}$ denotes the 12 parameters of \autoref{eq:para10} and
$c_0$ are the corresponding central values, which are vanishing by construction
in our study. The uncertainties $\delta c_i$ and the correlation matrix $\rho$
can thus be obtained from $\sigma^{-2}_{ij} = {\partial^2
\,\chi^2} \big/ {\partial c_i \partial c_j}$.

It should also be noted that the measured Higgs decay width reported in the
corresponding documents of the colliders is a quantity derived (with certain
assumptions) from several measurements which are already included in the fit. We 
therefore do not include it in our fit as an additional independent measurement.

\paragraph{Global determinant parameter (GDP)}
\begin{figure}
\centering
\includegraphics[width=5cm]{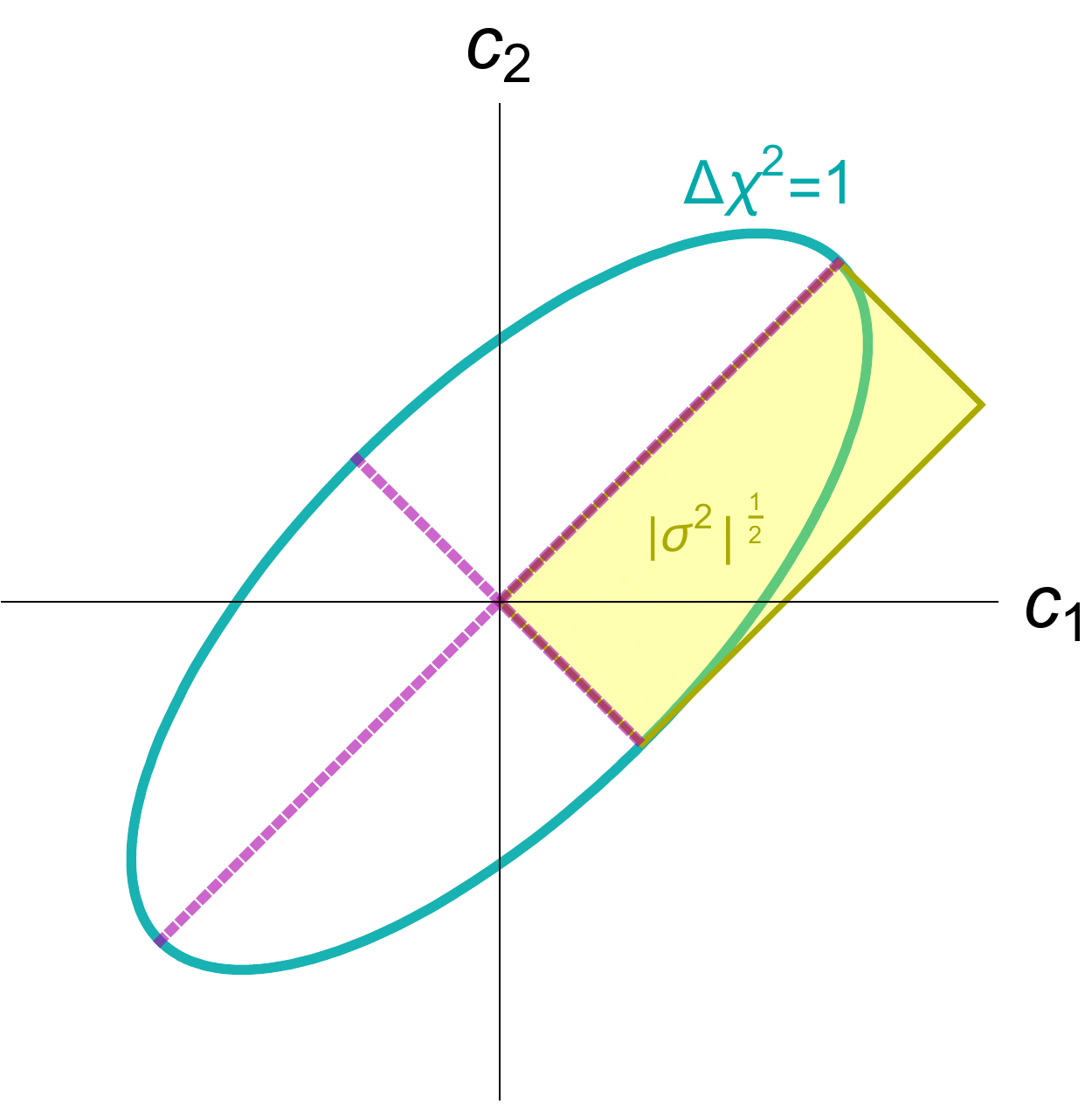}
\caption{In a two-dimensional parameter space, the area of the Gaussian
         one-sigma ellipse is proportional to the square root of the determinant
         of the covariance matrix, $\sqrt{\det \sigma^2}$. In $n$ dimensions,
         the $n$th root of this quantity or \emph{global determinant parameter}
         ($\GDP\equiv\sqrt[2n]{\det \sigma^2}$) provides an average of
         constraints strengths. GDP ratios measure improvements in global
         constraint strengths independently of an effective-field-theory operator
         basis.}
\label{fig:gdshow1}
\end{figure}
We introduce a metric, dubbed \emph{global determinant parameter}, for assessing
the overall strength of constraints. In a global analysis featuring $n$ degrees
of freedom, it is defined as the determinant of the covariance matrix raised to
the $1/2n$ power, $\GDP \equiv\sqrt[2n]{\det\sigma^2}$. In a multivariate
Gaussian problem, the square root of the determinant is proportional to the
volume of the one-sigma ellipsoid ($\pi^\frac{n}{2}/\Gamma(\frac{n}{2}+1)\:
\sqrt{\det\sigma^2}$) and therefore measures the allowed parameter space size
(see \autoref{fig:gdshow1}). Its $n$th root is the geometric average of the half
lengths of the ellipsoid axes and can thus serve as an average constraint
strength. Interestingly, the ellipsoid volume transforms linearly under
rescalings of the fit parameters. So, ratios of GDPs do not depend on
parameters' normalization. They are obviously also invariant under rotations in
the multidimensional parameters space. Such ratios are thus independent on the
choice of effective-operator basis used to describe the same underlying physics.
We therefore judge these quantities especially convenient to measure the
improvement in global constraints brought by different run scenarios of future
lepton colliders.  It is however to be noted that the GDP measure weights
equally all directions in the effective-field-theory parameter space, so that it
is on its own certainly not accounting for the fact some directions are
privileged by specific power countings or models.


\section{Results}
\label{sec:results}
We first discuss in this section the precision reach of the whole
program of each collider before examining, in subsequent subsections, the
impact of different measurements, center-of-mass energies, systematic
uncertainties, and beam polarization. The CEPC is then taken as an illustrative
example (except when studying polarization) and the corresponding figures for
the FCC-ee and ILC are provided in \autoref{app:more}.

\begin{figure}[t]
\centering
\includegraphics[width=.95\textwidth]{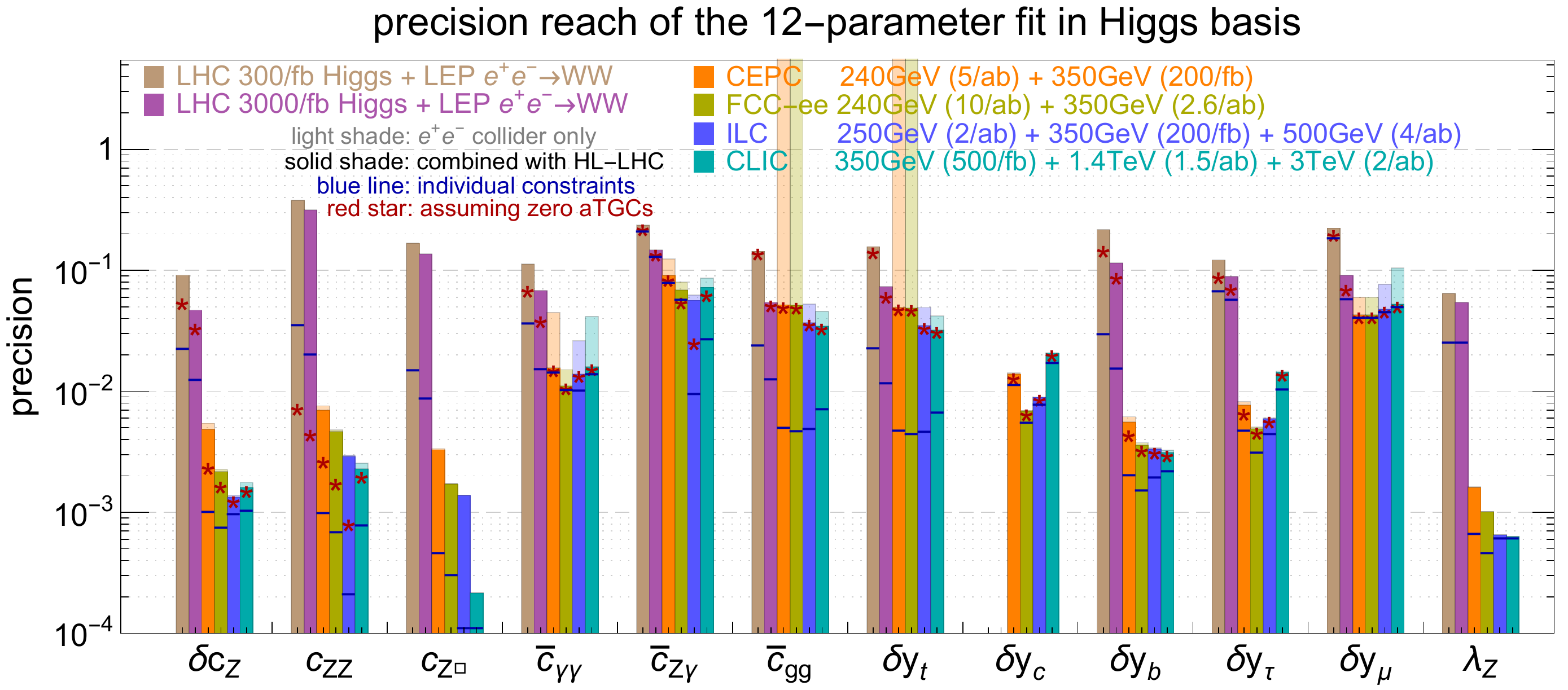}%
\raisebox{4.3mm}{\includegraphics[width=.075\textwidth]{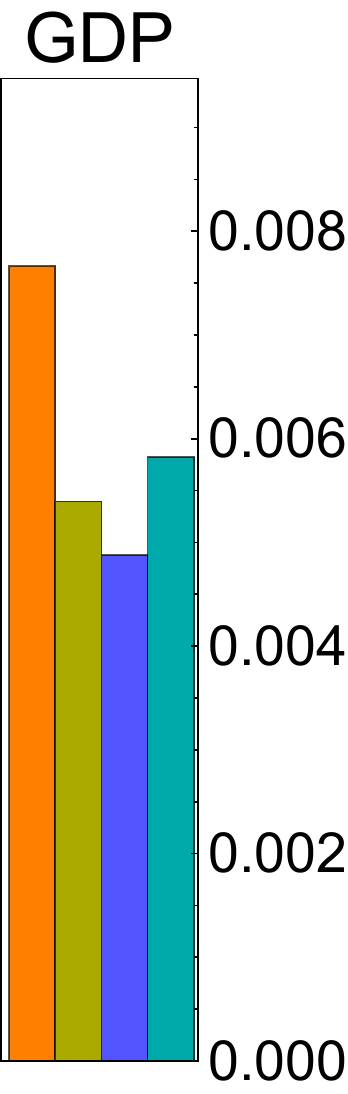}}
\caption{One-sigma precision reach of future lepton colliders on our
         effective-field-theory parameters. All results but the light-shaded
         columns include the $14$\,TeV LHC (with $3000\infb$) and LEP
         measurements. LHC constraints also include measurements carried out at
         $8$\,TeV. Note that, without run above the $t\bar{t}h$ threshold,
         circular colliders alone do not constrain the $\bar{c}_{gg}$ and
         $\delta y_t$ effective-field-theory parameter individually. The
         combination with LHC measurements however resolves this flat direction.
         The horizontal blue lines on each column correspond to the constraints
         obtained when one single parameter is kept at the time, assuming all
         others vanish. The red stars correspond to the constraints assuming
         vanishing aTGCs. The GDPs of future lepton colliders are shown on the
         right panel. See main text for comparisons with the LHC GDPs.}
\label{fig:fit0}
\end{figure}

We show in \autoref{fig:fit0} the one-sigma precision reach at various future
lepton colliders on our effective-field-theory parameters. These projections are
compared to the reach of the Higgs measurements at the $14$\,TeV LHC with $300\infb$ and
$3000\infb$ of integrated luminosity, combined with the diboson production
measurement at LEP.
The estimated reach of Higgs measurements at the high-luminosity LHC derives
from projection by the ATLAS collaboration~\cite{ATL-PHYS-PUB-2014-016} which
collected information from various other sources. Information about the
composition of each channel are extracted from Ref.~\cite{ATL-PHYS-PUB-2013-014,
ATL-PHYS-PUB-2014-012, ATL-PHYS-PUB-2014-006, ATL-PHYS-PUB-2014-011,
ATL-PHYS-PUB-2014-018}. Theory uncertainties on these LHC measurements are not
included in our estimations. In LHC results, we also assume the charm Yukawa to
be SM-like as Ref.~\cite{ATL-PHYS-PUB-2014-016} does not provide estimations on
the $h\to c\bar{c}$ branching fraction precision reach. The constraints from the
diboson measurements at LEP are obtained from Ref.~\cite{Falkowski:2015jaa}.
We do not include the LHC constraints arising from diboson production, as issues
related to the validity of the effective-field-theory~\cite{Contino:2016jqw,
Falkowski:2016cxu} and of the TGC dominance
assumption~\cite{Zhang:2016zsp} need to be simultaneously considered. A
dedicated study of the reach of the high-luminosity LHC on these processes
should be carried out. The constraints set at future lepton colliders are
however expected to be much more stringent.

Compared with LHC and LEP, future lepton colliders would improve the
measurements of effective-field-theory parameters by roughly one order of
magnitude. A combination with the LHC measurements provides a marginal
improvement for most of the parameters. For $\bar{c}_{\gamma\gamma}$,
$\bar{c}_{Z\gamma}$ and $\delta y_\mu$, the improvements are more significant,
as the small rates and clean signals make the LHC reaches comparable to that of
lepton colliders. It should be noted that the measurements of the $h\to gg$
branching fraction only constrain a linear combination of $\bar{c}_{gg}$ and
$\delta y_t$. These two parameters are thus only constrained independently by
lepton colliders when $t\bar{t}h$ production is measured. Therefore, the
combination with LHC measurements is required for CEPC and FCC-ee to constrain
$\bar{c}_{gg}$ and $\delta y_t$.  The resulting bounds on $\delta y_t$ are then even substantially better than that set by the LHC alone.

The twelve-parameter GDPs for the combination of future lepton collider,
LHC $3000\infb$ and LEP measurements are displayed on the right panel of
\autoref{fig:fit0}. Corresponding numerical values are 0.0077, 0.0054,
0.0049, 0.0058 for CEPC, FCC-ee, ILC and CLIC, respectively. Varying prospective
constraints on the charm Yukawa measurement complicate the comparison with the
high-luminosity LHC. The ATLAS collaboration estimated the $h\to
J\!/\!\psi\,\gamma$ branching fraction could be constrained to be smaller than
$15$ times its standard model value with $3\inab$ at
$14$\,TeV~\cite{ATL-PHYS-PUB-2015-043}. Such a constraint would translate into a
one-sigma precision reach on $\delta y_c$ of order one. To broadly cover
the range spent by other studies~\cite{Bodwin:2013gca, Perez:2015aoa,
Brivio:2015fxa, Bishara:2016jga, Carpenter:2016mwd}, we vary the expected
precision reach on $\delta y_c$ in the $0.01-10$ range. The combination of LHC
$300\infb$ ($3000\infb$) and LEP measurements only then leads to GDPs in
the $0.065-0.116$ ($0.039-0.069$) interval, one order of magnitude worst than
when future lepton collider measurements are included. On the other hand,
with $\delta y_c$ set to zero, the eleven-parameter GDP for the combination of
LHC $300\infb$ ($3000\infb$) and LEP measurements only is of $0.078$
($0.044$). In comparison, when future lepton collider measurements are also
included, the corresponding eleven-parameter GDP are 0.0073, 0.0053, 0.0046,
0.0052 for CEPC, FCC-ee, ILC and CLIC, respectively.

Let us also comment further on the impact of having discarded the quadratic
dependence on dimension-six operator coefficients. As stressed in
\autoref{sec:eft}, no significant effect is expected given the good precision
achieved at future lepton colliders in the measurement of most observable. Note
that even the branching ratios for rare Higgs decays like $h\to\gamma Z$ are
sufficiently well constrained for quadratic contributions to be subleading. Only
cases in which accidental suppressions of the standard-model interference with
effective-field-theory amplitudes require a case-by-case discussion. We identify
two such cases.
First, helicity selection rules are known to suppress the ratio of linear and
quadratic dependences on the $\lambda_Z$ aTGC at high energies. Reproducing the
analysis made at $250$\,GeV for a center-of-mass energy of $500$\,GeV and
$500\,\infb$ shared between two beam polarization configurations, with and
without quadratic aTGC contributions, we obtained differences in the derived
limits of $10\%$ at most. The linear approximation thus seems to be reasonably
accurate in that case and no strong quadratic aTGC dependence should affect the
bounds derived in Ref.~\cite{Marchesini:2011aka}. We also checked that quadratic
contributions would be subleading at $\sqrt{s}=3\,$TeV, provided the whole
differential information is included. The non-interference between
standard-model and dimension-six operator indeed does not hold when the
azimuthal angles of the $W$ decay products are not integrated over.
Secondly, as noted in \autoref{sec:hzrate}, the interference between the
$s$-channel photon and $Z$ amplitudes in the unpolarized Higgsstrahlung cross
section suffers from an accidental numerical suppression. Moreover, at high
energies, the Higgsstrahlung cross section goes down and so does the accuracy
with which it can be measured. Therefore, one can expect the quadratic
dependence on the operator modifying the $HZ\gamma$ vertex with an off-shell
photon to be important in that specific case. Although we use
unpolarized cross section measurements to determine CLIC reach on
effective-field-theory parameters to match experimental studies, beam
polarization would actually be available at CLIC and we checked explicitly that
the quadratic effective-field-theory contributions become unimportant once
measurements with polarized beams are performed.

\let\paragraph\subsection
\paragraph{Impact of the various measurements}

\begin{figure}[t]
\centering
\includegraphics[width=14cm]{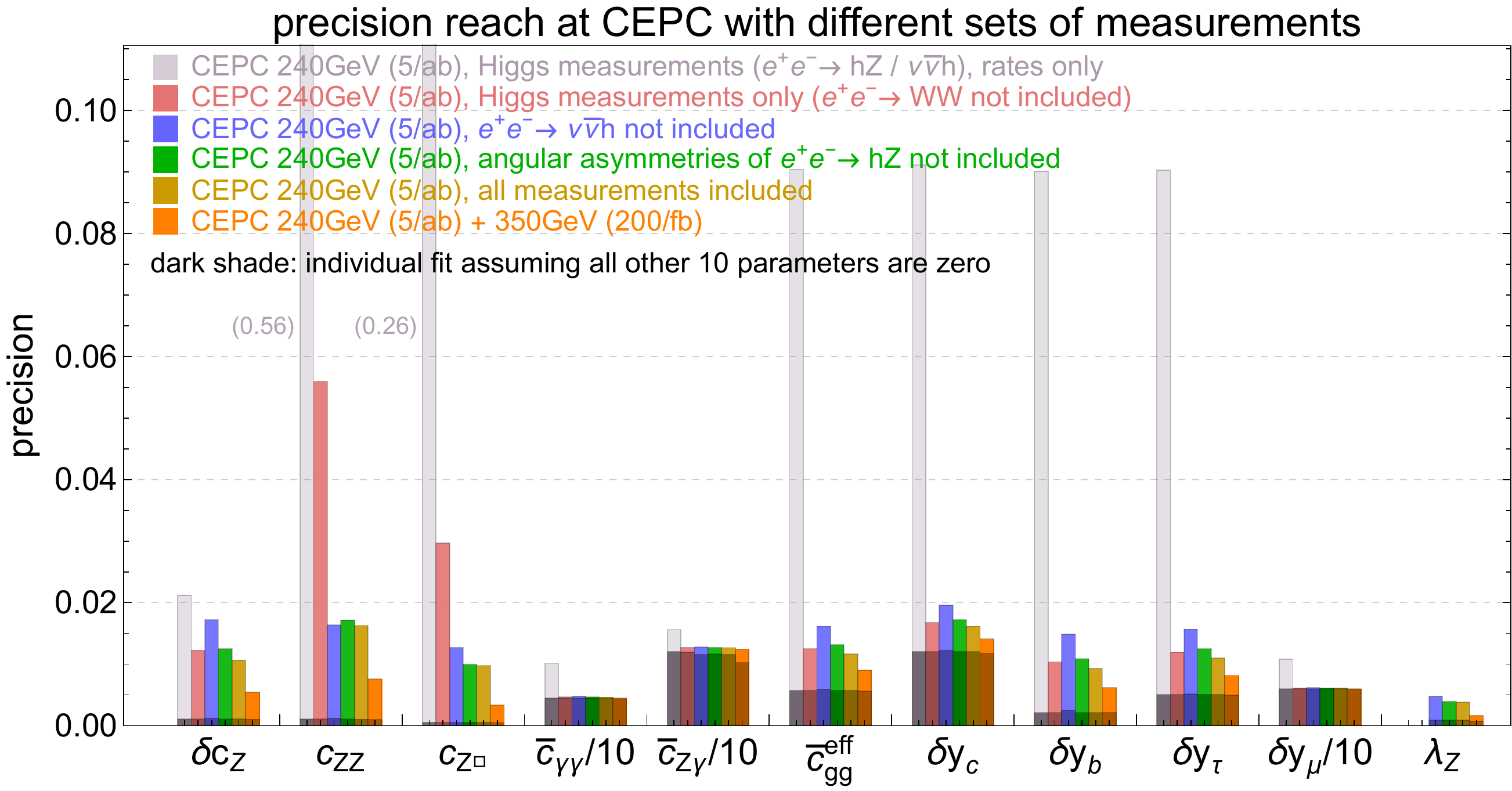}
\caption{One-sigma precision reach obtained with various combinations of
         measurements at the CEPC. The first five columns exploit $5\inab$ of
         $240$\,GeV data while the last column also includes $200\infb$ at
         $350$\,GeV. Only Higgs rate measurements ($\eehzvvh$) are included in
         the first column. One single measurement is excluded at the time in the
         three subsequent columns: $\eeww$ in the second, $\eevvh$ in the third,
         and the angular asymmetries of $\eehz$ in the fourth. Note that
         $\lambda_Z$ is left unconstrained by Higgs data. All measurements at
         $240$\,GeV are included to obtain the constraints in the fifth column.
         A run at $350$\,GeV is also included in the last, sixth, column. The
         dark shades correspond to the constraints obtained when one single
         parameter is kept at the time, assuming all other vanish.}
\label{fig:fitc1}
\end{figure}

We examine, in \autoref{fig:fitc1}, the impact of different measurements. The
one-sigma precision are displayed with one or more measurements removed from the
global fit, using CEPC as an example. Since the degeneracy between
$\bar{c}_{gg}$ and $\delta y_t$ can not be resolved with measurements at $240$
and $350$\,GeV, we display the constraint on $\bar{c}^{\rm \,eff}_{gg}$, defined
in \autoref{eq:barcgg}. The first five columns use the measurements at 240\,GeV
($5\inab$) only. The first column on the left shows the results from rate
measurements in Higgs processes ($\eehzvvh$) only. To obtain the second, third,
and fourth columns, one single measurement is excluded at the time: $\eeww$
(2nd), $\eevvh$ (3rd), and the angular asymmetries of $\eehz$ (4th),
respectively. The fifth column expresses the constraints deriving from all
measurements at $240$\,GeV. In the last column, $200\infb$ of data at $350$\,GeV
is also included. The dark shades finally display the constraints deriving when
one single effective-field-theory parameter is kept at a time.

\hyperref[fig:fitc1]{Figure~\ref{fig:fitc1}} transparently demonstrates that
the Higgs rate measurements alone are insufficient to constrain simultaneously
all parameters to a satisfactory degree. They leave poorly constrained some
directions of the multidimensional parameter space, thereby weakening the whole fit. As
already stressed, in such a global treatment, the combination of several
observables is capital to effectively bound all parameter combinations. The
global strength of constraints is dramatically improved by the first few
measurement which resolve approximate degeneracies. Once a sufficient number of
constraints is imposed, the exclusion of one single observable does not
dramatically affect the overall precision. The individual constraints (obtained
by switching on one parameter at a time), on the other hand, receive little
improvement from the additional measurements ---~a clear demonstration that
global constraints are driven by approximate degeneracies. A marginal
improvement of the constraints obtained for a given run would be obtained by
including a set of observables even more complete than the one we use.

\begin{figure}[t]
\centering
\includegraphics[width=14cm]{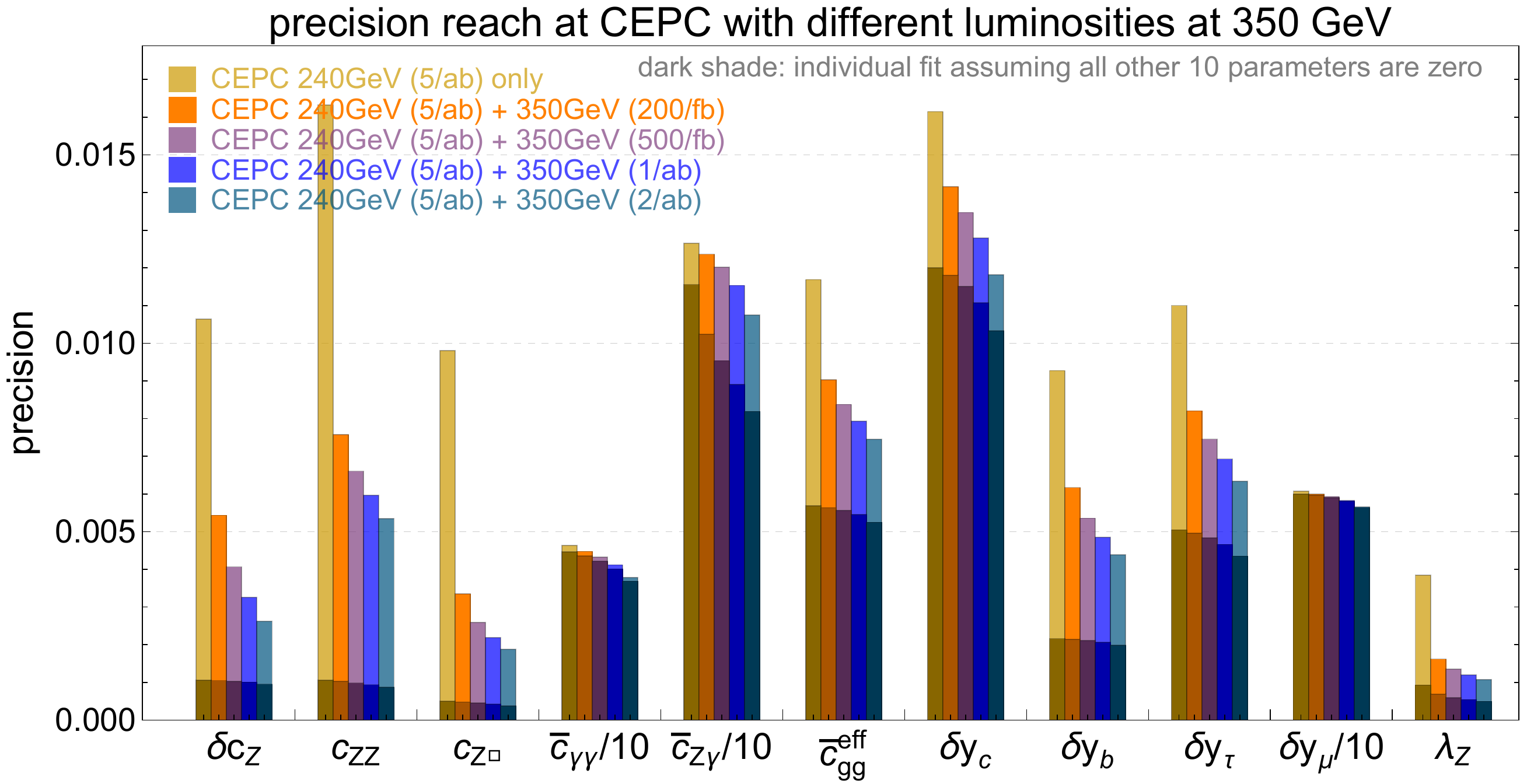}
\raisebox{0.105\height}{\includegraphics[width=1.37cm]{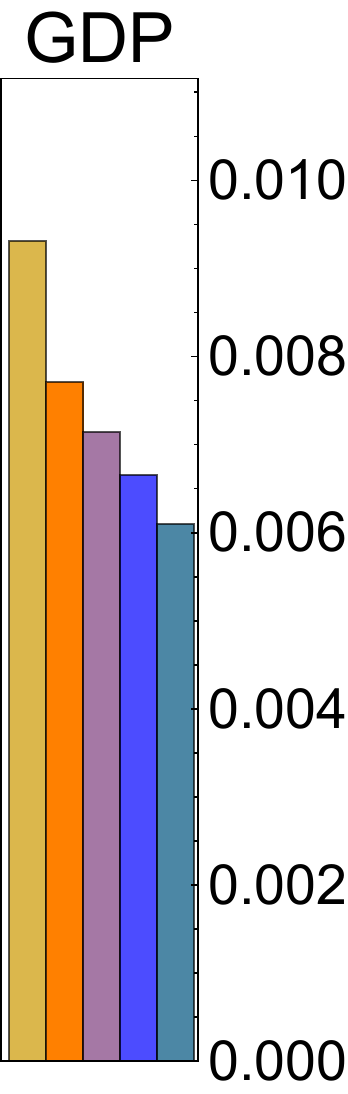}}
\caption{Precision reach (one standard deviation) of the 11-parameter fit in the
         Higgs basis at CEPC. Five different assumptions on the luminosity at
         350\,GeV are made, namely $0$ (240\,GeV run only), $200\infb$,
         $500\infb$, $1\inab$ and $2\inab$ (from left to right). The luminosity
         at 240\,GeV is fixed to $5\inab$. The parameters
         $\bar{c}_{\gamma\gamma}$, $\bar{c}_{Z\gamma}$ and $\bar{c}^{\rm
         \,eff}_{gg}$ are defined in \autoref{eq:barcvv} and
         \autoref{eq:barcgg}. The dark shades correspond to the constraints
         obtained when one single parameter is kept at the time, assuming all
         others vanish.}
\label{fig:fitclu1}
\end{figure}

\paragraph{Impact of a \texorpdfstring{$350$\,}{350 }GeV run at circular colliders}
As already visible in \autoref{fig:fitc1}, a $350$\,GeV run significantly
improves the strength of the constraints set by circular colliders. An important
question for their design is the optimal amount of luminosity to gather at that
energy, in view of the physics performance and the budget cost. In addition to the
top mass and electroweak coupling measurements, the improvement on the precision
of Higgs coupling could be considered too. This is addressed in
\autoref{fig:fitclu1} which shows the reach of the CEPC for increasing amounts
of integrated luminosity collected at $350$\,GeV, from $0$ to $2\inab$. It is clear
that a run at this energy is able to lift further approximate degeneracy among
effective-field-theory parameters. A GDP reduction of about $17\%$ is obtained
with only $200\infb$, and reaches about $34\%$ with $2\inab$. The improvements on the
$\bar{c}_{\gamma\gamma}$, $\bar{c}_{Z\gamma}$, and $\delta y_\mu$ effective
parameters are less significant. The Higgs decay channels which provide the
dominant constraints on these parameters suffer from small rates. These
constraints are thus mainly statistics limited and approximate degeneracies play
a secondary role. It should be noted that the estimations for Higgs measurements
at 350\,GeV for various luminosities are obtained by scaling from the ones in
\autoref{tab:higgsinputc}, assuming statistical uncertainties dominate. This
assumption may cease to be valid for large integrated luminosities.
%


\paragraph{Impact of beam polarization at linear colliders}
\begin{figure}[t]
\centering
\includegraphics[width=14cm]{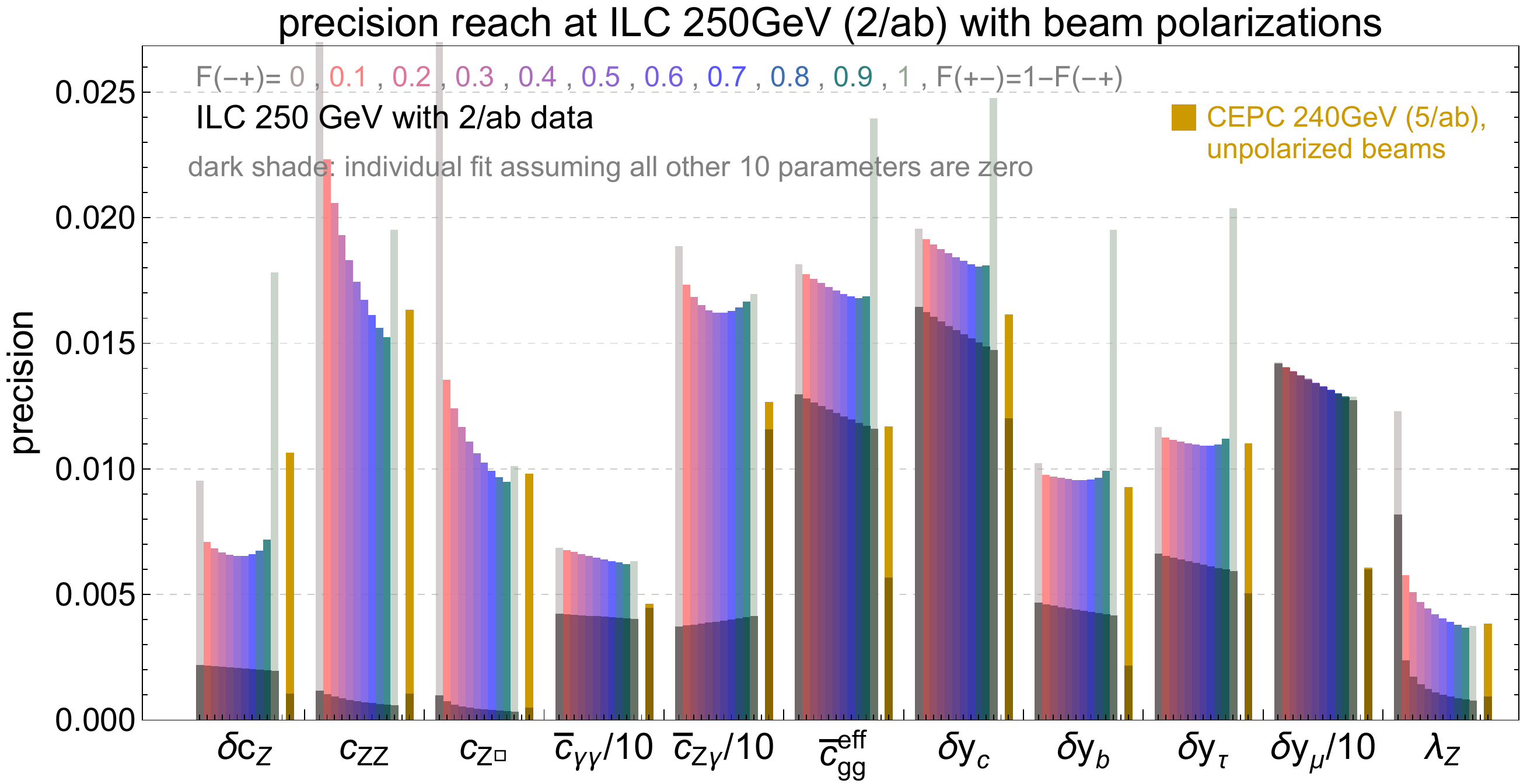}
\raisebox{0.08\height}{\includegraphics[width=1.65cm]{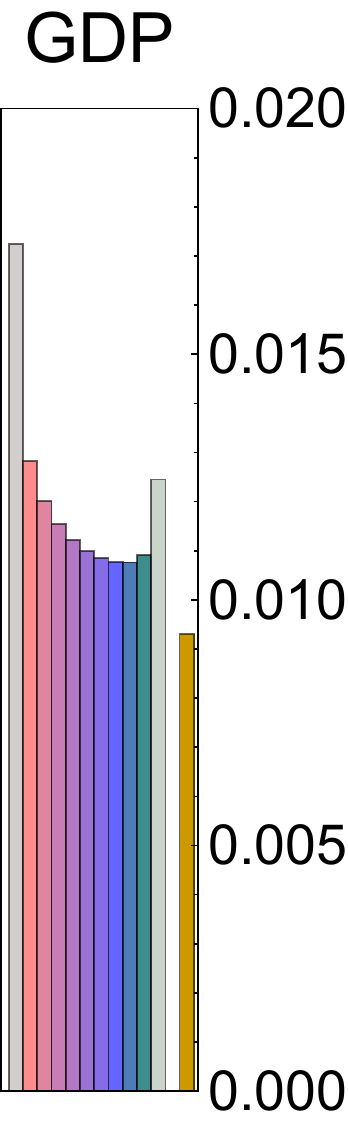}}
\caption{One-sigma precision reach of ILC runs at $250$\,GeV with $2\inab$ of
         integrated luminosity shared between $P(e^-,e^+)=(-0.8,+0.3)$ and
         $(+0.8,-0.3)$ beam polarization configurations. The corresponding
         fractions are denoted as $F_{(-+)}$ and $F_{(+-)} = 1- F_{(-+)}$. For
         the sake of comparison, the constraints resulting from a CEPC run at
         $240$\,GeV with $5\inab$ of integrated luminosity collected without
         beam polarization are also shown. The dark shades correspond to the
         constraints obtained when one single parameter is kept at the time,
         assuming all others vanish.
         }
\label{fig:fiti1}
\end{figure}
The possibility of longitudinal beam polarization constitutes a distinct
advantage for linear colliders. Implementing it at circular colliders may be
difficult (especially at high center-of-mass energies) and not economically
feasible~\cite{Gomez-Ceballos:2013zzn}. Dividing the total luminosity into
multiple runs of different polarization configurations effectively provides
several independent observables and helps constraining different direction of
the effective-theory parameter space.
In \autoref{fig:fiti1}, we examine what subdivision of the total ILC luminosity
at $250$\,GeV would optimize the final precision reach. We follow the ILC
TDR~\cite{Baer:2013cma} and assume that the ILC could achieve a maximum beam
polarization of $80\%$ for electrons and $30\%$ for positrons.
Ref.~\cite{Barklow:2015tja} proposes a run plan with four polarization
configurations $\sign\{P(e^-,e^+)\}=$ $(-,+)$, $(+,-)$, $(-,-)$, $(+,+)$ and
corresponding luminosity fractions of $67.5\%$, $22.5\%$, $5\%$, and $5\%$,
respectively. The $(-,-)$ and $(+,+)$ polarizations could serve to probe exotic
new physics, like electron dipole or Yukawa operators. They however suppress the
rate of Higgs and gauge boson production and are thus not very helpful for the
precision study of these processes. For simplicity, we will thus only consider
the $(-,+)$ and $(+,-)$ polarizations. Uncertainty estimates are often only
provided for an entire run in the $P(e^-,e^+)=(-0.8,+0.3)$ configuration.
Scaling with statistics is performed to obtain estimates for other scenarios,
assuming no correlation among the measurements carried out with different
polarizations.
In agreement with the proposal of Ref.~\cite{Barklow:2015tja},
\autoref{fig:fiti1} shows that the best overall constraints are obtained with
about $70\%$ of the $2\inab$ ILC luminosity at $250$\,GeV spent with
$P(e^-,e^+)=(-0.8,+0.3)$ beam polarization and $30\%$ with
$P(e^-,e^+)=(+0.8,-0.3)$. The $(-0.8,+0.3)$ polarization enhances the cross
section, while the $(+0.8,-0.3)$ one helps resolving degeneracies. In terms of
GDP, this optimal repartition of luminosity provides results that are $14\%$
better than a full run with $P(e^-,e^+)=(-0.8,+0.3)$ beam polarization.
For comparison, we also show the reach of a $240$\,GeV CEPC run with $5\inab$ of
integrated luminosity and unpolarized beams. The higher luminosity is able
compensate for the lack of polarization and comparable overall results are
obtained. This is further quantified by \autoref{fig:gdp_lu_log} which displays
the GDP of our eleven-parameter fit as a function of luminosity collected at
$250$\,GeV with polarized beams and at $240$\,GeV with unpolarized ones. It is
notably seen that only about $1.5\inab$ of additional luminosity are required
without polarized beams to match the overall performance obtained with $2\inab$
of polarized beams. With $5\inab$ and $10\inab$ collected at $240$\,GeV, the
CEPC and FCC-ee reach GDPs respectively $14\%$ and $34\%$ smaller
than that of the full ILC run ($2\inab$) at $250$\,GeV.

\begin{figure}\centering
\includegraphics[scale=.7]{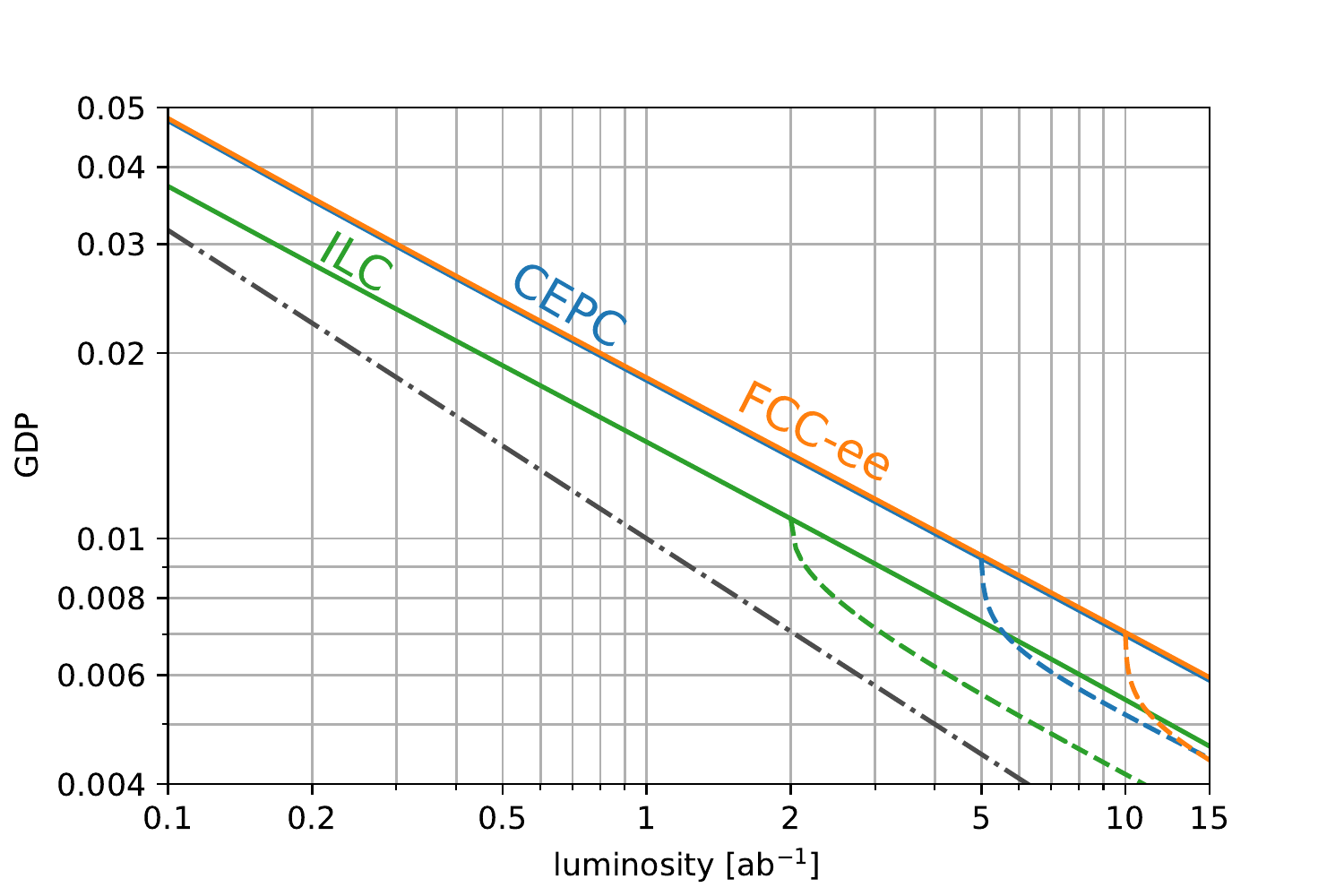}
\caption{Global GDP strength of the constraints in our eleven-dimensional
         parameter space as a function the luminosity collected, without beam
         polarization, at a center-of-mass energy of $240$\,GeV (for the CEPC
         and FCC-ee) and at $250$\,GeV, with polarized beams (for the ILC). The
         dashed lines show the improvements brought by subsequent runs at
         $350$\,GeV. A pure statistical scaling of constraints, in the absence
         of systematic uncertainties, would have led to the slope of the
         dot-dashed line.}
\label{fig:gdp_lu_log}
\end{figure}

\paragraph{Impact of systematic uncertainties in diboson production}
\begin{figure}
\centering
\includegraphics[width=13cm]{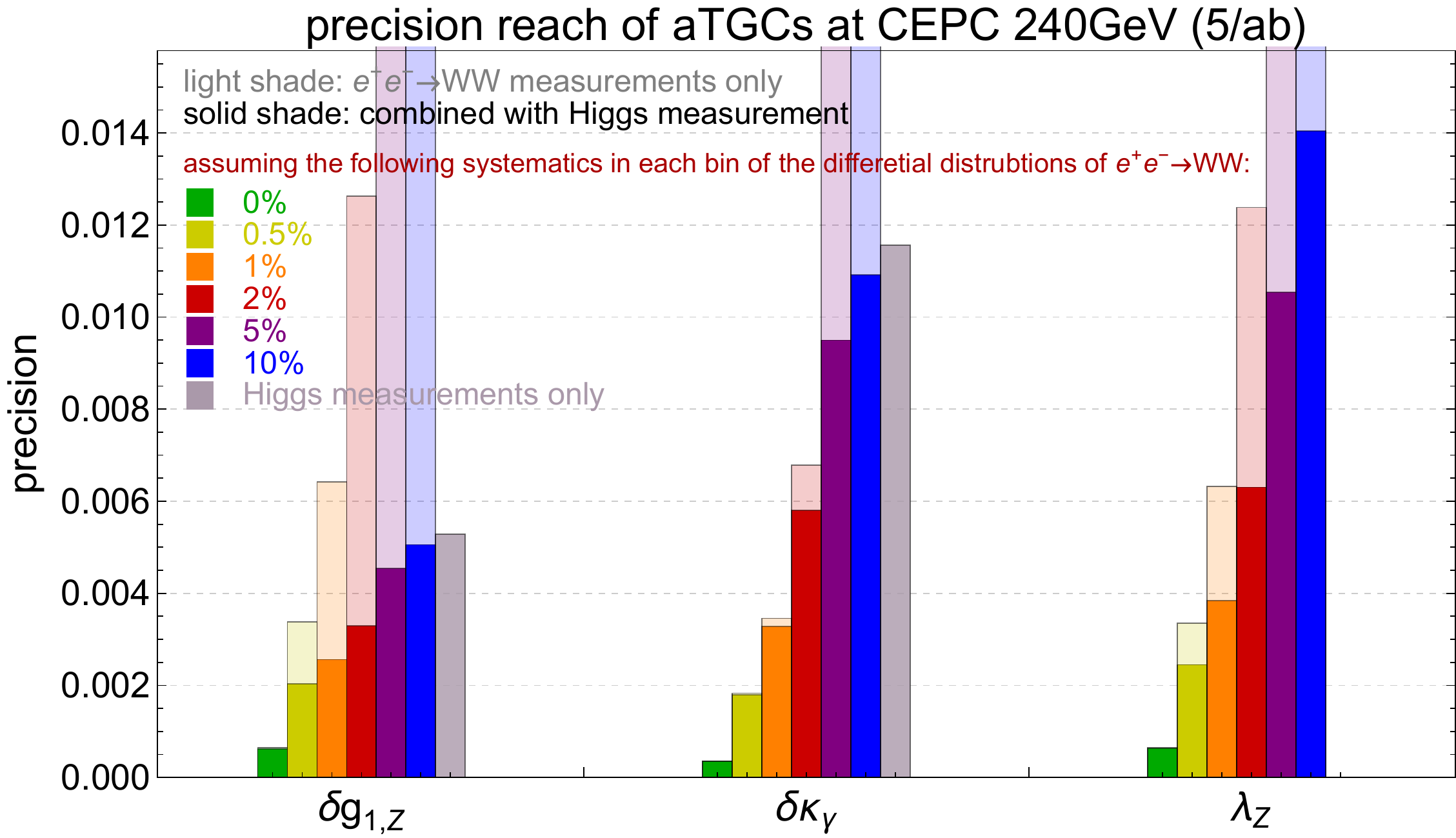} \\ \vspace{0.5cm}
\includegraphics[width=14cm]{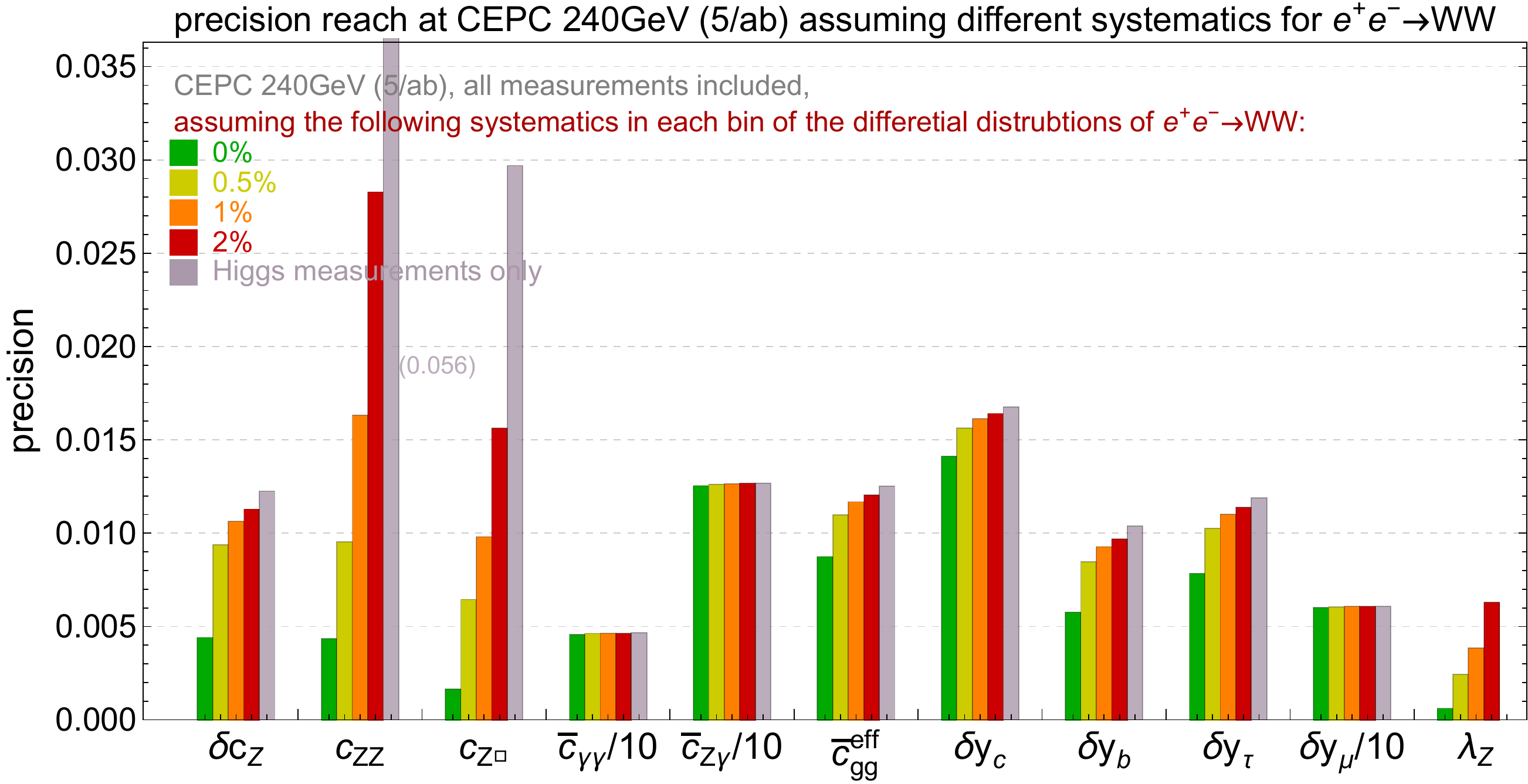}
\caption{\emph{Top:} One-sigma constraints on aTGCs parameters for different
         assumptions about the systematic uncertainties affecting the $\eeww$
         measurements at the CEPC. Each of the five angular distributions is
         divided into 20 bins (or 10 bins for the angles characterizing $W$
         decays in indistinguishable quark--antiquark pairs). We assume a fixed
         relative uncertainty each bin, and no correlation among them. A
         benchmark value of $1\%$ is used elsewhere in this paper, for CEPC and
         FCC-ee measurements. \emph{Bottom:} One-sigma reach of the $240$\,GeV
         CEPC run for different systematic uncertainties in the differential
         measurements of diboson production.}
\label{fig:fittgc1}
\end{figure}
Another important issue concerns the impact of systematic uncertainties on the
constraints deriving from $\eeww$ measurements. As discussed earlier, they have
not yet been determined by dedicated experimental studies except for the
500\,GeV ILC run. The top panel of \autoref{fig:fittgc1} focuses on the aTGC
parameters $\delta g_{1,Z}$, $\delta \kappa_\gamma$, and $\lambda_Z$. Systematic
uncertainties ranging between $0$ and $10\%$ are assumed in each bin of the
$\eeww\to4f$ angular distributions. The constraints derived from diboson
production only are shown in lighter shades. Darker shades show their
combination with Higgs measurements, which alone give the gray limits (leaving
$\lambda_Z$ unconstrained). It is noted that the Higgs measurement constraints
on these TGC parameters are improved as soon as uncertainties fall below $10\%$.
On the other hand, Higgs measurements still bring improvements to aTGCs
determination when systematic uncertainties fall to $0.5\%$. The improvement on
the whole fit brought by the combination of Higgs and $\eeww\to4f$ measurements
of varying systematic uncertainties is displayed in \autoref{fig:fittgc1}. As
shown in \autoref{eq:tgchb}, two of the TGC parameters, $\delta g_{1,Z}$, and
$\delta \kappa_\gamma$, are related to $c_{ZZ}$, $c_{Z\square}$,
$c_{\gamma\gamma}$, $c_{Z\gamma}$ in Higgs measurements. The constraints on the
$c_{ZZ}$, $c_{Z\square}$, and $\lambda_Z$ effective-field-theory parameters are
benefiting the most from a reduction of systematic uncertainties in diboson
production measurements. Improvements are limited for $c_{\gamma\gamma}$ and
$c_{Z\gamma}$ which are already well constrained by the measurements of the
$h\to\gamma\gamma$ and $h\to Z\gamma$ decays.

As our conservative estimate of systematic uncertainties often render the
$\eeww$ measurements systematics dominated, we show for comparison the results
that would have been obtained with perfect TGC constraints with red stars on
\autoref{fig:fit0}. The \autoref{fig:fit0tgc0} in \autoref{app:more} also
contains more detailed fit results under this assumption.

\paragraph{Comparison with previous global analyses} Comparing our results with
the ones in Refs.~\cite{Ellis:2015sca, Ellis:2017kfi}, a few important
differences should be noted. Different assumptions have been made for the run
plans of some colliders. In particular, we adopted the most recent ILC run
scenario for ILC described in Refs.~\cite{Fujii:2015jha, Barklow:2015tja}, while
Ref.~\cite{Ellis:2015sca} assumed only $250\infb$ would be collected at
$250$\,GeV. Contrarily to Refs.~\cite{Ellis:2015sca, Ellis:2017kfi}, we also
lift the flavor-universality assumption on Yukawa modifications. Information
about the $\eehz$ angular distributions were not included in these existing
global analyses. Our respective treatments for the TGC measurements, for which
no dedicated experimental study has been carried out, is also different.
Reference~\cite{Ellis:2017kfi} also addressed the potential of CLIC in probing
the Higgsstrahlung process and measuring TGCs at 1.4 and 3\,TeV. Their
estimations for the measurements of $\sigma(hZ)\times \br(h\to b\bar{b})$ at 1.4
and 3\,TeV are adopted in our study.


\section{Conclusions}
\label{sec:con}

Future lepton colliders running at center-of-mass energies of around $240$\,GeV
and above are ideal to narrow down the Higgs boson properties, examine the fine
details of the electroweak symmetry breaking mechanism and indirectly reveal new
physics. Applicable in a low-energy limit, the standard-model effective field
theory provides a consistent model-independent framework to parametrize
systematically the theory space in direct vicinity of the standard model. We
performed, in this paper, a global effective-field-theory analysis of
measurements planned at the CEPC, ILC, FCC-ee and CLIC. A basis-independent
metric, GDP ratios, was introduced to assess the global strengthening of
constraints obtained in different scenarios. We stressed that a
consistent effective-field-theory treatment should be global and that the
combination of various measurements is crucial to constrain effectively all
directions of its multidimensional parameter space. We considered the $\eehz$
rates in its various channels as well as angular distributions, the measurement
of Higgs production through weak-boson fusion, and that of weak-boson pair
production sensitive to anomalous triple gauge couplings which are related to
Higgs interactions. Under assumptions discussed in detail, a twelve-dimensional
parameter space describes effective-field-theory contributions to the above
observables. We demonstrated that measurements carried out at different
center-of-mass energies, or with different beam polarizations, are very
effective in resolving approximate degeneracies among effective-field-theory
parameters. While circular colliders could collect more luminosity, their linear
analogues can reach higher center-of-mass energies and implement longitudinal
beam polarizations. High luminosities collected at hadron collider where
production rates are often much larger also help constraining rare but clean
processes. In that matter, future circular lepton colliders are likely to give
way to the next generation of proton colliders reaching $100$\,TeV
center-of-mass energies (the SppC after CEPC~\cite{CEPC-SPPCStudyGroup:2015csa}
and FCC-hh after FCC-ee \cite{fccplan}).

Several improvements to the present analysis are possible. Considering a
more complete set of differential observables could obviously strengthen
slightly the overall constraints. Quantities that suffer from reduced
statistical or systematical uncertainties could also be studied. An important
issue which remains to be examined in details concerns the loop-level handles
available to constrain the top Yukawa operator below the $t\bar{t}h$ threshold,
and around Higgsstrahlung cross section peak in particular. What impact our
limited knowledge of electroweak precision observables would have in a global
analysis is also a question that should be addressed. Runs at the $Z$ pole or
$WW$ threshold may indeed be required to take full advantage of the
higher-energy ones. Finally, experimental studies of the weak-boson pair
production would be highly valuable to make realistic estimates of the physics
potential of future lepton colliders in probing electroweak symmetry breaking.


\subsection*{Acknowledgments}
We thank Adam Falkowski, Zhen Liu, Marc Montull, Michael Peskin, Manqi Ruan, Wei Su, Lian-Tao Wang and
Xiangpeng Wang for useful discussions, and James Brau, John Ellis, Jenny List,
Zhen Liu, Philipp Roloff, Veronica Sanz and Tevong You for valuable comments on
the manuscript. We thank J\"urgen Reuter and Jenny List for pointing out the
importance of the ILC runs at higher energy, Michelangelo Mangano for
suggestions on the normalization of the plots, and Thibaud Vantalon for providing us with quadratic EFT dependences.
CG is supported by the European
Commission through the Marie Curie Career Integration Grant 631962 and by the
Helmholtz Association through the recruitment initiative. JG and KW are
supported by an International Postdoctoral Exchange Fellowship Program between
the Office of the National Administrative Committee of Postdoctoral Researchers
of China (ONACPR) and DESY. We also thank the Collaborative Research Center
SFB676 of the Deutsche Forschungsgemeinschaft (DFG), Particles, Strings and the
Early Universe, for support.


\appendix


\section{Effective-field-theory parameter definitions}
\label{app:basis}

We define here our effective-field-theory parameters which are closely related
to that of the Higgs basis~\cite{Falkowski:2001958}. As explained in
\autoref{sec:eft}, our framework is based on that of
Ref.~\cite{Falkowski:2015fla, Falkowski:2015jaa} where electroweak precision
observables are assumed to be standard-model like, and where fermion dipole as
well as CP-odd operators are discarded. The assumption of flavor universality
for Yukawa operators is however relaxed and we include possible modifications of
that of the top, charm, bottom, tau, and muon. The expression of our twelve
effective-field-theory parameters in the SILH' basis of dimension-six operators
is provided at the end of this section.

The relevant terms in the potential are
\begin{equation}
\La \supset \La_{hVV} + \La_{hff} + \La_{\rm tgc}\,,  \label{eq:latot}
\end{equation}
where the coupling of Higgs boson to a pair of SM gauge bosons are given by
\begin{align}
\La_{hVV} = &~ \frac{h}{v} \bigg[ (1+\delta c_W) \frac{g^2 v^2}{2} W^+_\mu W^-_\mu + (1+\delta c_Z) \frac{(g^2+g'^2)v^2}{4} Z_\mu Z_\mu \nonumber\\
&~~~ + c_{WW} \, \frac{g^2}{2}W^+_{\mu\nu}W^-_{\mu\nu} +c_{W\square}\, g^2(W^-_\mu \partial_\nu W^+_{\mu\nu} +{\rm h.c.}) \nonumber\\
&~~~ + c_{gg} \, \frac{g^2_s}{4} G^a_{\mu\nu} G^a_{\mu\nu} + c_{\gamma\gamma} \, \frac{e^2}{4} A_{\mu\nu} A_{\mu\nu} + c_{Z\gamma} \, \frac{e\sqrt{g^2+g'^2}}{2} Z_{\mu\nu} A_{\mu\nu} \nonumber\\
&~~~ + c_{ZZ} \, \frac{g^2+g'^2}{4} Z_{\mu\nu} Z_{\mu\nu} + c_{Z\square} \, g^2 Z_\mu \partial_\nu Z_{\mu\nu} + c_{\gamma\square} \, g g' Z_\mu \partial_\nu A_{\mu\nu} \, \bigg] \,. \label{eq:hvv}
\end{align}
Not all the couplings in \autoref{eq:hvv} are independent. In particular,
imposing gauge invariance, we rewrite the following couplings as
\begin{align}
\delta c_W =&~ \delta c_Z + 4\delta m \,,  \nonumber\\
c_{WW} =&~ c_{ZZ} + 2s^2_{\theta_W} c_{Z\gamma} + s^4_{\theta_W} c_{\gamma\gamma} \,, \nonumber\\
c_{W\square} =&~ \frac{1}{g^2-g'^2} \left[  g^2 c_{Z\square} + g'^2 c_{ZZ} - e^2 s^2_{\theta_W} c_{\gamma\gamma} - (g^2-g'^2) s^2_{\theta_W} c_{Z\gamma}  \right]  \,, \nonumber \\
c_{\gamma\square} =&~ \frac{1}{g^2-g'^2} \left[  2 g^2 c_{Z\square} + (g^2+g'^2)c_{ZZ} - e^2 c_{\gamma\gamma} - (g^2-g'^2) c_{Z\gamma}  \right]  \,, \label{eq:cwtocz}
\end{align}
where $\delta m$ parameterizes custodial symmetry breaking effects which are set
to zero in our framework. While the modifications of Yukawa couplings are, in
general, $3\times 3$ complex matrices in generation space, we only consider the
diagonal elements relevant for the measurements considered, which are
\begin{equation}
\La_{hff} = -\frac{h}{v} \sum_{f = t,c,b,\tau,\mu} m_{f} (1+\delta y_f ) \bar{f}_{R} f_{L} + {\rm h.c.} \,.  \label{eq:deltayf}
\end{equation}
Finally, the triple gauge couplings are given by
\begin{align}
\La_{\rm tgc}  ~=~&~~~ i g s_{\theta_W}  A^\mu (W^{-\nu} W^+_{\mu\nu} - W^{+\nu} W^-_{\mu\nu}) \nonumber\\
&~ + ig (1+\delta g^Z_1) c_{\theta_W} Z^\mu (W^{-\nu} W^+_{\mu\nu} - W^{+\nu} W^-_{\mu\nu}) \nonumber\\
&~ + ig\left[ (1+\delta \kappa_Z) c_{\theta_W} Z^{\mu\nu} + (1+ \delta \kappa_\gamma) s_{\theta_W} A^{\mu\nu}\right] W^-_\mu W^+_\nu  \nonumber\\
&~ + \frac{ig}{m^2_W} (\lambda_Z c_{\theta_W} Z^{\mu\nu} + \lambda_\gamma s_{\theta_W} A^{\mu\nu}) W^{-\rho}_v W^+_{\rho\mu} \,, \label{eq:tgc}
\end{align}
where $V_{\mu\nu} \equiv \partial_\mu V_\nu - \partial_\nu V_\mu$ for $V =
W^\pm,\, Z, \, A$. Imposing gauge invariance, one obtains $\delta \kappa_Z =
\delta g_{1,Z} - t^2_{\theta_W} \delta \kappa_\gamma$ and $\lambda_Z =
\lambda_\gamma$, and the contribution from NP can be parameterized by 3 aTGCs,
$\delta g_{1,Z}$, $\delta \kappa_\gamma$ and $\lambda_Z$.
$\delta g_{1,Z}$ and $\delta \kappa_\gamma$ are related to the Higgs observables and can be expressed as
\begin{align}
\delta g_{1,Z} =&~ \frac{1}{2(g^2-g'^2)} \left[  -g^2(g^2+g'^2) c_{Z\square} -g'^2(g^2+g'^2)c_{ZZ} + e^2 g'^2 c_{\gamma\gamma} + g'^2(g^2-g'^2)c_{Z\gamma}  \right]  \,, \nonumber \\
\delta \kappa_\gamma =&~ -\frac{g^2}{2} \left(  c_{\gamma\gamma} \frac{e^2}{g^2+g'^2} + c_{Z\gamma}\frac{g^2-g'^2}{g^2+g'^2} - c_{ZZ}   \right) \, . \label{eq:tgchb}
\end{align}

To summarize, under the assumptions we make, the contribution from dimension-six operators to the potential in \autoref{eq:latot} can be parameterized by the following non-redundant set of 12 parameters:
\begin{equation}
	\delta c_Z		\,,~~
	c_{ZZ}			\,,~~
	c_{Z\square}		\,,~~
	c_{\gamma\gamma}	\,,~~
	c_{Z\gamma}		\,,~~
	c_{gg}			\,,~~
	\delta y_t		\,,~~
	\delta y_c		\,,~~
	\delta y_b		\,,~~
	\delta y_\tau		\,,~~
	\delta y_\mu		\,,~~
	\lambda_Z		\,.
\label{eq:para10ap}
\end{equation}

It is straightforward to translate results obtained in the Higgs basis to other
bases of dimension-six operators. While all non-redundant basis
are equivalent, we found the one listed in \autoref{tab:op1} particularly
convenient under our assumption that the to $Z$-pole and
$W$-mass measurements are perfectly standard-model like. In this basis, the 12
parameters of \autoref{eq:para10ap} are replaced by the following ones,
\begin{align}
\La_{\rm D6} =&~ \frac{c_{H}}{v^2} \mathcal{O}_{H} + \frac{\kappa_{WW}}{m^2_W} \mathcal{O}_{WW} +  \frac{\kappa_{BB}}{m^2_W} \mathcal{O}_{BB} +  \frac{\kappa_{HW}}{m^2_W} \mathcal{O}_{HW} +  \frac{\kappa_{HB}}{m^2_W} \mathcal{O}_{HB}  \nonumber\\
&~~~~ +  \frac{\kappa_{GG}}{m^2_W} \mathcal{O}_{GG}
+ \frac{\kappa_{3W}}{m^2_W} \mathcal{O}_{3W} + \!\!\! \sum_{f=t,c,b,\tau,\mu} \! \frac{c_{y_f}}{v^2} \mathcal{O}_{y_f}  \,,  \label{eq:lad6}
\end{align}
where the normalization of the parameters are also defined. To go from the SILH'
basis~\cite{Elias-Miro:2013mua, Pomarol:2013zra} to the one in
\autoref{tab:op1}, one simply trades $\mathcal{O}_{W} \,, \mathcal{O}_{B}\to
\mathcal{O}_{WW}\,, \mathcal{O}_{WB}$ using
\begin{align}
\mathcal{O}_B = &~ \mathcal{O}_{HB} + \frac{1}{4} \mathcal{O}_{BB} + \frac{1}{4} \mathcal{O}_{WB}  \,, \nonumber \\
\mathcal{O}_W = &~ \mathcal{O}_{HW} + \frac{1}{4} \mathcal{O}_{WW} + \frac{1}{4} \mathcal{O}_{WB}  \,,
\end{align}
where $\mathcal{O}_{WB}$ is directly related to the $Z$-pole measurements
($S$-parameter) and is thus eliminated. The basis in \autoref{tab:op1} is also
used in Ref.~\cite{Butter:2016cvz}
 with a different notation. In particular, the $\mathcal{O}_{HW}$ and $\mathcal{O}_{HB}$ in
\autoref{tab:op1} are denoted as $\mathcal{O}_{W}$ and $\mathcal{O}_{B}$ in
those references, which are different from the $\mathcal{O}_{W}$ and
$\mathcal{O}_{B}$ in the SILH convention.

\begin{table}[t]
\centering
\begin{tabular}{l|l} \hline\hline
$\mathcal{O}_H = \frac{1}{2} (\partial_\mu |H^2| )^2$ &  $\mathcal{O}_{GG} =  g_s^2 |H|^2 G^A_{\mu\nu} G^{A,\mu\nu}$  \\ 
$\mathcal{O}_{WW} =  g^2 |H|^2 W^a_{\mu\nu} W^{a,\mu\nu}$  & $\mathcal{O}_{y_u} = y_u |H|^2 \bar{Q}_L \tilde{H} u_R$  \\
$\mathcal{O}_{BB} =  g'^2 |H|^2 B_{\mu\nu} B^{\mu\nu}$ &  $\mathcal{O}_{y_d} = y_d |H|^2 \bar{Q}_L H d_R$    \\
$\mathcal{O}_{HW} =  ig(D^\mu H)^\dagger \sigma^a (D^\nu H) W^a_{\mu\nu}$  &  $\mathcal{O}_{y_e} = y_e |H|^2 \bar{L}_L H e_R$  \\
$\mathcal{O}_{HB} =  ig'(D^\mu H)^\dagger  (D^\nu H) B_{\mu\nu}$ &      $\mathcal{O}_{3W} = \frac{1}{3!} g \epsilon_{abc} W^{a\,\nu}_\mu W^b_{\nu \rho} W^{c\,\rho\mu} $  \\   \hline\hline
\end{tabular}
\caption{A complete set of CP-even dimension-six operators that contribute to
         the Higgs and TGC measurements, assuming there is no correction to the
         $Z$-pole and $W$~mass measurements and no dipole interaction. We only
         consider the flavor-conserving component of $\mathcal{O}_{y_u}$,
         $\mathcal{O}_{y_d}$ and $\mathcal{O}_{y_e}$ contributing to the top,
         charm, bottom, tau, and muon Yukawa couplings.
}
\label{tab:op1}
\end{table}

The aTGCs in this basis are given by
\begin{align}
\delta g_{1,Z} =&~ - \frac{\kappa_{HW}}{c^2_{\theta_W}} \,, \nonumber\\
\delta \kappa_\gamma =&~  -  \kappa_{HW} - \kappa_{HB} \,, \nonumber\\
\lambda_Z =&~ - \kappa_{3W} \,,  \label{eq:tgctrans}
\end{align}
which are obtained from the general results in Ref.~\cite{Falkowski:2014tna}.
Finally, the expression of our effective-field-theory parameters in terms of the
operators in \autoref{tab:op1} are:
\begin{align}
\delta c_Z
	=&~ -\frac{1}{2}\, c_H
	\,, \nonumber \\[0.1cm]
c_{ZZ} 
	=&~  \frac{4}{g^2+g'^2} (
		-\kappa_{HW}
		- t^2_{\theta_W} \kappa_{HB}
		+ 4\, c^2_{\theta_W} \kappa_{WW}
		+ 4 \, t^2_{\theta_W} s^2_{\theta_W} \kappa_{BB}
		)	\,, \nonumber\\[0.1cm]
c_{Z\square} 
	=&~  \frac{2}{g^2} (
		\kappa_{HW}
		+ t^2_{\theta_W} \kappa_{HB}
		)	\,, \nonumber\\[0.1cm]
c_{\gamma\gamma} 
	=&~ \frac{16}{g^2} (
		\kappa_{WW} 
		+ \kappa_{BB}
		)	\,, \nonumber\\[0.1cm]
c_{Z\gamma}
	=&~ \frac{2}{g^2} (
		\kappa_{HB}
		- \kappa_{HW}
		+ 8\, c^2_{\theta_W} \kappa_{WW}
		-  8\, s^2_{\theta_W} \kappa_{BB}
		)	\,, \nonumber\\[0.1cm]
c_{gg}
	=&~  \frac{16}{g^2}\, \kappa_{GG}
	\,,	\nonumber\\[0.1cm]
\delta y_f
	=&~ -\frac{1}{2}\, c_H - c_{y_f}
	\,.
\label{eq:basistrans}
\end{align}
It should be noted that \autoref{eq:basistrans} is only valid under the
assumptions made in this paper. More general basis translations from the Higss
basis to the SILH' basis (and others) are provided in
Ref.~\cite{Falkowski:2001958}.


\section{Measurement inputs}
\label{app:input}
We provide here additional details about the input measurements used in our
study, including the Higgs production rates ($\eehz$ and $\eevvh$), the angular
asymmetries in $\eehz$ and TGC measurements from $\eeww$. The estimated
one-sigma precisions of the Higgs rate measurements are respectively displayed
in \autoref{tab:higgsinputc} for the CEPC and FCC-ee, in
\autoref{tab:higgsinputi} for ILC and, in \autoref{tab:higgsinputclic} for CLIC.
When provided, the are respectively extracted from Ref.~\cite{CEPCupdate2} for
the CEPC (which updates the preCDR~\cite{CEPC-SPPCStudyGroup:2015csa}),
Ref.~\cite{Gomez-Ceballos:2013zzn} for the FCC-ee, Ref.~\cite{Barklow:2015tja}
for the ILC and Ref.~\cite{Abramowicz:2016zbo} for CLIC. For CLIC, we
also include the estimations for $\sigma(hZ)\times \br(h\to b\bar{b})$ at 1.4
and 3 TeV from Ref.~\cite{Ellis:2017kfi}. While these measurements suffer from
smaller cross sections, they nevertheless significantly improve the
constraints on $c_{ZZ}$ and $c_{Z\square}$ due to the huge sensitivities at high
energies.\footnote{We thank Tevong You for pointing this out.} 
We also found the $ZZ$ fusion measurements at CLIC (with $\sigma(e^+e^-h)\times \br(h\to b\bar{b})$ measured to a precision
of $1.8\%$ (2.3\%) at 1.4\,TeV (3\,TeV) \cite{Abramowicz:2016zbo}) to have a negligible impact in our analysis.\footnote{It is nevertheless possible to further optimize the precision reach of the cross section measurements of $ZZ$ fusion using judicious kinematic cuts, as pointed out in Ref.~\cite{Han:2015ofa}.  For simplicity, we do not perform such optimizations in our study.}
The numbers highlighted in {\color{BlueGreen} green} are obtained by scaling with
luminosity when dedicated estimates are not available. For the ILC, the
estimations of signal strengths are summarized in Ref.~\cite{Barklow:2015tja}
(Table~13) but only for benchmark run scenarios with smaller
luminosities. These are scaled up to the current run plan. For the 350\,GeV run
of CEPC and FCC-ee, relative precision are rescaled from the 350\,GeV ILC
ones.\footnote{A statistical precision of 0.6\% is reported in
Ref.~\cite{Gomez-Ceballos:2013zzn} for the $\sigma(\vvh)\times {\rm BR}(h\to
b\bar{b})$ measurement at FCC-ee 350\,GeV, which is in good agreement with our
estimation from scaling ($0.71\%$).} The precision of $\sigma(hZ)\times {\rm
BR}(h\to Z\gamma)$ is not provided for the FCC-ee and ILC. We thus scale it from
the CEPC estimation.
While a statistical precision of 2.2\% is reported in
Ref.~\cite{Gomez-Ceballos:2013zzn} for the $\sigma(\vvh)\times {\rm BR}(h\to
b\bar{b})$ measurement at FCC-ee 240\,GeV, it is not clear what assumptions on
the $\eehz, Z\to \nu\bar{\nu}$ process are made in obtaining this estimation.
Therefore, we scale it with luminosity from the CEPC one.
The difference between unpolarized and polarized cross sections are
taken into account in these rescalings. Given the moderate statistics in most of
the relevant channels, it is reasonable to assume their precision is statistics
limited. Nevertheless, it is important for these estimations to be updated by
experimental groups in the future.
\begin{table}[t]\footnotesize
\centering
\begin{tabular}{|c||cc|cc||cc|cc|} \hline
& \multicolumn{4}{c||}{CEPC} & \multicolumn{4}{c|}{FCC-ee} \\  \hline
&   \multicolumn{2}{c|}{[240\,GeV, $5\inab$]} & \multicolumn{2}{c||}{[350\,GeV, $200\infb$]} &   \multicolumn{2}{c|}{[240\,GeV, $10\inab$]} & \multicolumn{2}{c|}{[350\,GeV, $2.6\inab$]}  \\  \hline 
production &  $Zh$  &  $\vvh$ & $Zh$ & $\vvh$ &  $Zh$  &  $\vvh$ & $Zh$ & $\vvh$ \\  \hline
$\sigma$  &  0.50\%  &  -  &  \hl{2.4\%} &  -    & 0.40\%  & - & \hl{0.67\%}   &  -       \\  \hline\hline
&  \multicolumn{4}{c||}{$\sigma \times {\rm BR}$} & \multicolumn{4}{c|}{$\sigma \times {\rm BR}$}      \\  \hline
$h\to b\bar{b}$  & $0.21\%^\bigstar$  & 0.39\%$^\diamondsuit$  & \hl{2.0\%}   &  \hl{2.6\%}  & 0.20\%  & \hl{0.28\%}$^\diamondsuit$    &  \hl{0.54\%}   & \hl{0.71\%}   \\ 
$h\to c\bar{c}$  & 2.5\%  & -   & \hl{15\%}   & \hl{26\%}                                & 1.2\%  & -  & \hl{4.1\%}    & \hl{7.1\%} \\
$h\to gg$          & 1.2\%   & -  & \hl{11\%}  & \hl{17\%}                                  & 1.4\% & -  &  \hl{3.1\%}   &  \hl{4.7\%}  \\
$h\to \tau\tau$  & 1.0\%   & -  & \hl{5.3\%}  &  \hl{37\%}                                & 0.7\%  & -  &  \hl{1.5\%}   & \hl{10\%}  \\
$h\to WW^*$    & 1.0\%   & -  & \hl{10\%}  &  \hl{9.8\%}                                & 0.9\%  & -  & \hl{2.8\%}    & \hl{2.7\%}  \\
$h\to ZZ^*$      & 4.3\%    & -  & \hl{33\%}  &  \hl{33\%}                                & 3.1\%  & -  &  \hl{9.2\%}   & \hl{9.3\%}  \\
$h\to \gamma\gamma$ & 9.0\% & - & \hl{51\%}  &  \hl{77\%}                       & 3.0\%  & -  & \hl{14\%}    & \hl{21\%}   \\
$h\to \mu\mu$  & 12\%    & -  & \hl{115\%}  & \hl{275\%}                              & 13\%  & -  & \hl{32\%}    & \hl{76\%}  \\ 
$h\to Z\gamma$ & 25\%  & -  & \hl{144\%}  &   -                                           & \hl{18\%}   & - &  \hl{40\%}   & -  \\  \hline
\end{tabular}
\caption{The estimated precision of CEPC and FCC-ee Higgs measurements. We
         gather the available estimations from
         Refs.~\cite{CEPC-SPPCStudyGroup:2015csa, Gomez-Ceballos:2013zzn,
         CEPCupdate2}, while the missing ones (highlighted in {\color{BlueGreen}
         green}) are obtained from scaling with luminosity. See
         \autoref{app:input} for more details. For $\sigma(\eevvh)$, the precisions marked with a diamond
         $^\diamondsuit$ are normalized to the cross section of the inclusive
         channel which includes both the $WW$ fusion and $\eehz, Z\to
         \nu \bar{\nu}$, while the unmarked precisions are normalized to the $WW$ fusion process only. 
         For the CEPC, the precision of the $\sigma(hZ) \times
         {\rm BR}(h\to b\bar{b})$ measurement (marked by a star $^\bigstar$)
         reduces to 0.24\% if one excludes the contribution from $\eehz, Z\to
         \nu\bar{\nu}, h\to b\bar{b}$ to avoid double counting with $\eevvh,
         h\to b\bar{b}$. The corresponding information is not available for the
         FCC-ee.}  
\label{tab:higgsinputc}
\end{table}
\begin{table}[t]\footnotesize
\centering
\begin{adjustbox}{max width=\textwidth}
\begin{tabular}{|c||cc|cc|ccc||cc|cc|} 
 \multicolumn{12}{c}{\normalsize ILC}  \\  \hline
&   \multicolumn{2}{c|}{[250\,GeV, $2\inab$]} & \multicolumn{2}{c|}{[350\,GeV, $200\infb$]} &   \multicolumn{3}{c||}{[500\,GeV, $4\inab$]}  &   \multicolumn{2}{c|}{\!\![1\,TeV, $1\inab$]\!\!} &   \multicolumn{2}{c|}{\!\!\![1\,TeV, $2.5\inab$]\!\!\!} \\  \hline 
production &  $Zh$  &  $\vvh$ & $Zh$ & $\vvh$ &  $Zh$  &  $\vvh$ & $t\bar{t}h$ &   $\vvh$ & $t\bar{t}h$ &   $\vvh$ & $t\bar{t}h$ \\  \hline
$\sigma$  & \hl{0.71\%}   &  -  &  \hl{2.1\%}    &  -  &  \hl{1.1\%}    &  -   &  -   &  -   &  - &  -   &  -     \\  \hline\hline
&  \multicolumn{11}{c|}{$\sigma \times {\rm BR}$}       \\  \hline
$h\to b\bar{b}$  & \hl{0.42\%}   &  \hl{3.7\%}  &  \hl{1.7\%}    &  \hl{1.7\%}   &  \hl{0.64\%} &  \hl{0.25\%}   &  \hl{9.9\%}  & 0.5\% & 6.0\% & 0.3\%  & 3.8\% \\ 
$h\to c\bar{c}$  & \hl{2.9\%}   &  -  &  \hl{13\%}    &  \hl{17\%} &  \hl{4.6\%}    &  \hl{2.2\%}   &  -    & 3.1\%  & - & 2.0\% & - \\ 
$h\to gg$          & \hl{2.5\%}   &  -  &  \hl{9.4\%}    &  \hl{11\%} &  \hl{3.9\%}    &  \hl{1.4\%}   &  -    & 2.3\%  & - & 1.4\% & - \\
$h\to \tau\tau$  & \hl{1.1\%}   &  -  &  \hl{4.5\%}    &  \hl{24\%} &  \hl{1.9\%}    &  \hl{3.2\%}  &  -     & 1.6\%  & - & 1.0\% & -  \\      
$h\to WW^*$    & \hl{2.3\%}   &  -  &  \hl{8.7\%}    &  \hl{6.4\%} &  \hl{3.3\%}    &  \hl{0.85\%}  &  -    & 3.1\%  & - & 2.0\% & - \\      
$h\to ZZ^*$      & \hl{6.7\%}   &  -  &  \hl{28\%}    &  \hl{22\%}  &  \hl{8.8\%}    &  \hl{2.9\%} &  -     & 4.1\%  & - & 2.6\%  & - \\
$h\to \gamma\gamma$ & \hl{12\%}   &  -  &  \hl{44\%}    &  \hl{50\%}  &  \hl{12\%}    &  \hl{6.7\%}   &  -   & 8.5\% & - & 5.4\% & -  \\
$h\to \mu\mu$  & \hl{25\%}   &  -  &  \hl{98\%}    &  \hl{180\%}  &  \hl{31\%}    &  \hl{25\%}  &  -   & 31\% & - & 20\%  & -  \\
$h\to Z\gamma$ & \hl{34\%}   &  -  &  \hl{145\%}    &  - &  \hl{49\%}    &  -  &  -   & - & - &-  & -  \\  \hline
\end{tabular}
\end{adjustbox}
\caption{The estimated precision of ILC Higgs measurements. For the 250\,GeV, 350\,GeV and 500\,GeV runs, all numbers are
         scaled from Ref.~\cite{Barklow:2015tja} (Table~13), except for
         $\sigma(hZ)\times {\rm BR}(h\to Z\gamma)$ which is scaled from the CEPC
         estimation. A beam polarization of $P(e^-,e^+)=(-0.8,+0.3)$ is
         assumed.  The 1\,TeV run is only included in \autoref{fig:fitilcn1} of \autoref{app:more}, while the estimations are taken from Ref.~\cite{Asner:2013psa}
         which assumes a polarization of $P(e^-,e^+)=(-0.8,+0.2)$.}
\label{tab:higgsinputi}
\end{table}
\begin{table}[ht!]\footnotesize \vspace{1cm}
\centering
\begin{tabular}{|c||cc|cc|c|} 
 \multicolumn{6}{c}{CLIC}  \\  \hline
&   \multicolumn{2}{c|}{[350\,GeV, $500\infb$]} & \multicolumn{2}{c|}{[1.4\,TeV, $1.5\inab$]} &   [3\,TeV, $2\inab$]   \\  \hline 
production &  $Zh$  &  $\vvh$ &  $\vvh$ & $t\bar{t}h$ &  $\vvh$  \\  \hline
$\sigma$                              & 1.6\%   & -  & -   &  -  & -  \\  \hline\hline
&  \multicolumn{5}{c|}{$\sigma \times {\rm BR}$}       \\  \hline
$h\to b\bar{b}$                     &  0.84\%  & 1.9\%  &  0.4\%  &  8.4\%   & 0.3\%  \\ 
$h\to c\bar{c}$                     &  10.3\%  & 14.3\%   &  6.1\%  &   -   & 6.9\%  \\ 
$h\to gg$                             &  4.5\%  & 5.7\%  &  5.0\%  &   -   & 4.3\%  \\ 
$h\to \tau\tau$                     &  6.2\%  & -   &  4.2\%  &   -   & 4.4\%  \\
$h\to WW^*$                       &  5.1\%  & -   & 1.0\%   &   -   & 0.7\%  \\
$h\to ZZ^*$                         &  -          & -   &  5.6\%  &   -   & 3.9\%  \\
$h\to \gamma\gamma$      &  -           &  -  &  15\%  &   -   &  10\%  \\
$h\to \mu\mu$                    &  -           &  -  &   38\%  &   -   &  25\%  \\
$h\to Z\gamma$                 &  -           &  -  &  42\%  &   -   & 30\%  \\ \hline
\end{tabular}
\caption{The estimated precision of CLIC Higgs measurements taken from
         Ref.~\cite{Abramowicz:2016zbo}, which assumes unpolarized beams and
         considers only statistical uncertainties. In addition, we also include
         the estimations for $\sigma(hZ)\times \br(h\to b\bar{b})$ at high
         energies in Ref.~\cite{Ellis:2017kfi}, which are 3.3\% (6.8\%) at
         1.4\,TeV (3\,TeV). 
         We find the inclusion of the $ZZ$ fusion
         ($\ee \to \ee h$) measurements to have little impact in our analysis.
         }
\label{tab:higgsinputclic}
\end{table}

The constraints from angular observables in $\eehz$ are obtained with the method
described in \autoref{sec:hzasy}, making use of the channels $\eehz\,,~Z\to
\ell^+ \ell^-\,,~h\to b\bar{b}$, $c\bar{c}$, $gg$. They are included for all the
$\ee$ colliders at all energies except for the 1.4\,TeV and 3\,TeV runs of CLIC.

The constraints on aTGCs derived from the $\eeww$ measurements are obtained
using the method described in \autoref{sec:ww}, for the CEPC and FCC-ee. In
particular, 1\% systematic uncertainties are assumed in each bin with the
differential distribution of each measured angle divided in 20 bins (10 bins if
the angle is \emph{folded}). The results, including the correlation matrices,
are shown in \autoref{tab:tgccepc} and \autoref{tab:tgcfcc}, which are fed into
the global fit. For ILC, the constraints are shown in \autoref{tab:tgcilc},
taken from Ref.~\cite{Marchesini:2011aka}, which assumes $500\infb$ data at
500\,GeV and four $P(e^-,e^+)=(\pm0.8,\pm0.3)$ beam polarization configurations.
For CLIC, we simply use the ILC results.

\begin{table}[ht]\footnotesize
\centering
\begin{tabular}{|c||c|ccc||c|ccc|} \hline
\multicolumn{9}{|c|}{CEPC}  \\  \hline
& \multicolumn{4}{c||}{ 240\,GeV($5\inab$)} &  \multicolumn{4}{c|}{ 240\,GeV($5\inab$)+350\,GeV($200\infb$) } \\  \cline{2-9}
& uncertainty &   \multicolumn{3}{c||}{correlation matrix}  &  uncertainty  &   \multicolumn{3}{c|}{correlation matrix}   \\ \cline{2-9}
&  &    $\delta g_{1,Z}$  &  $\delta \kappa_\gamma$  & $\lambda_Z$  &  &    $\delta g_{1,Z}$  &  $\delta \kappa_\gamma$  & $\lambda_Z$    \\ \hline
$\delta g_{1,Z}$                & 0.0064  &  1 & 0.068 & -0.93      &    0.0037          &    1 & -0.51 & -0.89    \\
$\delta \kappa_\gamma$  & 0.0035  &     & 1        & -0.40      &    0.0017          &       &  1      &  0.12    \\
$\lambda_Z$                     & 0.0063  &     &           &  1           &    0.0030          &       &          & 1          \\  \hline
\end{tabular}
\caption{The constraints on aTGCs from the $\eeww$ measurement at CEPC using the
         methods described in \autoref{sec:ww}. Both the results from the
         240\,GeV run alone and the ones from the combination of the 240\,GeV
         and 350\,GeV runs are shown.}
\label{tab:tgccepc}
\end{table}
\begin{table}[ht!]\footnotesize\vspace{1cm}
\centering
\begin{tabular}{|c||c|ccc||c|ccc|} \hline
\multicolumn{9}{|c|}{FCC-ee}  \\  \hline
& \multicolumn{4}{c||}{ 240\,GeV($10\inab$)} &  \multicolumn{4}{c|}{ 240\,GeV($10\inab$)+350\,GeV($2.6\inab$) } \\ \cline{2-9}
& uncertainty &   \multicolumn{3}{c||}{correlation matrix}  &  uncertainty  &   \multicolumn{3}{c|}{correlation matrix}   \\ \cline{2-9}
&  &    $\delta g_{1,Z}$  &  $\delta \kappa_\gamma$  & $\lambda_Z$  &  &    $\delta g_{1,Z}$  &  $\delta \kappa_\gamma$  & $\lambda_Z$    \\ \hline
$\delta g_{1,Z}$                & 0.0064  &  1 & 0.066 & -0.93      &    0.0029          &    1 & -0.61 & -0.88    \\
$\delta \kappa_\gamma$  & 0.0034  &     & 1        & -0.40      &    0.0014          &       &  1      &  0.19    \\
$\lambda_Z$                     & 0.0062  &     &           &  1           &    0.0022          &       &          & 1          \\  \hline
\end{tabular}
\caption{Same as \autoref{tab:tgccepc} but for FCC-ee.}
\label{tab:tgcfcc}
\end{table}
\begin{table}[ht!]\footnotesize\vspace{1cm}
\centering
\begin{tabular}{|c||c|ccc|} \hline
\multicolumn{5}{|c|}{ILC (CLIC)}  \\  \hline
& uncertainty &   \multicolumn{3}{c|}{correlation matrix}     \\   \cline{2-5}
&  &    $\delta g_{1,Z}$  &  $\delta \kappa_\gamma$  & $\lambda_Z$  \\ \hline
$\delta g_{1,Z}$                & $6.1\times 10^{-4}$  &  1 & 0.634 & 0.477         \\
$\delta \kappa_\gamma$  & $6.4\times 10^{-4}$  &     & 1        & 0.354          \\
$\lambda_Z$                     & $7.2\times 10^{-4}$  &     &           &  1               \\  \hline
\end{tabular}
\caption{The estimated statistical precision of aTGCs from the $\eeww$
         measurements at ILC in Ref.~\cite{Marchesini:2011aka}, assuming
         $500\infb$ of data equally shared between four
         $P(e^-,e^+)=(\pm0.8,\pm0.3)$ beam polarization configurations at
         500\,GeV. We use the same results for CLIC. No scaling with statistics
         or center-of-mass energy is performed, given that systematic
         uncertainties may become important.
         }
\label{tab:tgcilc}
\end{table}

While the measurement inputs of LHC and LEP measurements are too lengthy to be
reported in this paper, here we simply list the results from the global fits in
terms of one sigma constraints and the correlation matrix, which can be used to
reconstruct the chi-square. The chi-square can then be combined with the ones of
the future $\ee$ colliders to reproduce the results in \autoref{fig:fit0} and
\autoref{tab:result}. In \autoref{tab:cscurrent}, we list the current
constraints in Ref.~\cite{Falkowski:2015jaa}, obtained from the LHC 8\,TeV Higgs
measurements and LEP $\eeww$ measurements. While Ref.~\cite{Falkowski:2015jaa}
explicitly assumes flavor universality for the Yukawa couplings, it is a good
approximation to simply assume the constraints given there apply to
third-generation couplings. Since we explicitly assume the future results are
SM-like, for consistency, we also set the central values of current results to
zero when combining them with the future collider results. In
\autoref{tab:cslhc300} and \ref{tab:cslhc3000}, we list the results for the
14\,TeV LHC with $300\infb$ and $3000\infb$ luminosity, derived from projection
by the ATLAS collaboration~\cite{ATL-PHYS-PUB-2014-016} which collected
information from various other sources, while the information about the
composition of each channel are extracted from
Refs.~\cite{ATL-PHYS-PUB-2013-014, ATL-PHYS-PUB-2014-012, ATL-PHYS-PUB-2014-006,
ATL-PHYS-PUB-2014-011, ATL-PHYS-PUB-2014-018}. While $\delta y_c$ is set to zero
in obtaining these results (due to the fact that
Ref.~\cite{ATL-PHYS-PUB-2014-016} did not provide estimations for the decay
$h\to c\bar{c}$), it is not set to zero when the $\chi^2$ is combined with the
ones from future $\ee$ colliders. However, this has little impact on the results
of the combined fits.

\begin{table}[ht]\scriptsize
\centering
\begin{tabular}{|c||c|cccccccccc|} \hline
\multicolumn{12}{|c|}{LHC 8\,TeV Higgs + LEP $\eeww$}  \\  \hline
& uncertainty &   \multicolumn{10}{c|}{correlation matrix}     \\   \cline{2-12}
&  & $\delta c_Z$ & $c_{ZZ}$ & $c_{Z\square}$ & $c_{\gamma\gamma}$ & $c_{Z\gamma}$ & $c_{gg}$ & $\delta y_u$ & $\delta y_d$ & $\delta y_e$ &  $\lambda_Z$ \\ \hline
$\delta c_Z$ 			& 0.17       &  1 & -0.04 & -0.21 & -0.76 & -0.15 & 0.15 & 0.12 & 0.88 & 0.71 & -0.22 \\ 
$c_{ZZ}$      		        & 0.42       &     & 1 & -0.96 & 0.37 & 0.19 & 0.03 & 0.04 & -0.12 & -0.31 & -0.88 \\   
$c_{Z\square}$ 		& 0.19       &      &   & 1 & -0.17 & -0.10 & -0.07 & -0.06 & -0.10 & 0.12 & 0.93 \\ 
$c_{\gamma\gamma}$      & 0.015     &     &   &   & 1 & 0.20 & -0.12 & -0.07 & -0.79 & -0.74 & -0.13 \\ 
$c_{Z\gamma}$   	        & 0.098      &    &   &   &   & 1 & -0.01 & -0.01 & -0.15 & -0.18 & -0.10 \\ 
$c_{gg}$             		& 0.0027    &    &   &   &   &   & 1 & -0.87 & 0.26 & 0.17 & -0.07 \\ 
$\delta y_u$        		& 0.30        &     &   &   &   &   &   & 1 & 0.13 & 0.11 & -0.06 \\ 
$\delta y_d$       		& 0.35        &     &   &   &   &   &   &   & 1 & 0.81 & -0.11 \\ 
$ \delta y_e$  		        & 0.20         &       &   &   &   &   &   &   &   & 1 & 0.09 \\ 
$\lambda_Z$      	        & 0.073       &     &   &   &   &   &   &   &   &   & 1 \\   \hline
\end{tabular}
\caption{Current constraints on the Higgs basis parameters from
         Ref.~\cite{Falkowski:2015jaa}, obtained from the LHC 8\,TeV Higgs
         measurements and LEP $\eeww$ measurements. Flavor universality is
         imposed. To transform it into our framework we simply take $\delta y_u
         \to \delta y_t$, $\delta_d \to \delta y_b$, $\delta y_e \to \delta
         y_\tau$. For consistency we also set the central values to zero.}
\label{tab:cscurrent}
\end{table}
\begin{table}[ht]\scriptsize
\centering
\begin{tabular}{|c||c|cccccccccc|} \hline
\multicolumn{12}{|c|}{LHC 14\,TeV Higgs measurements ($300\infb$)}  \\  \hline
& uncertainty &   \multicolumn{10}{c|}{correlation matrix}     \\   \cline{2-12}
&  & $\delta c_Z$ & $c_{ZZ}$ & $c_{Z\square}$ & $c_{\gamma\gamma}$ & $c_{Z\gamma}$ & $c_{gg}$ & $ \delta y_t$ & $\delta y_b$ & $\delta y_\tau$ & $\delta y_\mu$   \\ \hline
$\delta c_Z$ 			&  0.116        &      1 & -0.029 & -0.037 & -0.61 & -0.29 & 0.037 & 0.36 & 0.88 & 0.50 & 0.43  \\ 
$c_{ZZ}$      		        &  0.960          &          & 1 & -0.996 & 0.77 & 0.52 & -0.17 & 0.43 & -0.21 & -0.73 & -0.55   \\   
$c_{Z\square}$ 		&  0.419        &          &   & 1 & -0.73 & -0.50 & 0.17 & -0.46 & 0.15 & 0.70 & 0.52   \\ 
$c_{\gamma\gamma}$      &  0.0156     &         &   &   & 1 & 0.57 & -0.18 & 0.13 & -0.69 & -0.85 & -0.68 \\ 
$c_{Z\gamma}$   	        &  0.0164      &           &   &   &   & 1 & -0.10 & 0.070 & -0.41 & -0.54 & -0.41 \\ 
$c_{gg}$             		&   0.00137  &           &   &   &   &   & 1 & -0.74 & 0.063 & 0.14 & 0.26   \\ 
$\delta y_t$        		&  0.220       &           &   &   &   &   &   & 1 & 0.42 & -0.094 & -0.20 \\ 
$\delta y_b$       		&   0.303      &             &   &   &   &   &   &   & 1 & 0.61 & 0.47 \\ 
$ \delta y_\tau$  		&   0.196      &          &   &   &   &   &   &   &   & 1 & 0.60 \\ 
$ \delta y_\mu$     		&   0.271      &             &   &   &   &   &   &   &   &   & 1   \\ \hline
\end{tabular}
\caption{One sigma constraints and the correlation matrix of the Higgs basis
         parameters from the LHC 14\,TeV Higgs measurements with $300\infb$
         data, using the ATLAS projection with no theory
         error~\cite{ATL-PHYS-PUB-2014-016}. $\delta y_c$ to set to zero since
         Ref.~\cite{ATL-PHYS-PUB-2014-016} did not provide estimations for the
         decay $h\to c\bar{c}$.}
\label{tab:cslhc300}
\end{table}
\begin{table}[ht]\scriptsize
\centering
\begin{tabular}{|c||c|cccccccccc|} \hline
\multicolumn{12}{|c|}{LHC 14\,TeV Higgs measurements ($3000\infb$)}  \\  \hline
& uncertainty &   \multicolumn{10}{c|}{correlation matrix}     \\   \cline{2-12}
&  & $\delta c_Z$ & $c_{ZZ}$ & $c_{Z\square}$ & $c_{\gamma\gamma}$ & $c_{Z\gamma}$ & $c_{gg}$ & $ \delta y_t$ & $\delta y_b$ & $\delta y_\tau$ & $\delta y_\mu$   \\ \hline
$\delta c_Z$ 			&  0.0500        &      1 & 0.0015 & -0.045 & -0.54 & -0.22 & 0.034 & 0.38 & 0.87 & 0.40 & 0.45  \\ 
$c_{ZZ}$      		        &  0.495          &          & 1 & -0.998 & 0.81 & 0.46 & -0.19 & 0.55 & -0.22 & -0.62 & -0.68   \\   
$c_{Z\square}$ 		&  0.214        &             &   & 1 & -0.78 & -0.44 & 0.20 & -0.57 & 0.18 & 0.60 & 0.66   \\ 
$c_{\gamma\gamma}$      &  0.00738     &           &   &   & 1 & 0.50 & -0.20 & 0.27 & -0.66 & -0.73 & -0.81 \\ 
$c_{Z\gamma}$   	        &  0.00935      &            &   &   &   & 1 & -0.099 & 0.13 & -0.34 & -0.39 & -0.42 \\ 
$c_{gg}$             		&   0.000462  &            &   &   &   &   & 1 & -0.65 & 0.086 & 0.13 & 0.26   \\ 
$\delta y_t$        		&  0.0856       &            &   &   &   &   &   & 1 & 0.39 & -0.15 & -0.30 \\ 
$\delta y_b$       		&   0.125      &              &   &   &   &   &   &   & 1 & 0.50 & 0.54  \\ 
$ \delta y_\tau$  		&  0.114       &              &   &   &   &   &   &   &   & 1 & 0.61  \\ 
$ \delta y_\mu$     		&   0.108      &          &   &   &   &   &   &   &   &   & 1   \\ \hline
\end{tabular}
\caption{Same as \autoref{tab:cslhc300} but for $14$\,TeV LHC $3000\infb$. Note
         that while $\delta y_c$ is set to zero in obtaining these results, it
         is not set to zero when the $\chi^2$ is combined with the ones from
         future $\ee$ colliders. This has little impact on the results of the
         combined fits.}
\label{tab:cslhc3000}
\end{table}
%


\section{Additional figures}
\label{app:more}

\begin{figure}[ht]
\centering
\includegraphics[width=.90\textwidth]{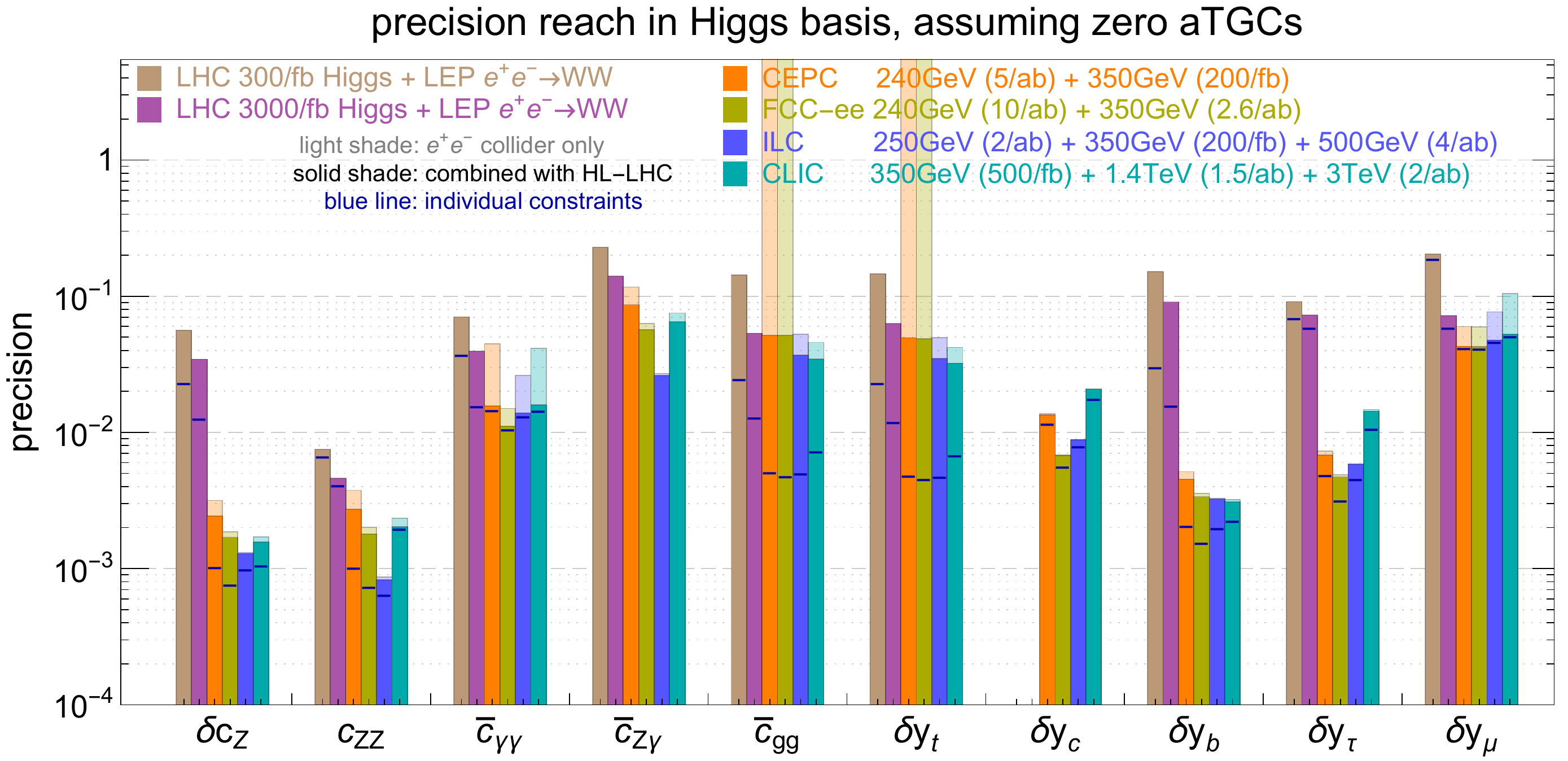}%
\raisebox{4mm}{\includegraphics[width=.08\textwidth]{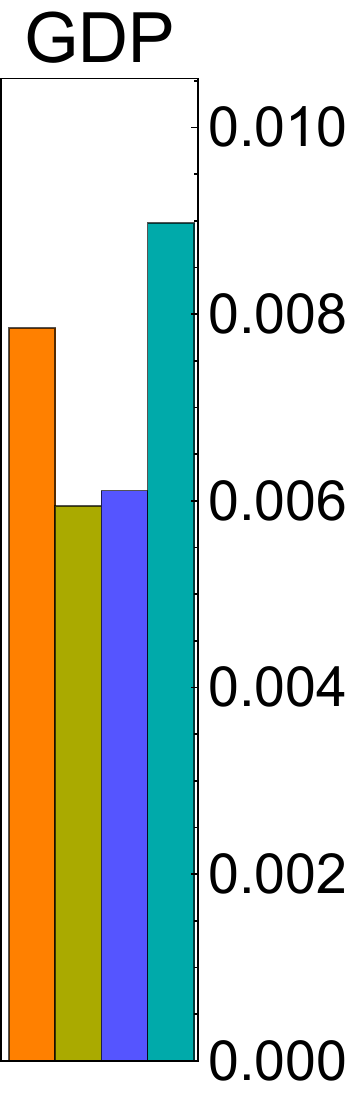}}
\caption{Same as \autoref{fig:fit0} but assuming vanishing aTGCs.  Imposing $\delta g_{1,Z} = \delta \kappa_\gamma = \lambda_Z  = 0$, both $\lambda_Z$ and $c_{Z\square}$ are eliminated, while the relation $e^2 c_{\gamma\gamma} + (g^2-g'^2)c_{Z\gamma}  - (g^2+g'^2)c_{ZZ}=0$ is imposed among $c_{ZZ}$, $c_{\gamma\gamma}$ and $c_{Z\gamma}$.  Note that the individual constraints are basis dependent.  We use the above relation to eliminate $c_{Z\gamma}$, hence its individual constraints are not shown.}
\label{fig:fit0tgc0}
\end{figure}

Here we provide additional results of the global fits. In our study,
conservative estimates have been made for the measurements of the diboson
process ($\eeww$) which often end up being systematics dominated. To give a
sense of the impact of these systematic uncertainties we show, in
\autoref{fig:fit0tgc0}, global fit results in which aTGCs are
assumed to be perfectly constrained.

\begin{figure}[ht]
\centering
\includegraphics[width=16cm]{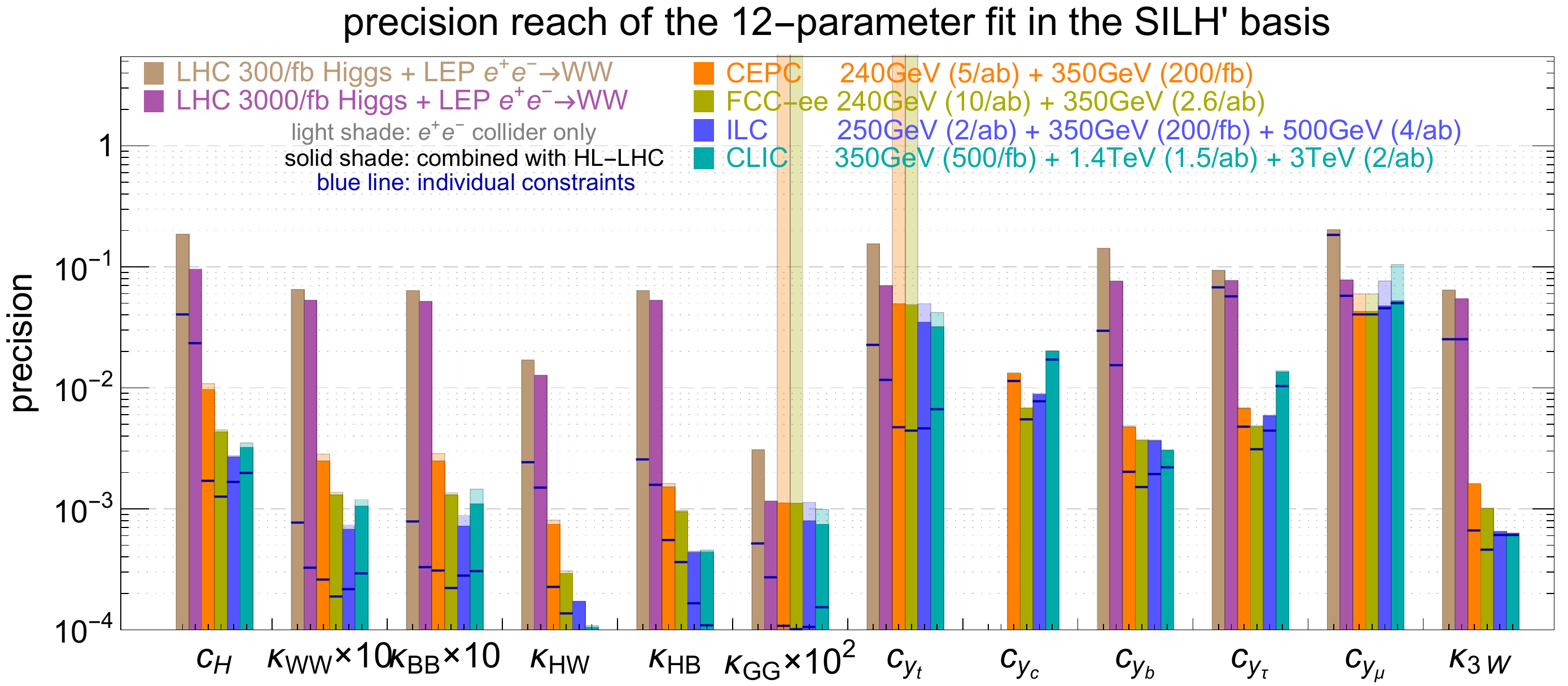}
\caption{Same as \autoref{fig:fit0} but is in the SILH'(-like) basis defined by \autoref{eq:lad6} and \autoref{tab:op1}.  }
\label{fig:fit0d6}
\end{figure}

\begin{figure}[ht]
\centering
\includegraphics[width=14cm]{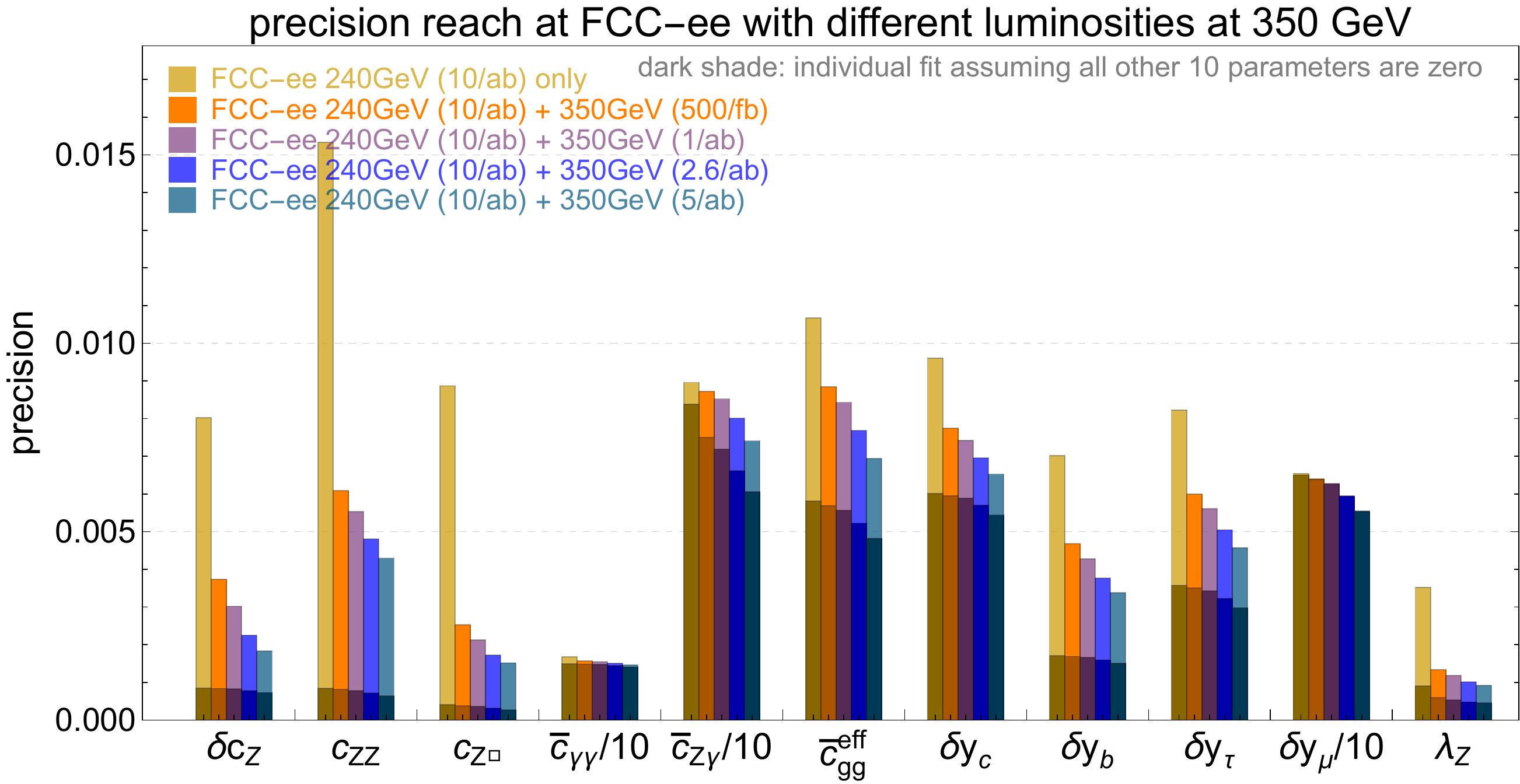}
\raisebox{0.105\height}{\includegraphics[width=1.37cm]{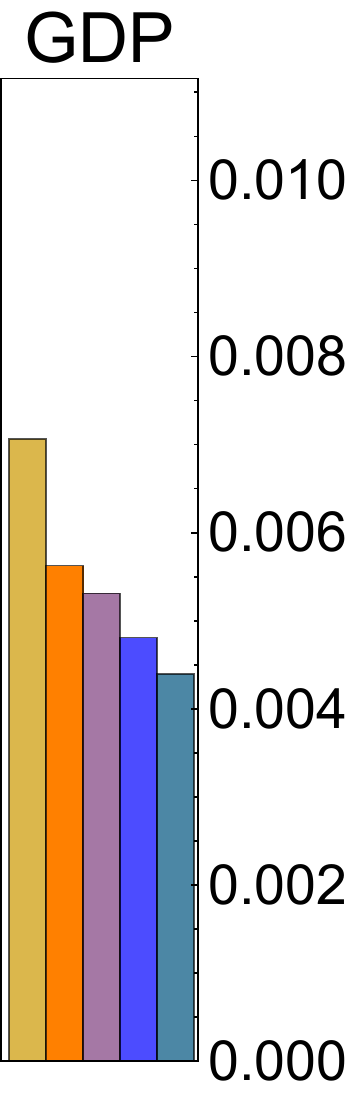}}
\caption{Same as \autoref{fig:fitclu1} but for FCC-ee with $10\inab$ at 240\,GeV and $500\infb$, $1\inab$, $2.6\inab$ and $5\inab$.}
\label{fig:fitflu2}
\end{figure}

\begin{figure}
\centering
\includegraphics[width=14cm]{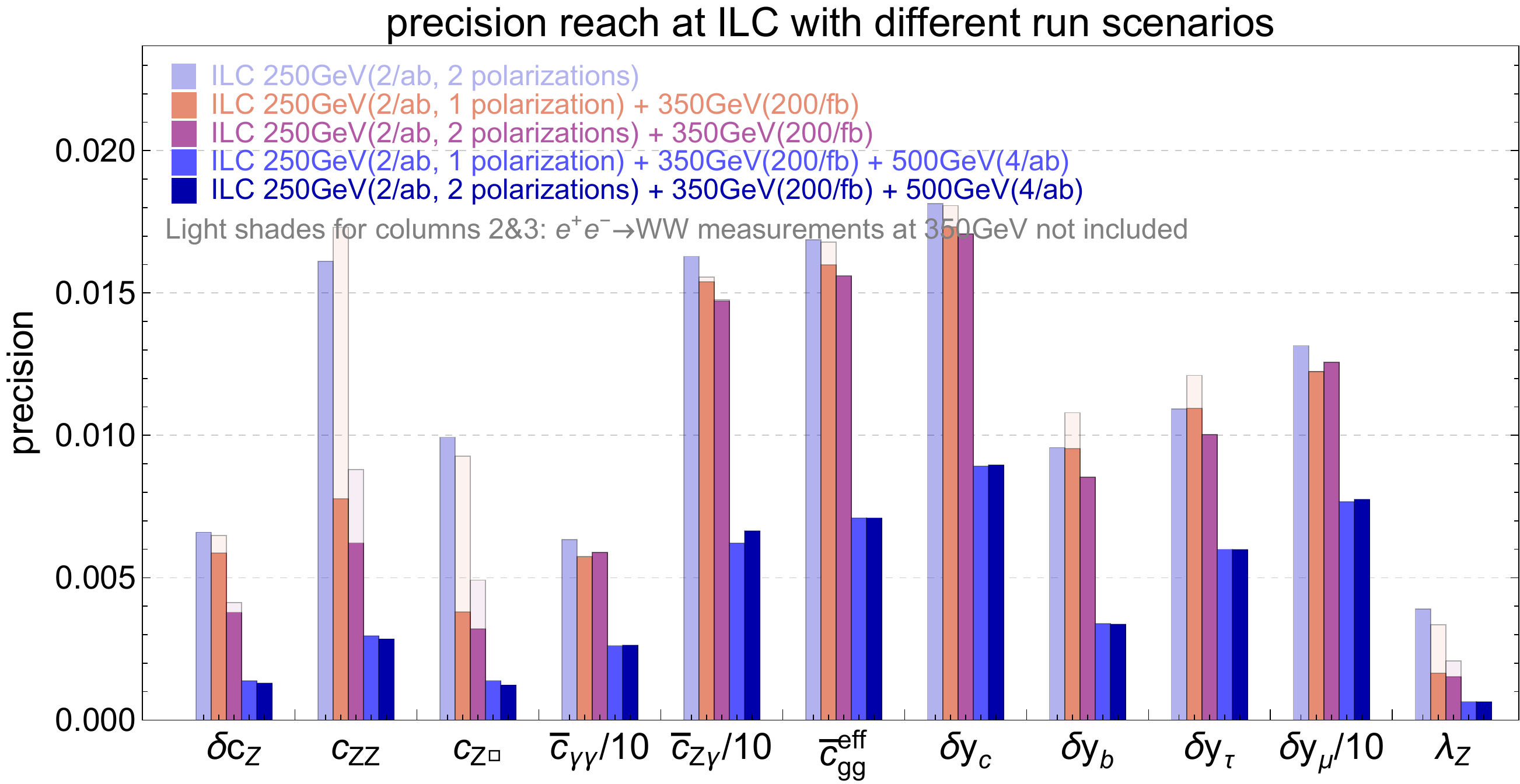} 
\raisebox{0.09\height}{\includegraphics[width=1.55cm]{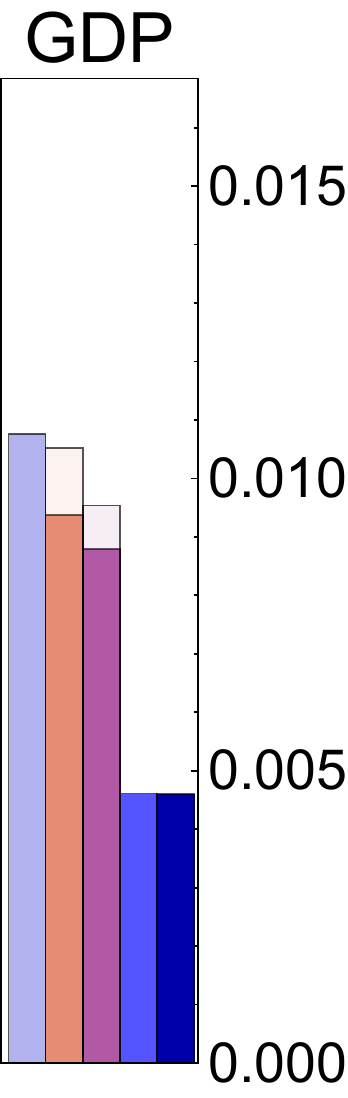}}
\caption{The precision reach for ILC with different scenarios. The 1st column
         corresponds to the ILC 250\,GeV run with $2\inab$ luminosity which is
         divided into two runs with polarizations $P(e^-,e^+)=(-0.8,+0.3)$ and
         $(+0.8,-0.3)$, and fractions $0.7$ and $0.3$, respectively (see
         \autoref{fig:fiti1}). The 2nd and 3rd columns include ILC 250\,GeV
         ($2\inab$) and 350\,GeV ($200\infb$). For the 2nd column, only the
         $(-0.8,+0.3)$ polarization is used for the 240\,GeV run, while for the
         3rd column the 240\,GeV run is divided in the same way as for the 1st
         column.
         The results of the ILC full run ($2\inab$ at 250\,GeV, $200\infb$ at
         350\,GeV and $4\inab$ at 500\,GeV) are shown in the 4th and 5th
         columns, while single polarization (two polarizations) at 250\,GeV has
         been assumed for the 4th (5th) column, analogous to the 2nd and 3rd
         columns. $P(e^-,e^+)=(-0.8,+0.3)$ is assumed for the 350\,GeV and
         500\,GeV runs. We found that dividing the runs at 350\,GeV and 500\,GeV
         into multiple polarization does not improve the results.
         For the ILC full program, we still show the constraint
         of $\bar{c}^{\rm \,eff}_{gg}$ instead of $c_{gg}$ and $\delta y_t$ in
         order to compare with other scenarios. For the full program
         only the 500\,GeV TGC results are used for consistency with the main
         results in \autoref{fig:fit0}.
}
\label{fig:fitii1}
\end{figure}

\begin{figure}
\centering
\includegraphics[width=14cm]{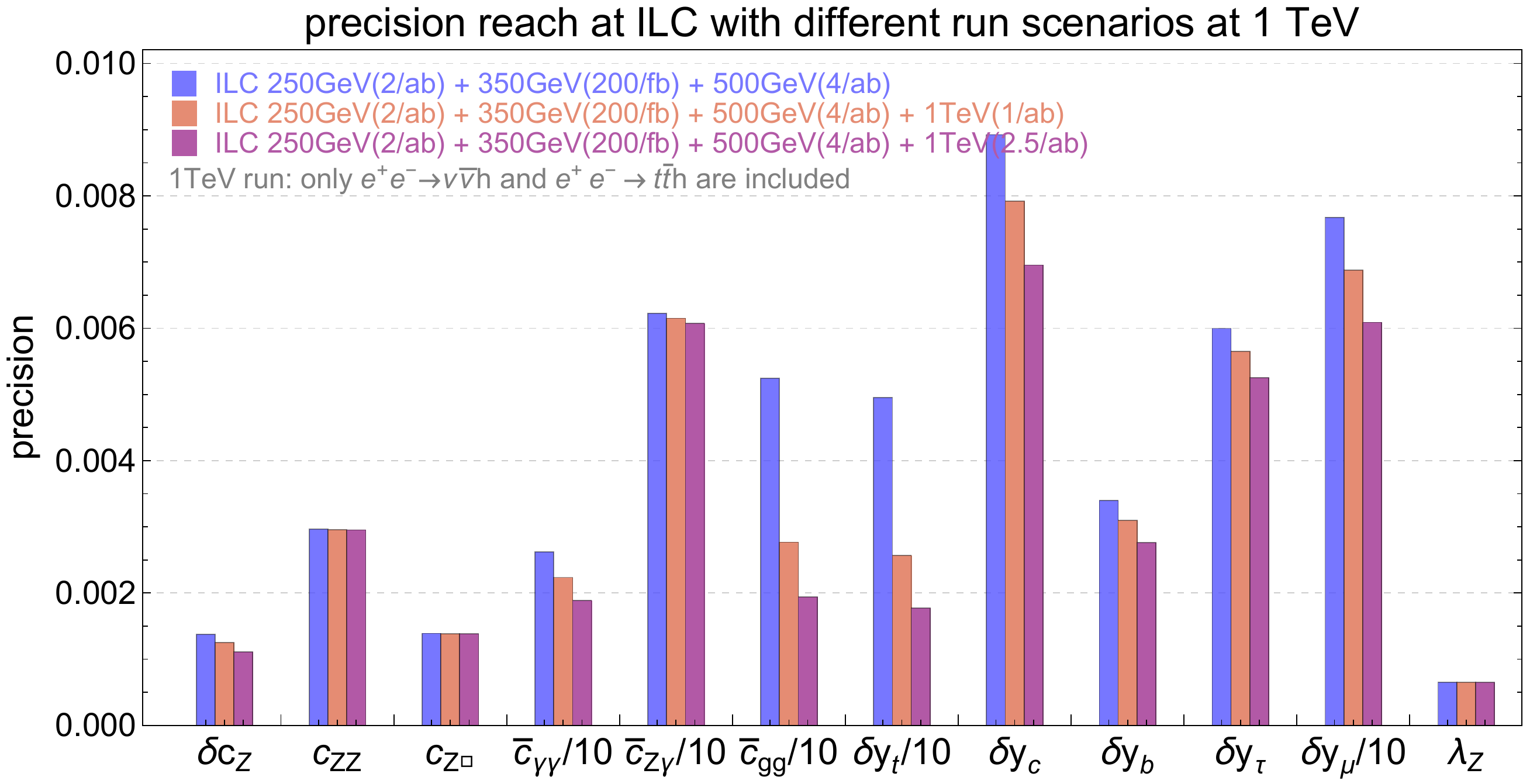} 
\raisebox{0.09\height}{\includegraphics[width=1.36cm]{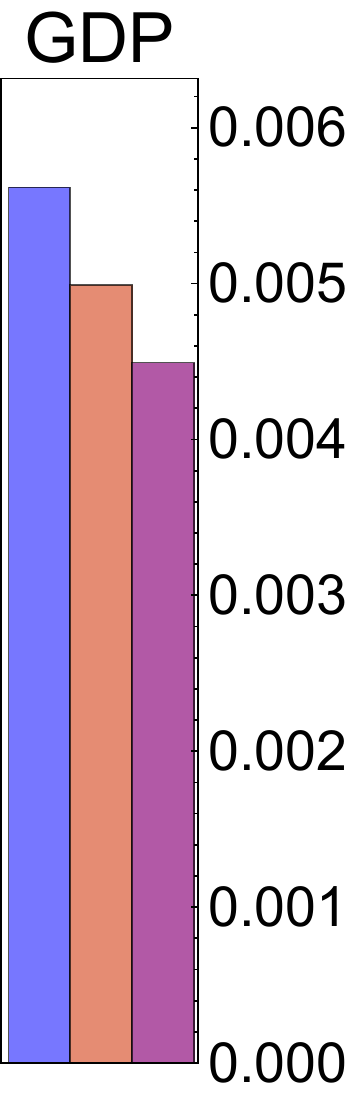}}
\caption{The precision reach for ILC with different scenarios for the 1\,TeV run. The 1st column
         corresponds to the ILC full run considered in our study with $2\inab$ at 250\,GeV, $200\infb$ at
         350\,GeV and $4\inab$ at 500\,GeV and a fixed polarization of $P(e^-,e^+)=(-0.8,+0.3)$.  For the
         2nd (3rd) column, an additional run at 1\,TeV with an integrated luminosity of $1\inab$ ($2.5\inab$) 
         and polarization $P(e^-,e^+)=(-0.8,+0.2)$ is also included.  For the 1\,TeV run, the estimated 
         measurement precisions in Ref.~\cite{Asner:2013psa} are used.   Only the measurements of the 
         $\eevvh$ and $\eetth$ processes are included at 1\,TeV, as the ones for $\eehz$ and $\eeww$ are not 
         provided.  In particular, the precision of $\sigma(t\bar{t}h)\times \br(h\to b\bar{b})$ at 1\,TeV is estimated 
         to be 6.0\% with $1\inab$ data and 3.8\% with $2.5\inab$ data, which significantly improves the precision at 500\,GeV.
         As such, the constraints on both $\bar{c}_{gg}$ and $\delta y_t$ are greatly improved.  It should be noted that
         the $\eehz$ and $\eeww$ processes are more sensitive to some of the EFT parameters at higher energies.  The inclusion of 
         their measurements could potentially further improve the overall reach of the global fit.
}
\label{fig:fitilcn1}
\end{figure}

\begin{figure}[ht]
\centering
\includegraphics[width=14cm]{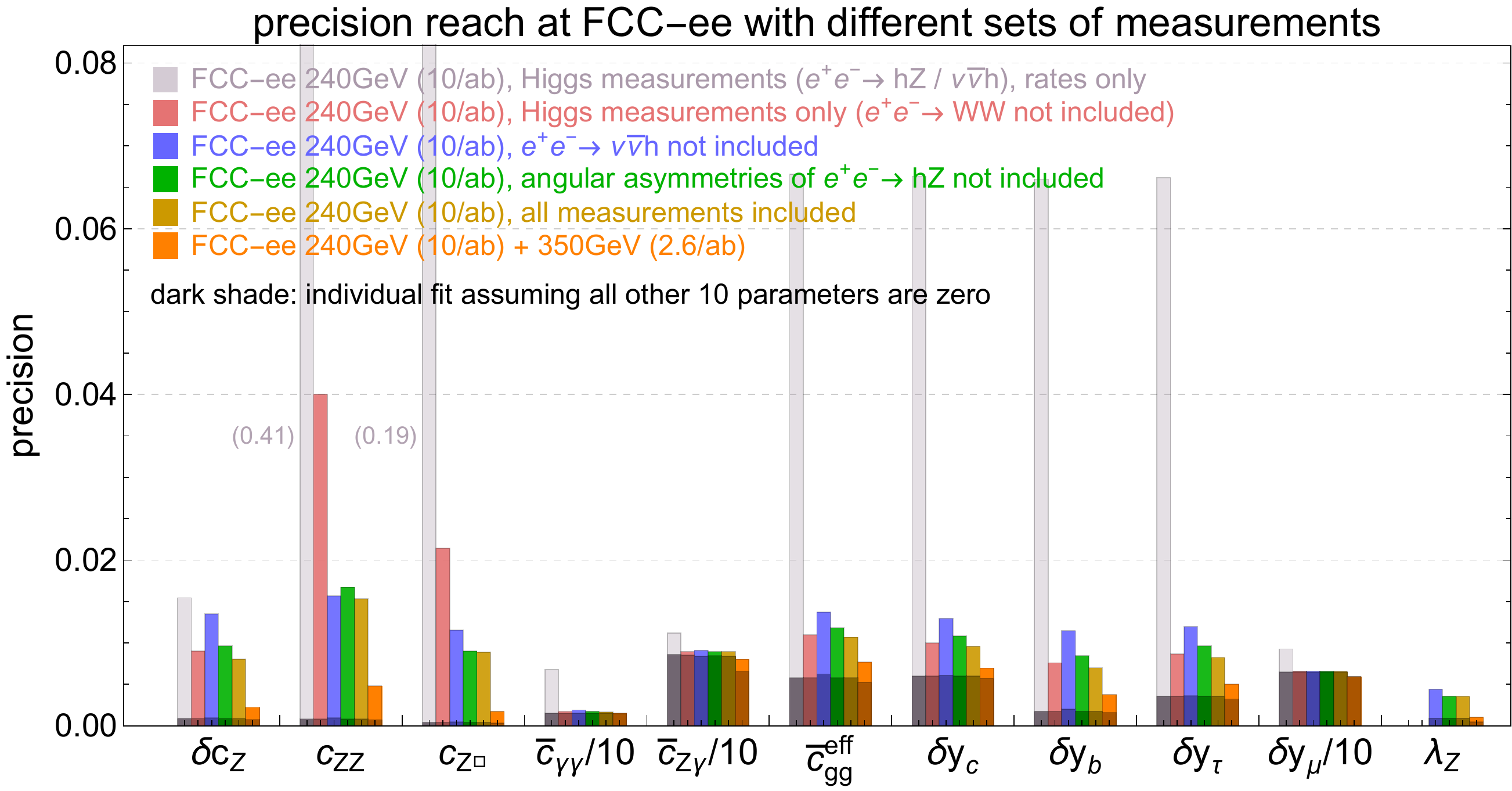}
\caption{Same as \autoref{fig:fitc1} but for FCC-ee.}
\label{fig:fitf1}
\end{figure}

\begin{figure}
\centering
\includegraphics[width=14cm]{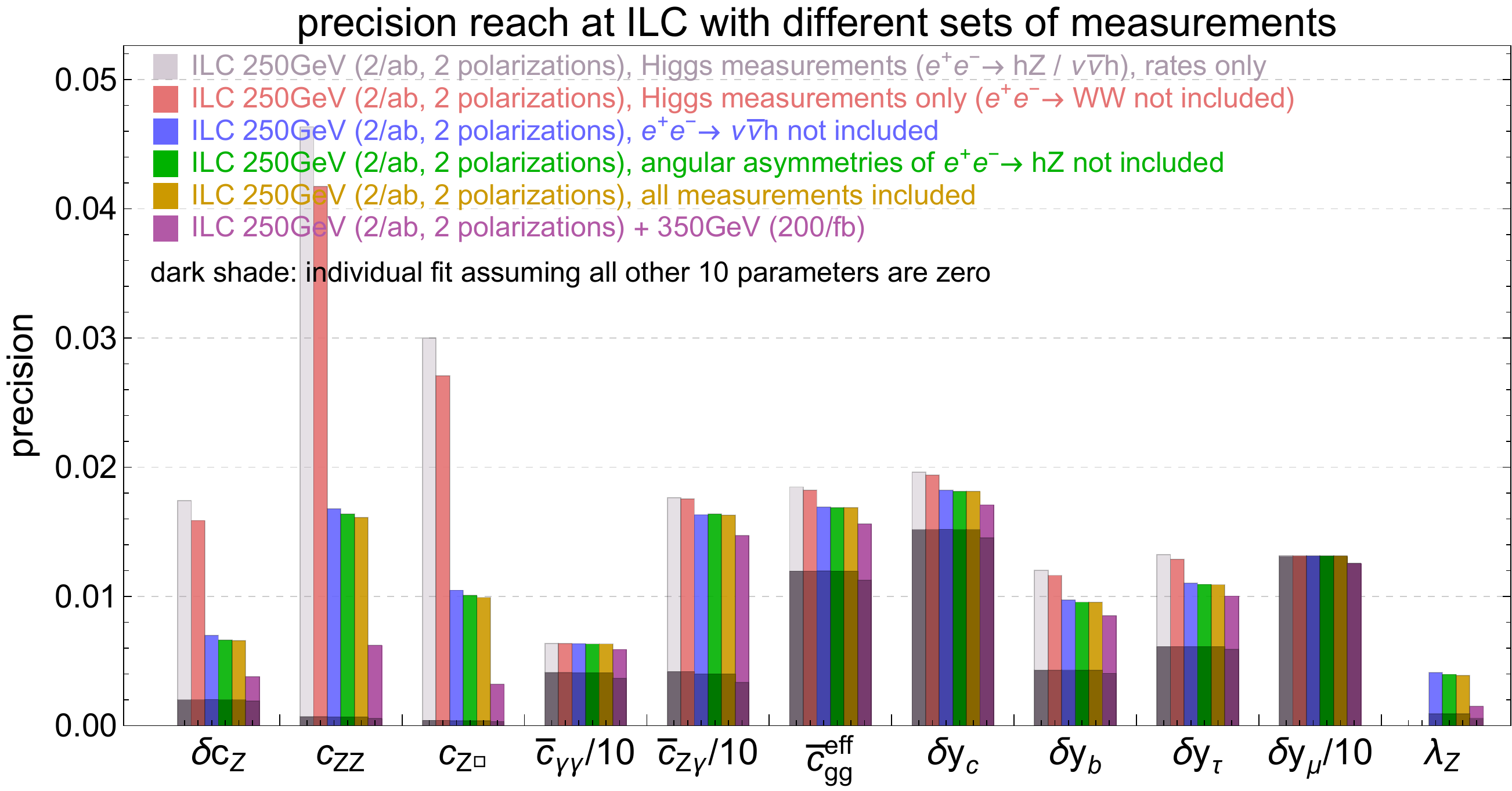}
\caption{Same as \autoref{fig:fitc1} but for ILC with $2\inab$ at 250\,GeV and
         $200\infb$ at 350\,GeV. The 250\,GeV run is divided into two runs with
         polarizations $P(e^-,e^+)=(-0.8,+0.3)$ and $(+0.8,-0.3)$, and fractions
         $0.7$ and $0.3$, respectively (see \autoref{fig:fiti1}).}
\label{fig:fitii2}
\end{figure}

\begin{figure}
\centering
\includegraphics[width=13cm]{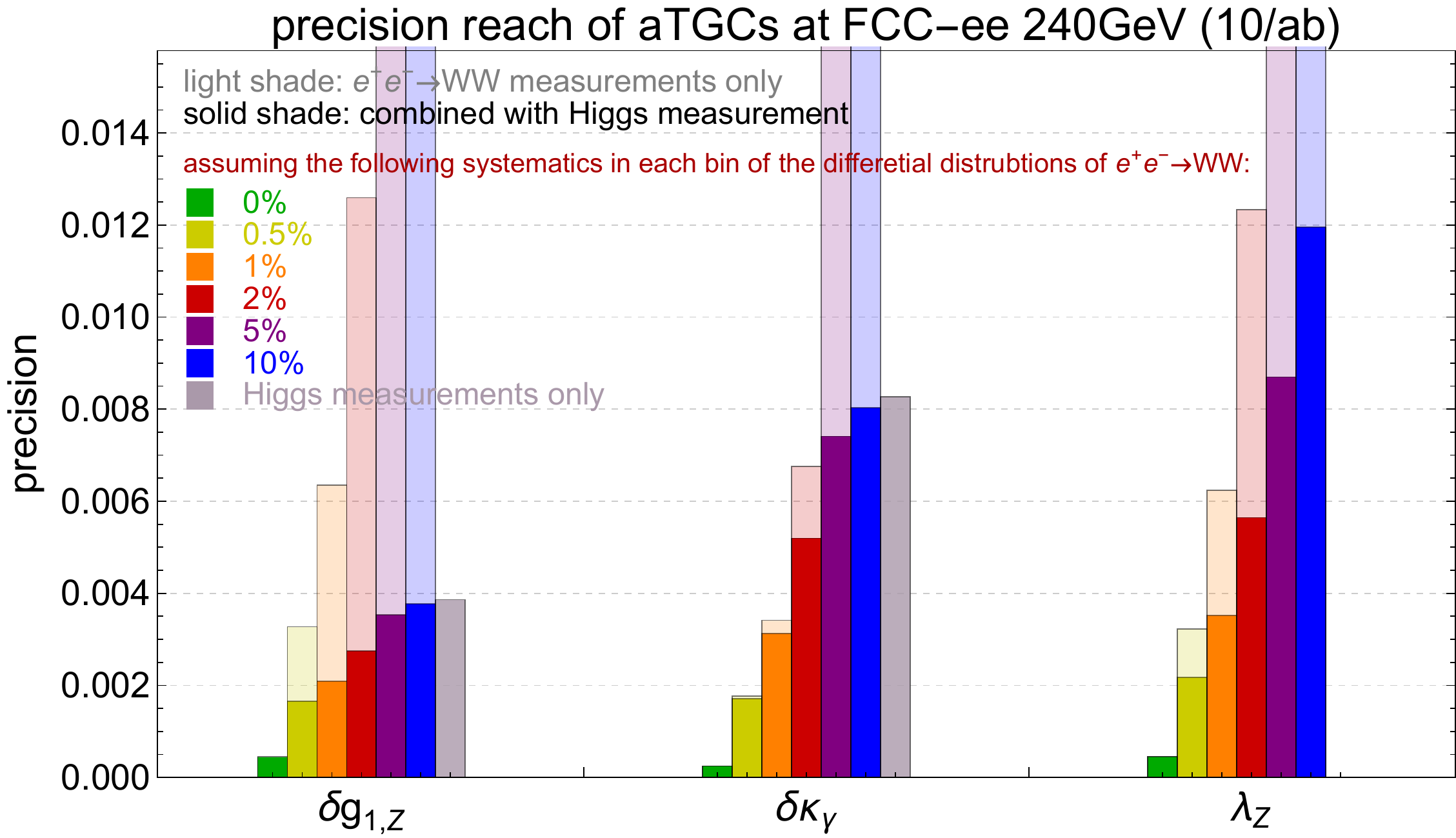} \\ \vspace{0.5cm}
\includegraphics[width=14cm]{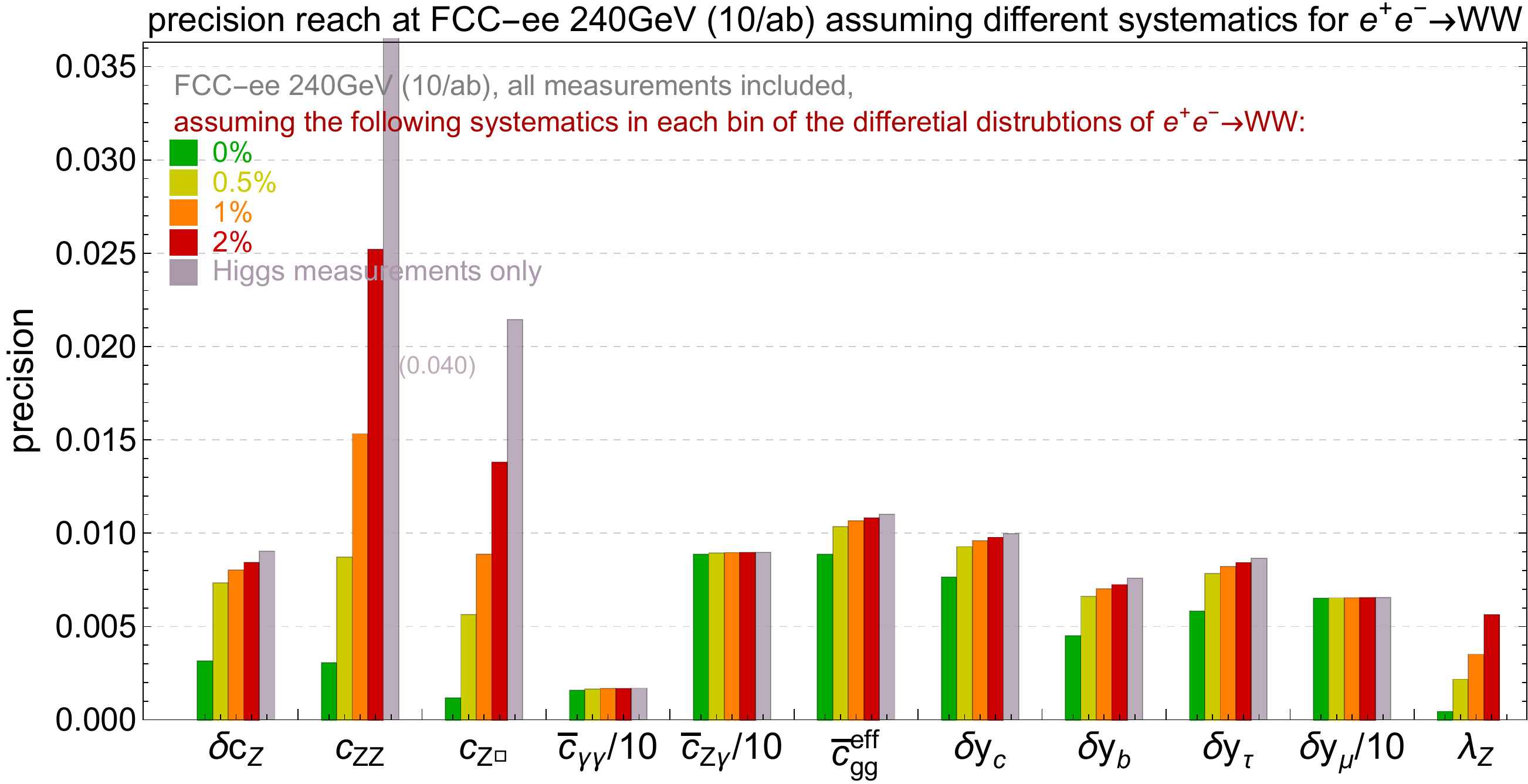}
\caption{Same as \autoref{fig:fittgc1} but for FCC-ee 240\,GeV with $10\inab$.}
\label{fig:fittgf1}
\end{figure}

\begin{figure}
\centering
\includegraphics[width=13cm]{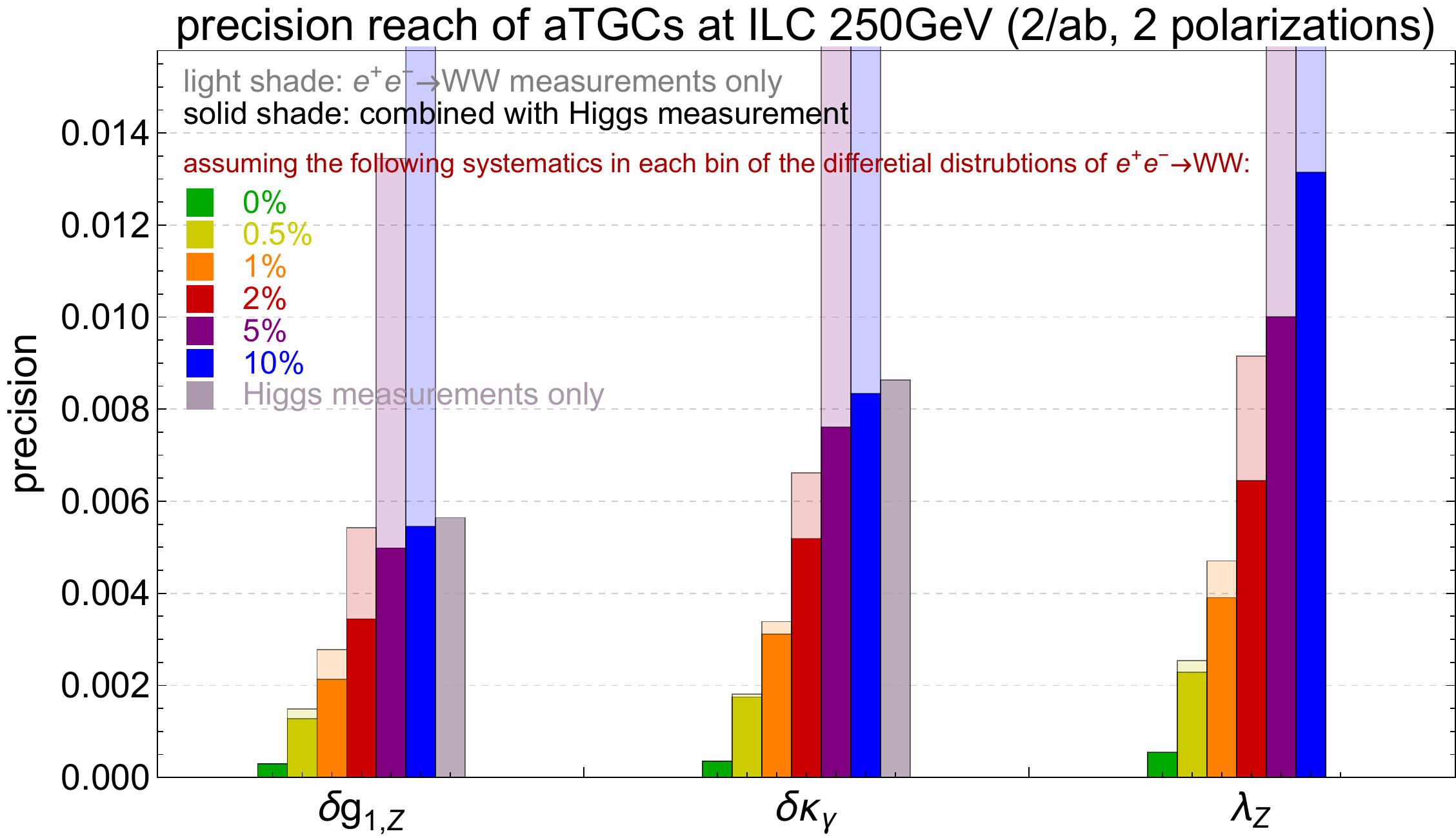} \\ \vspace{0.5cm}
\includegraphics[width=14cm]{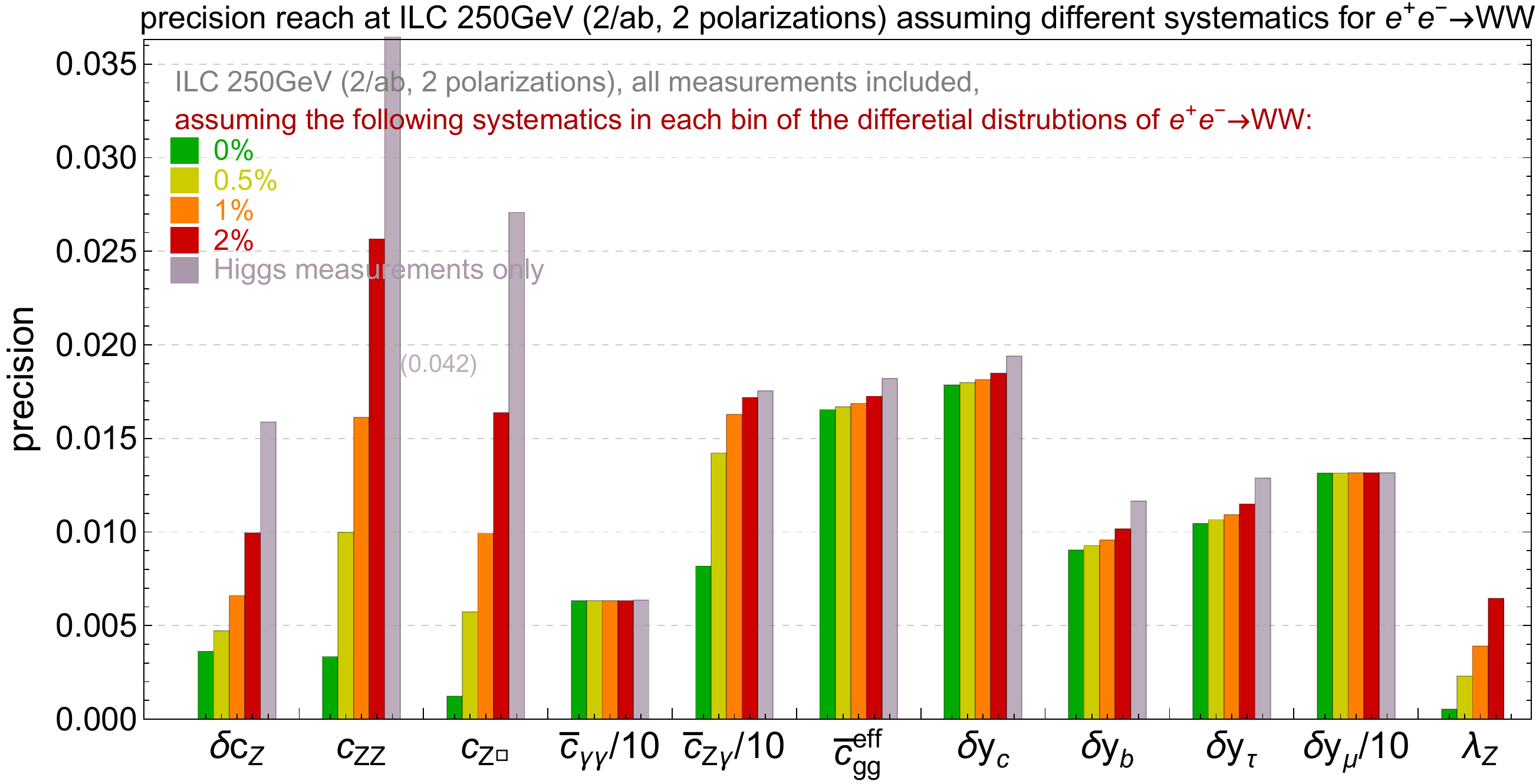}
\caption{Same as \autoref{fig:fittgc1} but for ILC 250\,GeV with $2\inab$
         luminosity, divided into two runs with polarizations
         $P(e^-,e^+)=(-0.8,+0.3)$ and $(+0.8,-0.3)$, and fractions $0.7$ and
         $0.3$, respectively (see \autoref{fig:fiti1}).}
\label{fig:fittgi1}
\end{figure}

Figure~\ref{fig:fit0d6} reproduces the results in \autoref{fig:fit0} in the basis
defined by \autoref{eq:lad6} and \autoref{tab:op1}. The analogues to the figures
presented in the main text for the CEPC, \autoref{fig:fitc1}--\ref{fig:fittgc1},
are given here for the FCC-ee and ILC in
\autoref{fig:fitflu2}--\ref{fig:fittgi1}. In particular, \autoref{fig:fitii1}
shows the precision reach for ILC with different scenarios including runs at
250\,GeV, 350\,GeV and 500\,GeV, while \autoref{fig:fitilcn1} further shows the
potential improvement with the inclusion of a 1\,TeV run.

%


\clearpage

\section{Numerical expressions for the observables}
\label{app:express}

We express some of the important observables as numerical functions of the
parameters in \autoref{eq:para10}, which is fed into the chi-square in
\autoref{eq:chia}--\autoref{eq:chic}. The SM input parameters we use in our
analytical expressions are $G_F=1.1663787 \times 10^{-5} \,\rm{GeV}^{-2}$, $m_Z=91.1876$\,GeV, $\alpha_{\rm em}(m^2_Z)=1/127.940$ and $m_h=125.09$\,GeV.
For the rate of $\eehz$, the measurements with the
following energies and polarizations $P(e^-,e^+)$ are used, 
\begin{footnotesize}%
\begin{equation}
\left. \frac{\sigma_{hZ}}{\sigma^{\rm SM}_{hZ}} \right\vert 
{\scriptstyle \bpm 240\,{\rm GeV} ~{\rm unpolarized} \\  250\,{\rm GeV} ~{(-0.8,+0.3)} \\  250\,{\rm GeV} ~{(+0.8,-0.3)} 
\\  350\,{\rm GeV} ~{\rm unpolarized}  \\  350\,{\rm GeV} ~{(-0.8,+0.3)}  \\  500\,{\rm GeV} ~{(-0.8,+0.3)} \\ 1.4\,{\rm TeV} ~{\rm unpolarized} \\ 3\,{\rm TeV} ~{\rm unpolarized}  \epm } 
\simeq 
1 + 2\,\delta c_Z
+{\scriptstyle \bpm       1.8  \\  5.6 \\ -2.9 \\ 2.8       \\  11   \\  21 \\ 14 \\ 52 \epm } \, c_{ZZ}  
+{\scriptstyle \bpm       3.7  \\ 9.8 \\ -3.2 \\  7.5         \\ 20  \\ 41 \\ 115 \\  526 \epm } \, c_{Z\square} 
+{\scriptstyle  \bpm -0.048 \\ -0.73 \\ 0.79 \\ -0.11  \\ -1.5  \\  -3.3 \\ -1.9 \\ -8.8  \epm  } \, c_{\gamma\gamma} 
+ {\scriptstyle \bpm -0.087 \\  -1.3  \\  1.5 \\  -0.24  \\ -3.3   \\  -8.1  \\ -5.5  \\ -26  \epm } \, c_{Z\gamma} \,.
\label{eq:hz_numerical}
\end{equation}\end{footnotesize}%
As noted in \autoref{sec:hzrate}, the interferences between $s$-channel $Z$ and
photon amplitudes are accidentally suppressed in the unpolarized total cross
section. On the contrary, they have a significant impact when polarized beams
are used, flipping for instance the sign of the $c_{ZZ}$ prefactor as
polarization is reversed at $\sqrt{s}=250\,$GeV.\footnote{For simplicity, the
one-loop standard-model contributions to the $hZ\gamma$ vertex are not included
in the expressions above. They have a relatively large impact on the numerical
prefactors of the $c_{\gamma\gamma}$ and $c_{Z\gamma}$ coefficients which are
accidentally suppressed in the unpolarized cross section, at $240\,$GeV in
particular. Given that this measurement has little sensitivity to these
coefficients, such contributions do however not affect the results of our global
analysis. Note that the $c_{\gamma \square}$ parameter, directly related to the
$hZ\gamma$ vertex, is written in terms of $c_{ZZ}$, $c_{Z\square}$,
$c_{\gamma\gamma}$ and $c_{Z\gamma}$ using \autoref{eq:cwtocz}.} The relevant
expressions for the WW fusion process are
\begin{equation}
\left. \frac{\sigma_{WW\to h}}{\sigma^{\rm SM}_{WW\to h}} \right\vert 
{\scriptstyle \bpm 240\,{\rm GeV}  \\  250\,{\rm GeV}  \\  350\,{\rm GeV}
\\  500\,{\rm GeV}  \\  1\,{\rm TeV}  \\  1.4\,{\rm TeV}  \\  3\,{\rm TeV}   \epm } 
\simeq 
1 + 2\,\delta c_Z
+{\scriptstyle \bpm -0.25 \\  -0.27 \\ -0.40  \\ -0.53 \\  -0.76    \\ -0.86  \\ -1.1 \epm } \, c_{ZZ}  
+{\scriptstyle \bpm -0.68 \\ -0.72 \\ -1.1   \\ -1.5 \\       -2.2    \\ -2.5 \\  -3.4  \epm } \, c_{Z\square} 
+{\scriptstyle  \bpm 0.035 \\ 0.037 \\ 0.056  \\ 0.075\\  0.12   \\  0.14 \\  0.18  \epm  } \, c_{\gamma\gamma} 
+ {\scriptstyle \bpm 0.090 \\  0.097  \\  0.14  \\  0.20 \\   0.32  \\ 0.37   \\  0.52  \epm } \, c_{Z\gamma} \,,
\end{equation}
which are obtained from \texttt{MadGraph5}~\cite{Alwall:2014hca} with the BSMC
model~\cite{BSMC, Falkowski:2015wza} as functions of $\delta c_W$, $c_{WW}$
and $c_{W\square}$ and then transformed into the basis in \autoref{eq:para10}
with \autoref{eq:cwtocz}. The default input parameters are used for these
numerical computations. They apply to any polarizations since only the initial
states with helicities $H(e^-, e^+) = (-,+)$ contribute to this process.

For the $\eetth$ process, we only consider the dominate NP contribution which is
from the modification of the top Yukawa, $\delta y_t$. It is therefore straight
forward to write down the rate of the $t\bar{t}h$ process as
\begin{equation}
\frac{\sigma_{t\bar{t}h}}{\sigma^{\rm SM}_{t\bar{t}h}} \simeq 1+ 2\, \delta y_t \,.
\end{equation}

For Higgs decays, we make use of the results in Ref.~\cite{Falkowski:2015fla}.  The Decay widths to a pair of fermions are
\begin{equation}
\frac{\Gamma_{cc}}{\Gamma^{\rm SM}_{cc}} \simeq 1+ 2\, \delta y_c \,, \hspace{0.7cm}
\frac{\Gamma_{bb}}{\Gamma^{\rm SM}_{bb}} \simeq 1+ 2\, \delta y_b \,, \hspace{0.7cm}
\frac{\Gamma_{\tau\tau}}{\Gamma^{\rm SM}_{\tau\tau}}  \simeq 1+ 2\, \delta y_\tau \,,  \hspace{0.7cm}
\frac{\Gamma_{\mu\mu}}{\Gamma^{\rm SM}_{\mu\mu}}  \simeq 1+ 2\, \delta y_\mu \,.
\end{equation}
The decay width to $WW^*$ $ZZ^*$ (with 4$f$ final states) are given by
\begin{align}
\frac{\Gamma_{WW^*}}{\Gamma^{\rm SM}_{WW^*}} 
\simeq &~ 1+  2\,\delta c_Z + 0.05 \, c_{ZZ} +0.67\, c_{Z\square} -0.05\, c_{\gamma\gamma} -0.17\, c_{Z\gamma} \,,  \\
\frac{\Gamma_{ZZ^*}}{\Gamma^{\rm SM}_{ZZ^*}} \simeq &~   1 + 2\,\delta c_Z -0.15 \, c_{ZZ}  + 0.41 \, c_{Z\square}  \,,
\end{align}
where we assume there is no NP correction to the gauge couplings of fermions.  As stated in \autoref{sec:eft}, we do not consider contribution from off-shell photons that gives the same final states as $ZZ^*$, as they can be relatively easily removed by kinematic cuts.

The decay of Higgs to $gg$, $\gamma\gamma$ and $Z\gamma$ are generated at one-loop level in the SM.  The leading EFT contribution could either be at tree level (which are generated in the UV theory by new particles in the loop) or come at loop level by modifying the couplings in the SM loops.
As mentioned in \autoref{sec:eft}, we follow Ref.~\cite{Falkowski:2015fla} and include both the tree level EFT contribution ($c_{gg}$) and the one-loop contribution (from $\delta y_t$ and $\delta y_b$) for $h\to gg$, while only keeping the tree level EFT contribution ($c_{\gamma\gamma}$ and $c_{Z\gamma}$) for $h \to \gamma\gamma$ and $h \to Z\gamma$.
The decay widths are given by \footnote{The choices of the bottom mass value would change the numerical values in \autoref{eq:dhgg}, but has little impact on the global fit results.}
\begin{equation}
\frac{\Gamma_{gg}}{\Gamma^{\rm SM}_{gg}} \simeq 1+ 241 \, c_{gg} + 2.10 \, \delta y_t -0.10 \, \delta y_b \,, \label{eq:dhgg}
\end{equation}
and
\begin{align}
\frac{\Gamma_{\gamma\gamma}}{\Gamma^{\rm SM}_{\gamma\gamma}} \simeq &~ (1+\frac{c_{\gamma\gamma}}{-8.3\times 10^{-2}})^2  \,, \nonumber\\
\frac{\Gamma_{Z\gamma}}{\Gamma^{\rm SM}_{Z\gamma}} \simeq &~ (1+\frac{c_{Z\gamma}}{-5.9\times 10^{-2}})^2  \,.  \label{eq:dhvv}
\end{align}

The branching ratio can be derived from the total decay width, which can be obtained from
\begin{equation}
\frac{\Gamma_{\rm tot}}{\Gamma^{\rm SM}_{\rm tot}} = \sum_i \frac{\Gamma_i}{\Gamma^{\rm SM}_i} {\rm Br}^{\rm SM}_i \,.
\end{equation}
In practice, one only needs to include the BSM effects of the main channels in the calculation of the total width.  Finally, the physical observables in the form of $\sigma \times {\rm BR}$ can be constructed from the above information.

For the aTGCs, we also express \autoref{eq:tgchb} numerically as
\begin{align}
\delta g_{1,Z} \simeq &~ -0.120 \, c_{ZZ}  -0.392 \, c_{Z\square} +0.0215 \, c_{\gamma\gamma} +0.0637 \, c_{Z\gamma} \,, \nonumber \\
\delta \kappa_\gamma \simeq &~ 0.208 \, c_{ZZ}  -0.0373 \, c_{\gamma\gamma} - 0.11 \, c_{Z\gamma} \,. 
\end{align}
The expressions for the differential distributions in $\eeww$ for all energies and polarizations are too lengthy to be reported here.  The necessary information can be conveniently reconstructed from the constraints on aTGCs in \autoref{tab:tgccepc}, \ref{tab:tgcfcc} and \ref{tab:tgcilc}.

Finally, the CP-even angular observables in $\eehz$ are given by
\begin{align}
A_{\theta_1}
\simeq &~
{\scriptstyle \bpm -0.45 \\ -0.46 \\ -0.46 \\  -0.57  \\  -0.57  \\  -0.65  \epm } 
+{\scriptstyle \bpm 0.050  \\ 0.074  \\ 0.074 \\ 0.33  \\  0.33  \\  0.62  \epm } \, c_{ZZ}  
+ {\scriptstyle \bpm 0.0019 \\  0.039  \\ -0.042  \\  0.013  \\  0.19  \\  0.37  \epm } \, c_{Z\gamma} \,,  \nonumber\\ \nonumber\\
A^{(3)}_\phi
\simeq &~
{\scriptstyle \bpm 0.0093  \\ 0.069  \\ -0.065 \\  0.0092 \\  0.068  \\  0.058  \epm } 
+{\scriptstyle \bpm 0.32  \\ 0.059  \\ 0.10  \\  0.75  \\  0.15  \\  0.34  \epm } \, c_{ZZ}  
+{\scriptstyle \bpm  0.50 \\ 0.096  \\ 0.15  \\  1.14  \\  0.20  \\  0.36  \epm } \, c_{Z\square}  
+ {\scriptstyle \bpm -0.058  \\  -0.011  \\  -0.017 \\  -0.13  \\  -0.023  \\  -0.042   \epm } \, c_{\gamma\gamma}
+ {\scriptstyle \bpm -0.11  \\  -0.023  \\ -0.034  \\  -0.28  \\  -0.036  \\  -0.025  \epm } \, c_{Z\gamma} \,,   \nonumber\\ \nonumber\\
A^{(4)}_\phi  
\simeq &~
{\scriptstyle \bpm 0.096 \\ 0.092 \\ 0.092 \\ 0.061  \\  0.061  \\  0.035  \epm } 
+{\scriptstyle \bpm 0.015  \\ 0.022  \\ 0.022 \\ 0.098  \\  0.098  \\  0.19  \epm } \, c_{ZZ}  
+ {\scriptstyle \bpm 0.00057 \\  0.012  \\  - 0.013  \\  0.0040  \\  0.056  \\  0.11   \epm } \, c_{Z\gamma} \,,   \nonumber\\ \nonumber\\
A_{c\theta_1,c\theta_2}  
\simeq &~
{\scriptstyle \bpm  -0.0052 \\ -0.037 \\ 0.034 \\ -0.0033  \\  -0.024  \\  -0.014   \epm } 
+{\scriptstyle \bpm -0.18  \\ -0.042  \\ -0.043 \\  -0.27  \\  -0.085  \\  -0.13   \epm } \, c_{ZZ}  
+{\scriptstyle \bpm  -0.28 \\ -0.050  \\ -0.078 \\  -0.40  \\  -0.070  \\  -0.086 \epm } \, c_{Z\square}  
+ {\scriptstyle \bpm  0.032 \\  0.0059  \\ 0.0092  \\  0.047  \\  0.0082  \\  0.010  \epm } \, c_{\gamma\gamma}
+ {\scriptstyle \bpm  0.054  \\  0.0051  \\  0.010 \\  0.074  \\  -0.0092  \\  -0.028  \epm } \, c_{Z\gamma} \,,
\end{align}
where the six entries in each column correspond to the center of energies and
beam polarizations $P(e^-,e^+)$ in the following order: 240\,GeV unpolarized,
250\,GeV $(-0.8,+0.3)$, 250\,GeV $(+0.8,-0.3)$, 350\,GeV unpolarized, 350\,GeV
$(-0.8,+0.3)$ and 500\,GeV $(-0.8,+0.3)$.


\section{Numerical results of the global fit}
\label{app:rho}

We hereby list the numerical results of the global fit for the future $\ee$
colliders. The one standard deviation constraints on each of the 12 parameters
in \autoref{eq:para10} are listed in \autoref{tab:result}, and the corresponding
correlation matrices are shown in \autoref{tab:rhocepc}--\ref{tab:rhoclic}. For
each collider, the LHC $3000\infb$  (including 8\,TeV results)  + LEP measurements are also combined in the
total $\chi^2$, so that the results represent the ``best reach'' for each
scenario. With this information, the corresponding chi-squared can be
reconstructed using \autoref{eq:chipara}, which can be used to constrain any
particular model that satisfies the assumptions of the 12-parameter framework,
where the 12 parameters in EFT are functions of a usually much smaller set of
model parameters.
To minimize the numerical uncertainties, three significant figures are provided
for the one standard deviation constraints, which is likely more than sufficient
for the level of precision of our estimations.
For easy mapping to dimension-6 operators and new physics models, we also switch
back to the original definitions of $c_{\gamma\gamma}$, $c_{Z\gamma}$ and
$c_{gg}$ (instead of $\bar{c}_{\gamma\gamma}$, $\bar{c}_{Z\gamma}$ and
$\bar{c}_{gg}$).

\begin{table}[ht]
\centering
\begin{tabular}{|c||c|c|c|c|} \hline
&  \multicolumn{4}{|c|}{precision (one standard deviation)} \\  \hline
&   CEPC & FCC-ee &  ILC & CLIC  \\ \hline\hline
$\delta c_Z  $                &  0.00485     &   0.00216     &  0.00134     &  0.00161    \\ \hline
$c_{ZZ}$                       &  0.00701     &    0.00466    &  0.00291     &  0.00229    \\ \hline
$c_{Z\square}$             &   0.00328    &  0.00171      &   0.00139    &   0.000215   \\ \hline
$c_{\gamma\gamma}$ & 0.00130      &   0.000922    &  0.00115     &  0.00132    \\ \hline
$c_{Z\gamma}$            &  0.00537     &   0.00406     &   0.00332      &   0.00426   \\ \hline
$c_{gg}$                       &  0.000430   &  0.000427     &  0.000307    &   0.000286   \\ \hline
$\delta y_t$                  &  0.0495       &   0.0489       &   0.0349       &   0.0322   \\ \hline
$\delta y_c$                  &  0.0139       &  0.00686      &  0.00890        &   0.0208   \\ \hline
$ \delta y_b$                &  0.00559      &  0.00359     &  0.00334       &  0.00311    \\ \hline
$ \delta y_\tau$            &  0.00769       &  0.00490     &  0.00592     &   0.0143   \\ \hline
$ \delta y_\mu$             &  0.0429       &  0.0427       &  0.0476         &   0.0525   \\ \hline
$\lambda_Z$                 &  0.00161     &   0.00101    &  0.000652     &   0.000632   \\ \hline
\end{tabular}
\caption{Precision reach (one standard deviation bounds) at each of the four
         future $\ee$ colliders. For each collider, the LHC $3000\infb$
         (including 8\,TeV results) + LEP measurements are also
         combined in the total $\chi^2$.}
\label{tab:result}
\end{table}
\begin{table}[ht!]\scriptsize
\centering
\begin{tabular}{|c||c|c|c|c|c|c|c|c|c|c|c|c|} 
  \multicolumn{13}{c}{correlation matrix, CEPC} \\  \hline
 & $\delta c_Z$ & $c_{ZZ}$ & $c_{Z\square}$ & $c_{\gamma\gamma}$ & $c_{Z\gamma}$ & $c_{gg}$ & $\delta y_t$ & $\delta y_c$ & $\delta y_b$ & $\delta y_\tau$ & $\delta y_\mu$  & $\lambda_Z$ \\ \hline\hline
$\delta c_Z$ 			&  1 & -0.37 & -0.39 & -0.072 & -0.15 & 0.023 & 0.041 & 0.31 & 0.60 & 0.50 & 0.050 & -0.21 \\ \hline
$c_{ZZ}$      		        &     & 1 & -0.69 & 0.083 & 0.34 & -0.028 & -0.015 & -0.19 & -0.44 & -0.33 & -0.040 & -0.72 \\   \hline
$c_{Z\square}$ 		&       &   & 1 & -0.0092 & -0.18 & 0.0094 & -0.022 & -0.063 & -0.11 & -0.11 & -0.0076 & 0.89 \\ \hline
$c_{\gamma\gamma}$      &       &   &   & 1 & 0.028 & -0.17 & 0.19 & -0.049 & -0.069 & -0.063 & -0.044 & -0.016 \\ \hline
$c_{Z\gamma}$   	         &      &   &   &   & 1 & -0.014 & -0.011 & -0.11 & -0.25 & -0.18 & -0.021 & -0.11 \\ \hline
$c_{gg}$             		 &       &   &   &   &   & 1 & -0.99 & 0.014 & 0.035 & 0.025 & 0.16 & 0.011 \\ \hline
$\delta y_t$        		 &      &   &   &   &   &   & 1 & 0.039 & 0.077 & 0.061 & -0.16 & -0.012 \\ \hline
$\delta y_c$       		 &       &   &   &   &   &   &   & 1 & 0.48 & 0.37 & 0.036 & -0.011 \\ \hline
$ \delta y_b$  		         &      &   &   &   &   &   &   &   & 1 & 0.76 & 0.067 & -0.0012 \\ \hline
$ \delta y_\tau$  	         &      &  &   &   &   &   &   &   &   & 1 & 0.054 & -0.019 \\ \hline
$ \delta y_\mu$      	        &       &  &   &   &   &   &   &   &   &   & 1 & 0.0017 \\ \hline
$\lambda_Z$      		&  &   &   &   &   &   &   &   &   &   &   & 1 \\ \hline
\end{tabular}
\caption{The corresponding correlation matrix for the CEPC one sigma bounds in \autoref{tab:result}.}
\label{tab:rhocepc}
\end{table}
\begin{table}[ht!]\scriptsize
\centering
\begin{tabular}{|c||c|c|c|c|c|c|c|c|c|c|c|c|} 
  \multicolumn{13}{c}{correlation matrix, FCC-ee} \\  \hline
 & $\delta c_Z$ & $c_{ZZ}$ & $c_{Z\square}$ & $c_{\gamma\gamma}$ & $c_{Z\gamma}$ & $c_{gg}$ & $\delta y_t$ & $\delta y_c$ & $\delta y_b$ & $\delta y_\tau$ & $\delta y_\mu$  & $\lambda_Z$ \\ \hline\hline
$\delta c_Z$ 			&    1 & -0.49 & 0.073 & -0.055 & -0.22 & 0.026 & 0.0056 & 0.19 & 0.26 & 0.26 & 0.013 & 0.17 \\ \hline
$c_{ZZ}$      		        &        & 1 & -0.88 & 0.13 & 0.34 & -0.023 & -0.0025 & -0.24 & -0.44 & -0.33 & -0.027 & -0.81 \\ \hline
$c_{Z\square}$ 		&         &   & 1 & -0.077 & -0.20 & 0.013 & -0.0018 & 0.13 & 0.24 & 0.15 & 0.015 & 0.86 \\ \hline
$c_{\gamma\gamma}$     &          &   &   & 1 & 0.057 & -0.13 & 0.13 & -0.10 & -0.16 & -0.13 & -0.034 & -0.069 \\ \hline
$c_{Z\gamma}$   	        &          &   &   &   & 1 & -0.013 & -0.0038 & -0.15 & -0.27 & -0.20 & -0.015 & -0.081 \\ \hline
$c_{gg}$             		&         &   &   &   &   & 1 & -0.99 & 0.026 & 0.049 & 0.036 & 0.16 & 0.012 \\ \hline
$\delta y_t$        		&        &   &   &   &   &   & 1 & 0.020 & 0.034 & 0.027 & -0.16 & 0.0008 \\ \hline
$\delta y_c$       		&        &   &   &   &   &   &   & 1 & 0.55 & 0.43 & 0.023 & 0.13 \\ \hline
$ \delta y_b$  		         &       &   &   &   &   &   &   &   & 1 & 0.75 & 0.038 & 0.25 \\ \hline
$ \delta y_\tau$  		&       &   &   &   &   &   &   &   &   & 1 & 0.031 & 0.17 \\ \hline
$ \delta y_\mu$      		&       &   &   &   &   &   &   &   &   &   & 1 & 0.016 \\ \hline
$\lambda_Z$      		&      &   &   &   &   &   &   &   &   &   &   & 1 \\ \hline
\end{tabular}
\caption{The corresponding correlation matrix for the FCC-ee one sigma bounds in \autoref{tab:result}.}
\label{tab:rhofcc}
\end{table}
\begin{table}[ht!]\scriptsize
\centering
\begin{tabular}{|c||c|c|c|c|c|c|c|c|c|c|c|c|} 
  \multicolumn{13}{c}{correlation matrix, ILC} \\  \hline
 & $\delta c_Z$ & $c_{ZZ}$ & $c_{Z\square}$ & $c_{\gamma\gamma}$& $c_{Z\gamma}$ & $c_{gg}$  & $\delta y_t$  & $\delta y_c$ & $\delta y_b$ & $\delta y_\tau$ & $\delta y_\mu$  & $\lambda_Z$ \\ \hline\hline
$\delta c_Z$ 			&     1 & -0.40 & 0.27 & 0.039 & -0.26 & 0.020 & 0.0087 & 0.091 & -0.059 & 0.13 & -0.0085 & -0.069 \\ \hline
$c_{ZZ}$      		        &         & 1 & -0.89 & 0.071 & 0.36 & -0.027 & 0.0020 & -0.13 & -0.29 & -0.24 & -0.014 & 0.23 \\ \hline
$c_{Z\square}$ 		&        &   & 1 & -0.020 & 0.08 & 0.019 & -0.00026 & 0.087 & 0.22 & 0.15 & 0.0094 & -0.25 \\ \hline
$c_{\gamma\gamma}$     &         &   &   & 1 & -0.024 & -0.10 & 0.13 & -0.030 & -0.046 & -0.043 & -0.025 & 0.0061  \\ \hline
$c_{Z\gamma}$   	        &       &   &   &   & 1 & -0.0049 & -0.013 & -0.084 & -0.17 & -0.20 & -0.0066 & 0.0096 \\ \hline
$c_{gg}$             		&       &   &   &   &   & 1 & -0.98 & 0.029 & 0.081 & 0.046 & 0.12 & -0.0049  \\ \hline
$\delta y_t$        		&       &   &   &   &   &   & 1 & 0.014 & 0.022 & 0.019 & -0.13 & -0.0000 \\ \hline
$\delta y_c$       		&       &   &   &   &   &   &   & 1 & 0.40 & 0.26 & 0.013 & -0.023  \\ \hline
$ \delta y_b$  		        &       &   &   &   &   &   &   &   & 1 & 0.60 & 0.024 & -0.058 \\ \hline
$ \delta y_\tau$  		&       &   &   &   &   &   &   &   &   & 1 & 0.019 & -0.039   \\ \hline
$ \delta y_\mu$      		&     &   &   &   &   &   &   &   &   &   & 1 & -0.0025 \\ \hline
$\lambda_Z$      		&      &   &   &   &   &   &   &   &   &   &   & 1 \\ \hline
\end{tabular}
\caption{The corresponding correlation matrix for the ILC one sigma bounds in \autoref{tab:result}.}
\label{tab:rhoilc}
\end{table}
\begin{table}[ht!]\scriptsize
\centering
\begin{tabular}{|c||c|c|c|c|c|c|c|c|c|c|c|c|} 
  \multicolumn{12}{c}{correlation matrix, CLIC} \\  \hline
 & $\delta c_Z$ & $c_{ZZ}$ & $c_{Z\square}$ & $c_{\gamma\gamma}$ & $c_{Z\gamma}$ & $c_{gg}$ & $\delta y_t$ & $\delta y_c$ & $\delta y_b$ & $\delta y_\tau$ & $\delta y_\mu$  & $\lambda_Z$ \\ \hline\hline
$\delta c_Z$ 			&      1 & -0.0065 & -0.14 & 0.089 & -0.17 & 0.088 & 0.048 & 0.36 & 0.30 & 0.49 & -0.019 & -0.013 \\ \hline
$c_{ZZ}$      		        &         & 1 & -0.46 & 0.051 & 0.73 & -0.013 & -0.0070 & -0.054 & -0.13 & -0.083 & -0.0014 & -0.0083 \\ \hline
$c_{Z\square}$ 		&          &   & 1 & 0.022 & 0.14 & -0.0014 & 0.0064 & 0.0027 & -0.073 & 0.0050 & -0.0054 & 0.020 \\ \hline
$c_{\gamma\gamma}$     &           &   &   & 1 & -0.023 & -0.041 & 0.14 & -0.032 & 0.013 & -0.052 & -0.043 & 0.0032  \\ \hline
$c_{Z\gamma}$   	        &            &   &   &   & 1 & -0.012 & -0.015 & -0.056 & -0.22 & -0.084 & -0.0033 & 0.027 \\ \hline
$c_{gg}$             		&           &   &   &   &   & 1 & -0.95 & 0.013 & 0.10 & 0.016 & 0.11 & -0.0004 \\ \hline
$\delta y_t$        		&          &   &   &   &   &   & 1 & 0.023 & 0.035 & 0.031 & -0.13 & 0.0005 \\ \hline
$\delta y_c$       		 &         &   &   &   &   &   &   & 1 & 0.30 & 0.095 & 0.011 & -0.0008 \\ \hline
$ \delta y_b$  		         &          &   &   &   &   &   &   &   & 1 & 0.42 & 0.0057 & -0.0082  \\ \hline
$ \delta y_\tau$  		&           &   &   &   &   &   &   &   &   & 1 & 0.016 & -0.0012 \\ \hline
$ \delta y_\mu$      		&      &   &   &   &   &   &   &   &   &   & 1 & -0.0005 \\ \hline
$\lambda_Z$      		&      &   &   &   &   &   &   &   &   &   &   & 1 \\ \hline
\end{tabular}
\caption{The corresponding correlation matrix for the CLIC one sigma bounds in \autoref{tab:result}.}
\label{tab:rhoclic}
\end{table}
%

\clearpage

\providecommand{\href}[2]{#2}\begingroup\raggedright\endgroup

\end{document}